\documentclass[12pt]{article}

\pdfoutput=1 
\usepackage{graphicx} 
\usepackage{color}                                                                                                                                                                                                                                                                                                                                                                                                                                                                                                                                                                                                                                                                                                                                                                                                                                                                                                                                                                                                                        
\usepackage{hyperref}                                                                                                                                                                                                                                                                                                                                                                                                                                                                                                                                                                                                                                                                                                                                                                                                                                                                                                                                                                                                                                                                                                                                                                                                                                                                                                                                                                                                                                                                                                                                                                                                                                                                                                                                                                                                                                                                                                                                                                                                                                                                                                                                                                                                                                                                                                                                                                                                                                                                                                                                                                                                                                                                                                                                                                                                                                                                                                                                                                                                                                                                                                                                                                                                                                                                                                                                                                                                                                                                                                                                                                                                                                                                                                                                                                                                                                                                                                                                                                                                                                                                                                                                                                                                                                          
\usepackage{cancel,tabularx,moreverb,fancybox,amsmath,float,bm,braket,slashbox,txfonts,amssymb,bm,accents}
\usepackage[top=30truemm,bottom=30truemm,left=25truemm,right=25truemm]{geometry}
\usepackage{latexsym}
\usepackage{here}

\newcommand{\ex}[1]{\mathrm{e}^{#1}}

\newcommand{\pa}[1]{\left(#1 \right)}

\newcommand{\BR}[1]{\Biggl[#1 \Biggr]}

\newcommand{\ca}[1]{\mathcal{#1}}

\newcommand{\abs}[1]{\left|#1\right|}
\newcommand{\ave}[1]{\langle #1\rangle}
\newcommand{\ar}[1]{\xrightarrow[#1]{}}

\newcommand{\lleq}[1]{\underaccent{\ \ \  #1}{\ \ \leq \ \ }}

\newcommand{\ti}[1]{\tilde{#1}}

\newcommand{\fr}{\frac}
\newcommand{\s}[1]{\sqrt{#1}}

\def\be{\begin{equation}}
\def\ee{\end{equation}}
\def\ba{\begin{eqnarray}}
\def\ea{\end{eqnarray}}

\def\m{{\mu}}
\def\la{{\lambda}}
 
 \def\n{{\nu}}
 \def\ep{{\epsilon}}
 \def\d{{\delta}}

 \def\a{{\alpha}}
 
 \def\l{{\lambda}}
 
 \def\D{{\Delta}}
 \def\g{{\gamma}}
 
 \def\b{{\beta}}
 \def\e{{\epsilon}}
 
 \def\L{{\Lambda}}

\def\dd{{\mathrm{d}}}
\def\sgn{{\text{sgn}}}

  \makeatletter
    
    \@addtoreset{equation}{section}
  \makeatother
\begin{document}

\begin{titlepage}
\thispagestyle{empty}

\begin{flushright}
YITP-18-22
\\

\end{flushright}

\bigskip

\begin{center}
\noindent{{\large \textbf{
New Properties of Large-$c$ Conformal Blocks\\
from Recursion Relation
}}}\\
\vspace{2cm}
Yuya Kusuki
\vspace{1cm}

{\it
Center for Gravitational Physics, \\
Yukawa Institute for Theoretical Physics (YITP), Kyoto University, \\
Kitashirakawa Oiwakecho, Sakyo-ku, Kyoto 606-8502, Japan
}
\vskip 2em
\end{center}

\begin{abstract}
We study large $c$ conformal blocks outside the known limits. This work seems to be hard, but it is possible numerically by using the Zamolodchikov recursion relation.
As a result, we find new some properties of large $c$ conformal blocks with a pair of two different dimensions for any channel and  with various internal dimensions. With light intermediate states, we find a Cardy-like asymptotic formula for large $c$ conformal blocks and also we find that the qualitative behavior of various large $c$ blocks drastically changes when  the dimensions of external primary states reach the value $c/32$. And we proceed to the study of blocks with heavy intermediate states $h_p$ and we find some simple dependence on heavy $h_p$ for large $c$ blocks. The results in this paper can be applied to, for example, the calculation of OTOC or Entanglement Entropy. In the end, we comment on the application to the conformal bootstrap in large $c$ CFTs.
\end{abstract}
 
\end{titlepage}

\restoregeometry

\tableofcontents

%%%%%%%%%%%%%%%%%%%%%%%%%%%%%%%%%%%%%%%%%%%%%%%%%%%%%%%%%%%%%%%%%%%%%%%%%%%%%%%%%%%%%%%%%%%%%%
%%%%%%%%%%%%%%%%%%%%%%%%%%%%%%%%%%%%%%%%%%%%%%%%%%%%%%%%%%%%%%%%%%%%%%%%%%%%%%%%%%%%%%%%%%%%%%
\section{Introduction \& Summary}
%%%%%%%%%%%%%%%%%%%%%%%%%%%%%%%%%%%%%%%%%%%%%%%%%%%%%%%%%%%%%%%%%%%%%%%%%%%%%%%%%%%%%%%%%%%%%%
%%%%%%%%%%%%%%%%%%%%%%%%%%%%%%%%%%%%%%%%%%%%%%%%%%%%%%%%%%%%%%%%%%%%%%%%%%%%%%%%%%%%%%%%%%%%%%
Conformal Field Theories (CFTs) in two dimensions have infinite symmetry group and, as a result, 2d CFTs are perfectly specified by a central charge, operator spectrum and OPE coefficients. Moreover, possible CFT data are limited by crossing symmetry and modular invariance, which come from the consistency requirements of CFTs. Recently the bootstrap program, which is based on crossing symmetry or modular invariance, attracts attention to classify CFTs  \cite{Ferrara1973, Polyakov1974, Belavin1984,Rattazzi2008}. Once we have CFT data, we can construct all the correlators in the CFT by taking a sum of conformal blocks weighted by the OPE coefficients. The conformal blocks correspond to a virtual exchange of a primary operator and its descendants, which are completely determined by conformal symmetry, that is, by using Virasolo algebra in principle  \cite{Belavin1984}. However,  we do not know the simple closed form of conformal blocks, except in special cases. Only recursion relations for conformal blocks are known  \cite{Zamolodchikov1984,Zamolodchikov1987}, which are very complicated. Therefore, we have not made much progress on the study of conformal blocks, despite decades of effort.

Conformal blocks play a very important role in some scenarios. For example, to solve the bootstrap program in an unknown CFT,  one has to know conformal blocks with a central charge and conformal dimensions in the CFT. This conformal bootstrap equation can be described by
\begin{equation}
\sum_p C_{12p}C_{34p} \ca{F}^{21}_{34}(h_p|z)\overline{\ca{F}^{21}_{34}}(\bar{h}_p|\bar{z})=\sum_p C_{14p}C_{23p} \ca{F}^{41}_{32}(h_p|1-z)\overline{\ca{F}^{41}_{32}}(\bar{h}_p|1-\bar{z}),
\end{equation}
where $C_{ijk}$ are OPE coefficients and $\ca{F}^{ij}_{kl}(h_p|z)$ are conformal blocks, which are usually expressed by using the Feynman diagram as

\newsavebox{\boxpc}
\sbox{\boxpc}{\includegraphics[width=130pt]{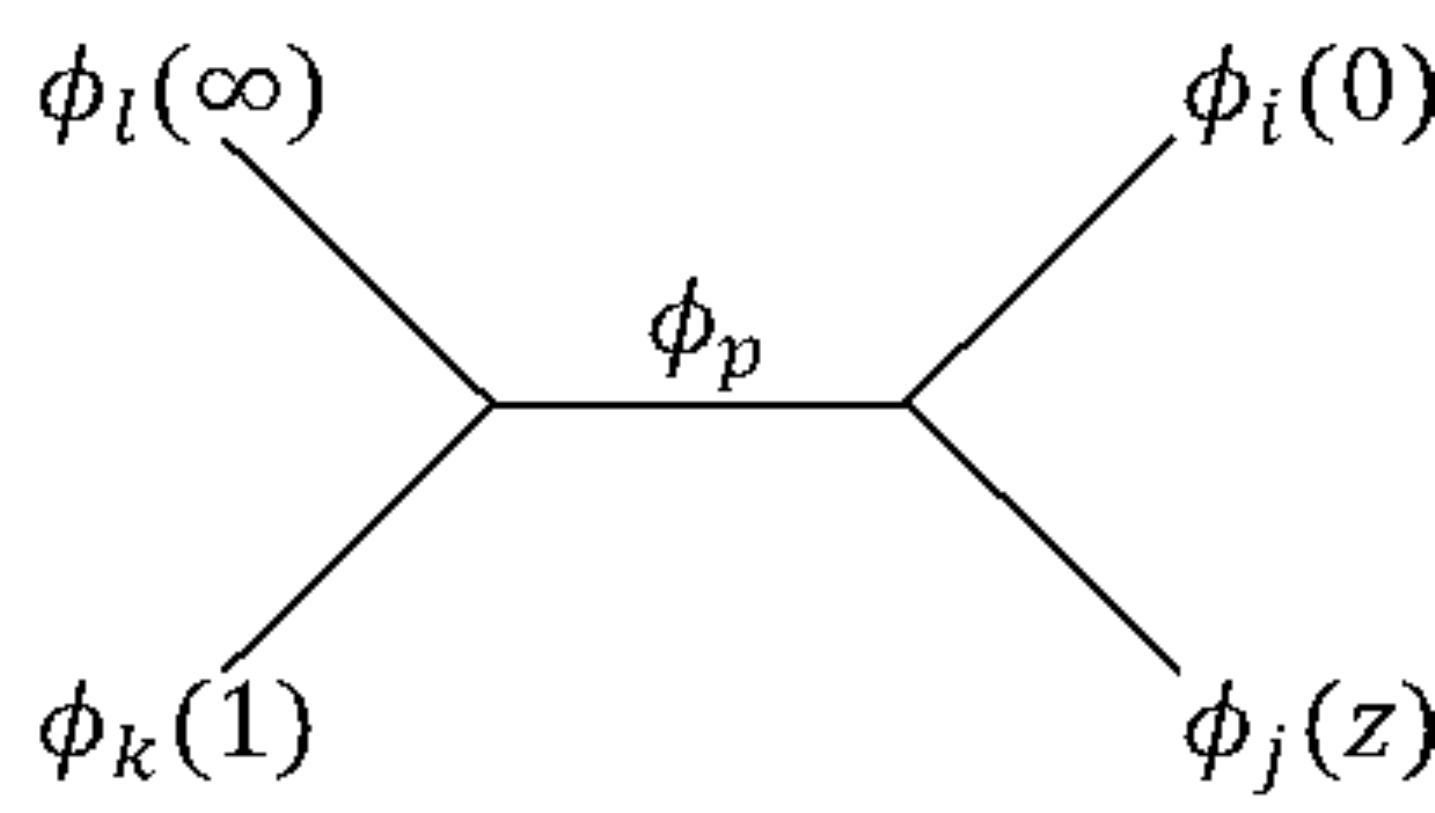}}
\newlength{\pcw}
\settowidth{\pcw}{\usebox{\boxpc}} 

\begin{equation*}
 \ca{F}^{ji}_{kl}(h_p|z) \equiv \parbox{\pcw}{\usebox{\boxpc}}.
\end{equation*}
And also in the context of AdS/CFT correspondence, conformal blocks receive attention recently \cite{Heemskerk2009,Heemskerk2010,El-Showk2012,Fitzpatrick2013,Fitzpatrick2013a,Hijano2016} and, in particular,  in AdS$_3$/CFT$_2$, the  semiclasical Virasoro blocks have been used to probe information loss, which appears  in CFT$_2$ as forbidden singularities and exponential decay at late times \cite{Fitzpatrick2014,Fitzpatrick2015,Fitzpatrick2016,Fitzpatrick2017}. This semiclassical blocks can be computed in the dual AdS$_3$ gravity  \cite{Alkalaev2015,Hijano2015,Hijano2015a,Alkalaev2016}. Some other progresses attributed to conformal blocks are the study of the dynamics of the Renyi entropy \cite{Caputa2015, Asplund2015,Kusuki2018} and out-of-time-ordered correlators (OTOCs)  \cite{Roberts2015}.

In this paper, we focus on the letter context, that is, we consider the CFT which is the dual of Einstein gravity in AdS$_3$, called { \it holographic CFT}. Unfortunately, there is a little known data for the holographic CFT for now. Nevertheless, we know that some constrains on a CFT data can be given by using the bootstrap, for example, the bound on spectrum density  \cite{Hartman2014}, the bound on the dimension of a first excited state  \cite{Hellerman2009,Friedan2013,Collier2016} and the universal formula for OPE coefficients  \cite{Pappadopulo2012,Kraus2016,Cardy2017,Keller2017,Cho2017}. We can extract CFT data from conformal blocks.
\footnote{Here, we interpret the modular bootstrap as a kind of the conformal bootstrap because the torus partition function can be explicitly given by the 4-pt function of twist-2 operators.}

We are interested in large $c$ conformal blocks because it is known that the holographic CFT has a large central charge. Actually in some special limits on external and internal dimensions of blocks, we have useful expressions of large $c$ conformal blocks. However, if one would try to go beyond the limits, even if focusing on the holographic CFT, no simple expression for conformal blocks is found. Nevertheless we can study any conformal block by using the {\it Zamolodchikov recursion relation} \cite{Zamolodchikov1984, Zamolodchikov1987}
\footnote{A good review is given by  \cite{Perlmutter2015}, which also explains the relation between various recursion relations. A generalization of the recursion relation to more general Riemann surfaces is given in \cite{Cho2017a} }.  
Recently this recursion relation is used to probe information loss non-perturbatively in central charge \cite{Chen2017}, and it shows that the exact conformal blocks in the $O_A O_A \to O_B O_B$ channel decay as $t^{-\fr{3}{2}}$ at late times, which is quite different behavior from the semiclassical block. It means that a non-perturbative correction in central charge is very important when one would try to probe information loss by using large $c$ conformal blocks.

In our recent paper  \cite{Kusuki2018}, we study large $c$ vacuum conformal blocks for the correlator $\braket{O_B(\infty)O_B(1)O_A(z)O_A(0)}$ in the  $O_A O_A \to O_B O_B$ channel. And we find that the qualitative behavior of large $c$ conformal blocks drastically changes  at $h_{A,B}=\fr{c}{32}$. This statement is interesting both in physical and mathematical contexts, for example, conformal bootstrap, physical meaning of this transition  and so on. And moreover we find the simple asymptotic form of the conformal blocks. More information are in \cite{Kusuki2018} and briefly summarized in  Section \ref{subsec:vacAABB}.

In  \cite{Kusuki2018}, we focused only on vacuum blocks for the correlator $\braket{O_B(\infty)O_B(1)O_A(z)O_A(0)}$ in the  $O_A O_A \to O_B O_B$ channel. But it is also interesting to investigate (i) whether similar properties also hold for conformal blocks with a non-zero (in particular, heavy) intermediate dimension and  (ii) whether the similar transition occurs in {\it ABBA block}, which is the Virasoro block for the correlator  $\braket{O_A(\infty)O_B(1)O_B(z)O_A(0)}$ in the $O_B(z)O_A(0)$ OPE channel, in that, 

\newsavebox{\boxpd}
\sbox{\boxpd}{\includegraphics[width=130pt]{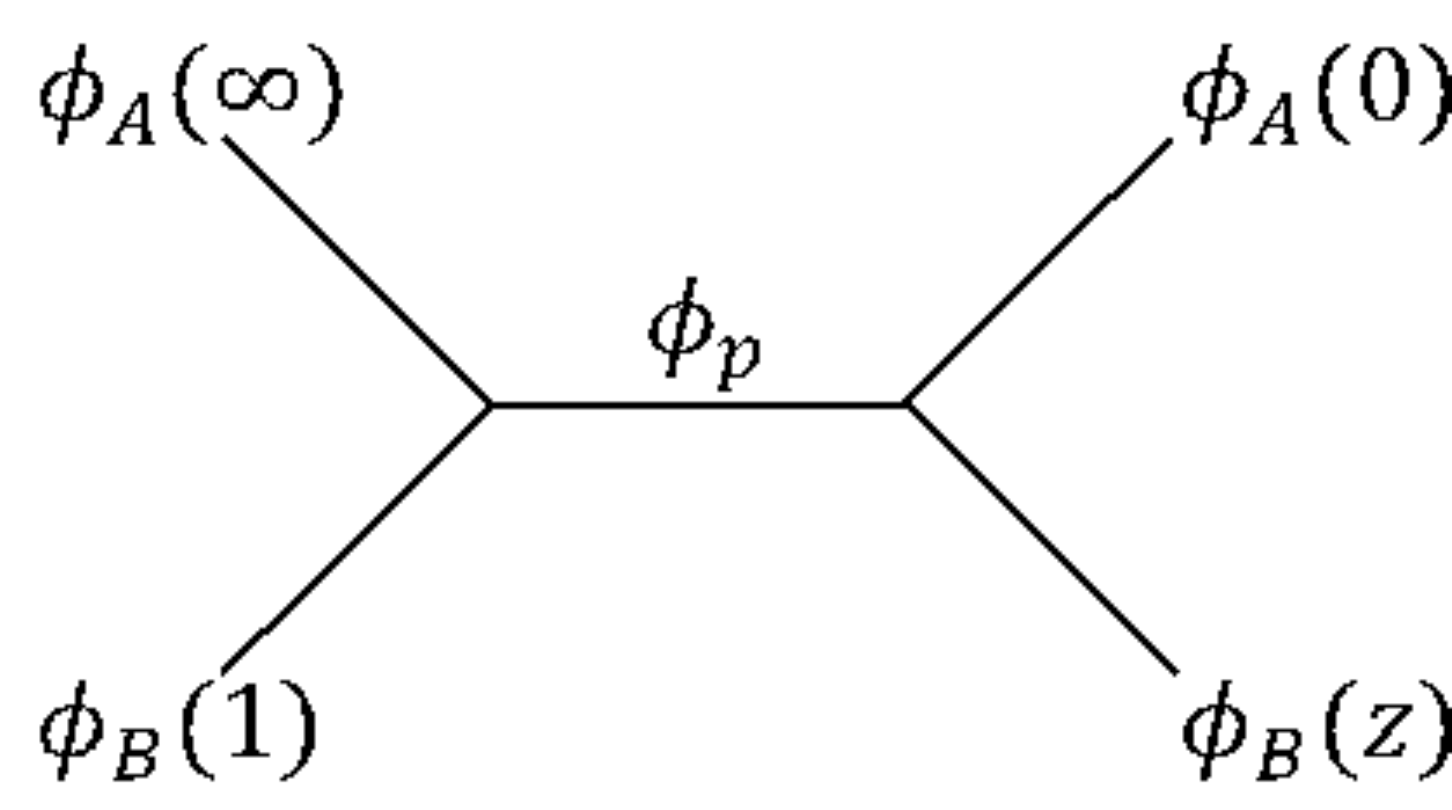}}
\newlength{\pdw}
\settowidth{\pdw}{\usebox{\boxpd}} 

\begin{equation*}
 \ca{F}^{BA}_{BA}(h_p|z) \equiv \parbox{\pdw}{\usebox{\boxpd}},
\end{equation*}
which is important because this channel also appears in the bootstrap equation for the correlator  $\braket{O_B(\infty)O_B(1)O_A(z)O_A(0)}$ as well as the  $O_A O_A \to O_B O_B$ channel. Note that in contrast with this ABBA block, we call the following blocks as {\it AABB blocks},
\newsavebox{\boxpa}
\sbox{\boxpa}{\includegraphics[width=130pt]{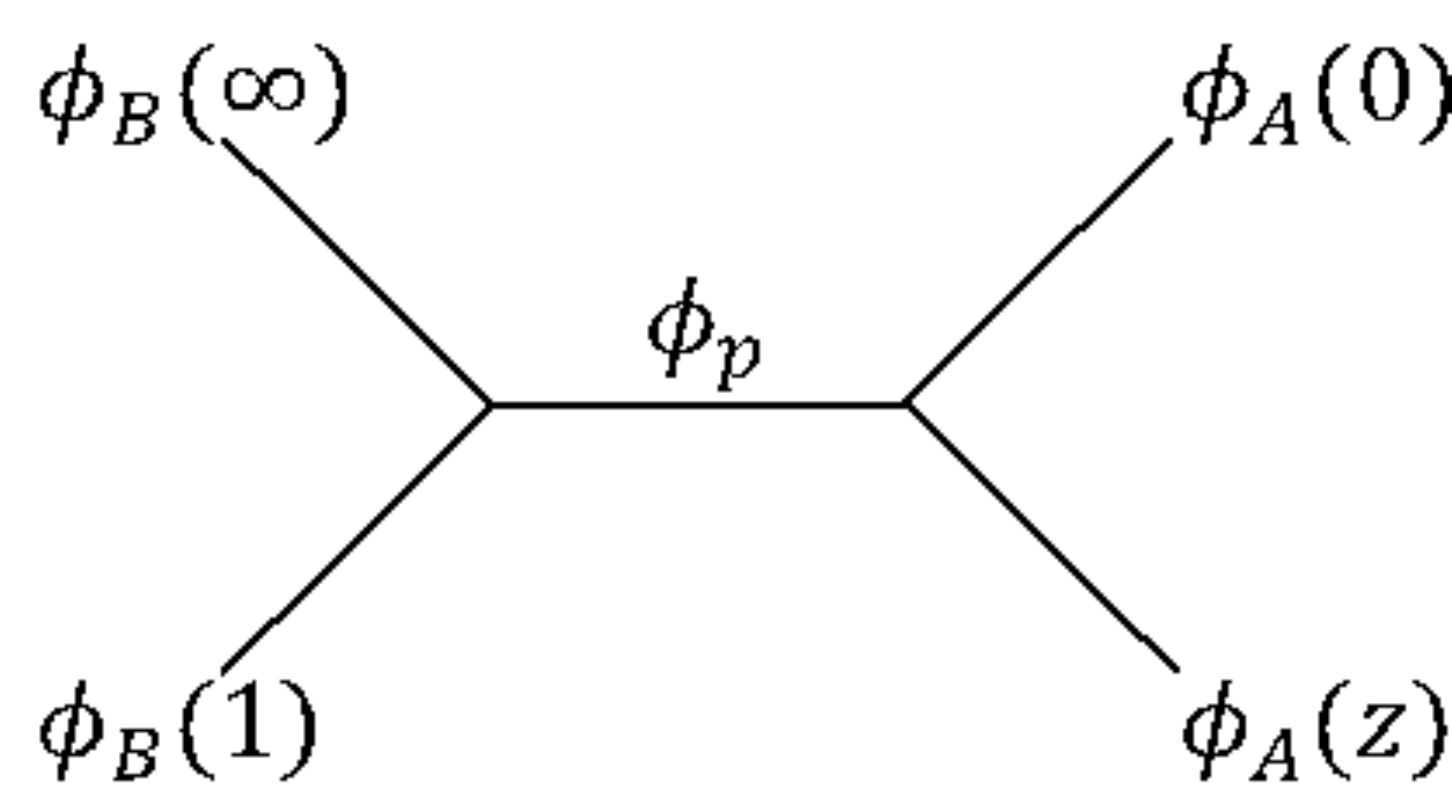}}
\newlength{\paw}
\settowidth{\paw}{\usebox{\boxpa}} 

\begin{equation*}
 \ca{F}^{AA}_{BB}(h_p|z) \equiv \parbox{\paw}{\usebox{\boxpa}},
\end{equation*}
and as {\it ABAB blocks}, we define
\newsavebox{\boxpg}
\sbox{\boxpg}{\includegraphics[width=130pt]{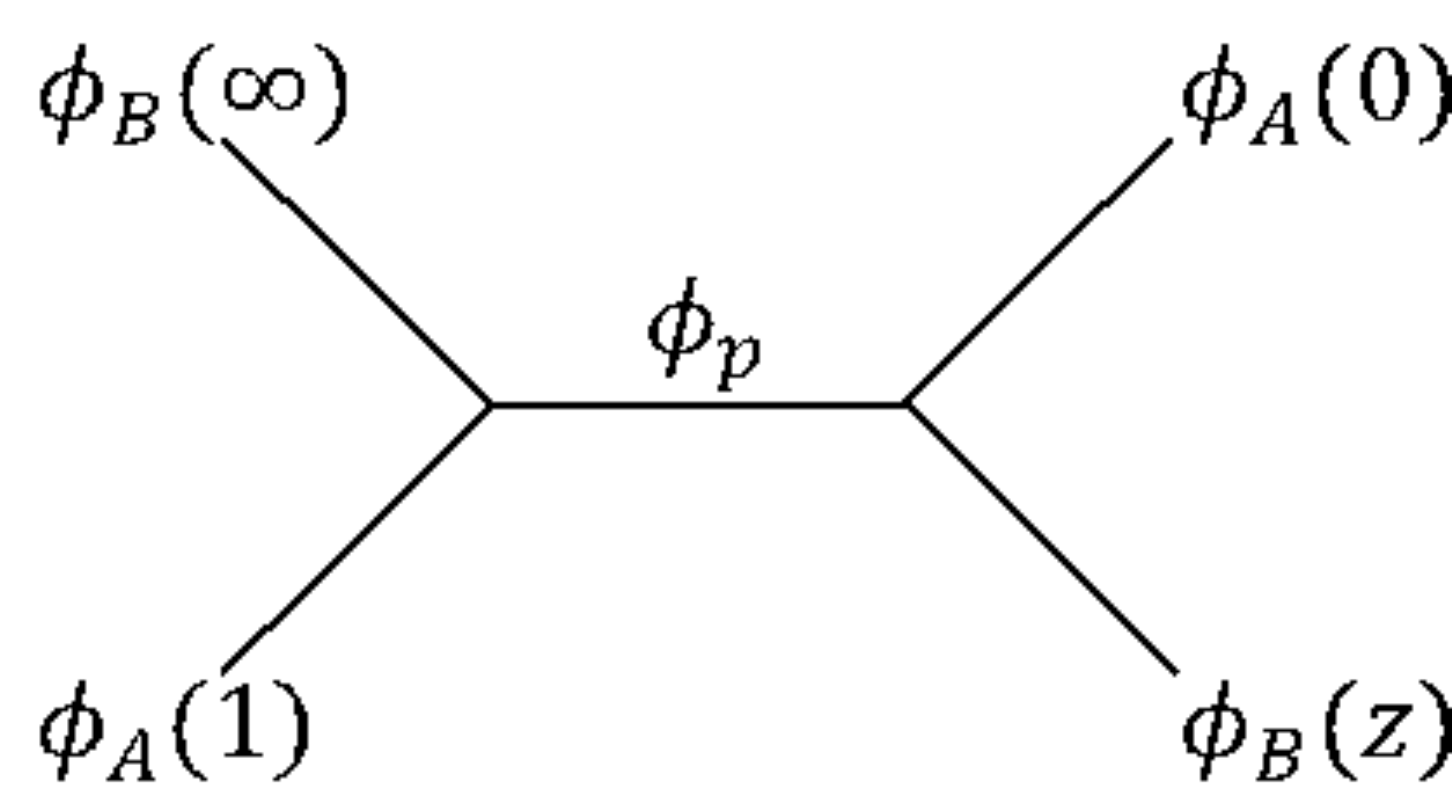}}
\newlength{\pgw}
\settowidth{\pgw}{\usebox{\boxpg}} 

\begin{equation*}
 \ca{F}^{BA}_{AB}(h_p|z) \equiv \parbox{\pgw}{\usebox{\boxpg}}.
\end{equation*}

In this paper, we try to address these questions, (i) and (ii). In the following, we summarize our main results: 

\begin{quote}
{\large \it Result (1):}
\\
Denoting ABBA blocks by
\begin{equation}
\ca{F}^{BA}_{BA}(h_p|z)=\Lambda(z)\sum_n c_n q^n,
\end{equation}
where $\Lambda(z)$ is an universal prefactor and $q$ is the elliptic nome defined as
\begin{equation}
q(z)=\ex{-\pi \fr{K(1-z)}{K(z)}},
\end{equation}
where $K(z)$ is an elliptic integral of the first kind, then we can see that the coefficients $c_n$ behave like
\footnote{In this paper, ``$\simeq$'' means an approximation by extracting a leading contribution and ``$\sim$'' means an approximation up to a constant factor.}
	\begin{equation}\label{eq:introcn}
		c_n\sim n^{\a} \ex{A \s{n}} \ \ \ \ \text{for } n \gg c,
	\end{equation}
where $c_n$ are always positive, which are different from those of AABB blocks.
(See Figure \ref{fig:HHLL-hAhBdep}, which shows the sign pattern of $c_n$ of AABB blocks.) 

For $h_p \ll c$, the values of $A$ and $\a$ in (\ref{eq:introcn}) are given by
\begin{enumerate}
\item In the {\it heavy-heavy} region ($h_A>\fr{c}{32} $ and $h_B>\fr{c}{32}$),
\begin{equation}\label{eq:AaHH}
\begin{aligned}
A&=0,\\
\a&=4(h_A+h_B)-\fr{c+9}{4}.
\end{aligned}
\end{equation}

\item In the region where any $h_A$ and $h_B \ll c$,
\begin{equation}\label{eq:AaHL}
\begin{aligned}
A&=2\pi \sqrt{ \frac{c-1}{24}-4h_B+\fr{c-1}{6}\pa{1-\s{1-\fr{24}{c-1}h_B}}},\\
\a&=2(h_A+h_B)-\fr{c+5}{8}.
\end{aligned}
\end{equation}
\end{enumerate}
Here we can assume $h_A>h_B$ without loss of generality because the coefficients $c_n$ is symmetric under $h_A \leftrightarrow h_B$ (see (\ref{eq:symhAhB})).
What we would like to emphasize here is that there is a transition of ABBA blocks at $h_{A,B}=\fr{c}{32}$ in the same way as AABB blocks. 
Accordingly, we will use the italic font for ``{\it heavy}'' to discriminate ``{\it heavy}'' from the usual definition of heavy, which means the order of $O(c)$. The italic font means larger than $\fr{c}{32}$. Similarly, we use ``{\it light}'' as smaller than $\fr{c}{32}$.
These results are summarized in Figure \ref{fig:HLHL-hAhBdep}.

Note that we can show a relation between the coefficients of ABBA blocks and ABAB blocks as
\begin{equation}
(c^{BA}_{AB})_n=(-1)^n (c^{BA}_{BA})_n.
\end{equation}
Therefore, it is straightforwardly shown that for ABAB blocks, the coefficients are given by
	\begin{equation}\label{eq:introcn2}
		c_n\sim  (-1)^n n^{\a} \ex{A \s{n}} \ \ \ \ \text{for } n \gg c
	\end{equation}
and the values of $A$ and $\a$ are given by the same expression as above.
\\
\\
\\
{\large \it Result (2):}

The conformal blocks with general intermediate dimension $h_p$ also have the simple asymptotic coefficients $c_n$ well-fitted by (\ref{eq:introcn}) and we find a transition at $h_{A,B}=\fr{c}{32}$ again. And moreover the values of $A$ and $\a$ of the coefficients $c_n$ are not sensitive to $h_p$ for large $n$. In other words, for any $h_p$, one can find one $N$ such that
\begin{equation}
\abs{c_n}
\sim \left\{
    \begin{array}{ll}
      n^\a   ,& \text{if } h_A, h_B > \fr{c}{32}  ,\\
     n^\a \ex{A \s{n}} \ \ \ \ (A>0)    ,& \text{otherwise }   ,\\
    \end{array}
  \right.\\
\ \ \ \ \ \text{for } n \gg N, c ,
\end{equation}
where $A$ and $\a$ are independent of $h_p$.
 These features are shown in Figure \ref{fig:HHLLhBhpdep} (for AABB blocks) and Figure \ref{fig:HLHLhBhpdep} (for ABBA blocks).
And actually we can suggest that this $N$ is of order $h_p$ from the form of the recursion relation (see Section \ref{subsec:AABBnon}).
For $n \ngg h_p$, we have the naive estimation both for AABB and ABBA blocks as
\begin{enumerate}
\item In the {\it heavy-heavy} region ($h_A>\fr{c}{32} $ and $h_B>\fr{c}{32}$),

\begin{equation}
\abs{c_n(h_p)} \sim \pa{\fr{1}{h_p}}^{const.} ,
\end{equation}

\item In the region where any $h_A$ and $h_B \ll \fr{c}{32}$,
\begin{equation}
\begin{aligned}
\abs{c_n(h_p)}& \sim \left\{
    \begin{array}{ll}
     const. \ \   ,& \text{if } h_p \lesssim n  ,\\
     \pa{\fr{1}{h_p}}^{const.}  ,& \text{if } h_p \gtrsim n  ,\\
    \end{array}
  \right.\\
\end{aligned}
\end{equation}
\end{enumerate}
which are also based on the recursion relation.
\\
\\
\\
{\large \it Result (3):}

As one of the applications of our results, the conformal bootstrap leads to the bound on the three point coefficients as
\begin{enumerate}
\item In the {\it heavy-heavy} region ($h_A>\fr{c}{32} $ and $h_B>\fr{c}{32}$),
\begin{equation}
\overline{C_{ABp}^2}  \ar{\D_p \to \infty} 16^{-\D_p} \ex{- 2\pi\s{\fr{c-1}{12}\pa{\D_p-\fr{c-1}{12}}}}.
\end{equation}

\item In the region where any $h_A$ and $h_B \ll \fr{c}{32}$,
\begin{equation}
\begin{aligned}
 16^{-\D_p} \ex{4\pi\s{\pa{\D_B-\fr{c-1}{12}\pa{1-\s{1-\fr{12}{c-1}\D_B}}}\pa{\D_p-\fr{c-1}{12}}} - 4\pi\s{\fr{c-1}{12}\pa{\D_p-\fr{c-1}{12}}}}
&\lleq{\D_p \to \infty} \overline{C_{ABp}^2} \\ 
&\lleq{\D_p \to \infty}  16^{-\D_p} \ex{- 2\pi\s{\fr{c-1}{12}\pa{\D_p-\fr{c-1}{12}}}}.
\end{aligned}
\end{equation}
\end{enumerate}
\end{quote}
Here the mean squared is over all primary operators of fixed dimension $\D_p$.

Note that the exponential suppression in the upper bound can be characterized by the entropy $S(E)= 2\pi \sqrt{\fr{c}{3} E}$ as $\ex{-\fr{1}{2}S(E)}$, which appears in the Cardy formula \cite{Cardy1986a}. This exponential suppression also can be seen in the asymptotics of the heavy-heavy-heavy OPE coefficients \cite{Cardy2017},
\begin{equation}
\overline{C^2_{ABC}}\ar{\D_A=\D_B=\D_C=\D_p\to\infty}\pa{\fr{16}{27}}^{-3\D_p} \ex{-\fr{1}{2} 3S(\D_p)},
\end{equation}
which is given by the modular bootstrap for 0-point correlators on a 2-genus surface (see also \cite{Keller2017,Cho2017}).
In addition, other universal formulas for OPE coefficients are also given by using the bootstrap approach for 1-point correlators on a torus \cite{Kraus2016} and 2-point correlators on a torus \cite{Hikida2018}.
The result in \cite{Hikida2018}  gives the heavy-heavy-light OPE coefficients as
\begin{equation}\label{eq:CAB}
\overline{C^2_{Apq}}\ar{\D_p=\D_q\to\infty} \ex{-\fr{1}{2} 2S(\D_p)}.
\end{equation}
One can again see the exponential suppression characterized by the entropy in this form.
\footnote{
 In \cite{Kraus2016}, one can also find the asymptotics of the heavy-heavy-light OPE coefficients. However, the result given in \cite{Kraus2016} is taken a mean squared over all primary operators of fixed dimension $\D_p$ with $A=B$ as microstates, which is different from (\ref{eq:CAB}).
}
Note also that this asymptotic behavior is different from the asymptotics for the OPE coefficients for any states \cite{PappadopuloRychkovEspinRattazzi2012} (see Section \ref{sec:descendants}). It is natural because the above result (3) is the mean over only primary states (and also the normalization is different).

From our results, we can construct the asymptotic form of large $c$ conformal blocks. When one considers the bootstrap equation for the correlator $\braket{O_B(\infty)O_B(1)O_A(z)O_A(0)}$, one has to know both AABB blocks and ABBA (correctly, not ABBA but BAAB)  blocks. Our results suggest that both AABB and ABBA large $c$ conformal blocks have simple form, therefore we expect that the conformal bootstrap in the holographic CFT can be solved by using our analysis or more information derived by the recursion relation in a similar way.

The outline of this paper is as follows. In Section \ref{sec:rec}, we review the Zamolodchikov recursion relation, which is a key tool of our strategy. In Section \ref{sec:AABB}, we revisit AABB blocks with vacuum intermediate states and moreover we extend our analysis to blocks with general intermediate states. In Section \ref{sec:ABBA}, we study ABBA blocks with light and heavy intermediate states. We  extract the simple properties of ABBA blocks and find similar transition to AABB blocks. However, in some cases, we can not find simple formula for blocks with heavy intermediate states. Nevertheless, we manage to extract some properties of the blocks with heavy intermediate states in Section \ref{sec:heavy}. In Section \ref{sec:block}, we apply our results in Section \ref{sec:AABB}, \ref{sec:ABBA} to estimating the asymptotic form of conformal blocks in some limits. In Section \ref{sec:OTOC}, we discuss correlators reproduced by our conformal blocks. In Section \ref{sec:bootstrap}, we comment on the future direction and application to the conformal bootstrap. We conclude with a discussion in Section \ref{sec:dis}. In Appendix \ref{app:fitting}, we show some more detailed numerical data from the Zamolodchikov recursion relation. In Appendix \ref{app:comparing}, we compare our results with the semiclassical blocks and comment on the consistency and inconsistency between them.  In Appendix \ref{sec:descendants}, we derive the asymptotic heavy-light-light coefficients with arbitrary operators.
%%%%%%%%%%%%%%%%%%%%%%%%%%%%%%%%%%%%%%%%%%%%%%%%%%%%%%%%%%%%%%%%%%%%%%%%%%%%%%%%%%%%%%%%%%%%%%
%%%%%%%%%%%%%%%%%%%%%%%%%%%%%%%%%%%%%%%%%%%%%%%%%%%%%%%%%%%%%%%%%%%%%%%%%%%%%%%%%%%%%%%%%%%%%%
\section{Recursion Relations for Conformal Blocks}\label{sec:rec}
%%%%%%%%%%%%%%%%%%%%%%%%%%%%%%%%%%%%%%%%%%%%%%%%%%%%%%%%%%%%%%%%%%%%%%%%%%%%%%%%%%%%%%%%%%%%%%
%%%%%%%%%%%%%%%%%%%%%%%%%%%%%%%%%%%%%%%%%%%%%%%%%%%%%%%%%%%%%%%%%%%%%%%%%%%%%%%%%%%%%%%%%%%%%%

Conformal blocks appear in the decomposition of the correlators as follows,
\begin{equation}
\braket{O_4(\infty)O_3(1)O_2(z,\bar{z})O_1(0)}=\sum_p C_{12p}C_{34p} \ca{F}^{21}_{34}(h_p|z)\overline{\ca{F}^{21}_{34}}(\bar{h}_p|\bar{z}),
\end{equation}
where the sum is taken over primary operators in the CFT. Conformal blocks $\ca{F}^{21}_{34}(h_p|z)$ can be split into two factors as
\begin{equation}
\ca{F}^{21}_{34}(h_p|z)=\Lambda^{21}_{34}(h_p|q)H^{21}_{34}(h_p|q),\ \ \ \ \ \ q(z)=\ex{-\pi \fr{K(1-z)}{K(z)}},
\end{equation}
where the function $\Lambda^{21}_{34}(h_p|q)$ is a universal prefactor, which is given by
\begin{equation}\label{eq:pre}
 \Lambda^{21}_{34}(h_p|q)=(16q)^{h_p-\frac{c-1}{24}}z^{\frac{c-1}{24}-h_1-h_2}(1-z)^{\frac{c-1}{24}-h_2-h_3}
(\theta_3(q))^{\frac{c-1}{2}-4(h_1+h_2+h_3+h_4)}
\end{equation}
and the function $H^{21}_{34}(h_p|q)$ can be calculated recursively by using the following relation,
\begin{equation}\label{eq:Hdef}
H^{21}_{34}(h_p|q)=1+\sum^\infty_{m=1,n=1}\frac{q^{mn}R_{m,n}}{h_p-h_{m,n}}H^{21}_{34}(h_{m,n}+mn|q),
\end{equation}
where $h_{m,n}$ is the zero of Kac determinant, that is,
\begin{equation}
\begin{aligned}
h_{m,n}&=\fr{1}{4}\pa{b+\fr{1}{b}}^2-\lambda_{m,n}^2,\\
\lambda_{m,n}&=\fr{1}{2} \pa{\fr{m}{b}+nb},
\end{aligned}
\end{equation}
and 
\begin{equation}\label{eq:Rmn}
R_{m,n}=2\fr{\prod_{p,q}\pa{\la_2+\la_1-\la_{p,q}}\pa{\la_2-\la_1-\la_{p,q}}\pa{\la_3+\la_4-\la_{p,q}}\pa{\la_3-\la_4-\la_{p,q}}}
{\prod_{k,l} ' \la_{k,l}}.
\end{equation}
Here the integers $p,q,k,l$ are defined as
\begin{equation}
\begin{aligned}
p&=-m+1,-m+3,\cdots,m-3,m-1,\\
q&=-n+1,-n+3,\cdots,n-3,n-1,\\
k&=-m+1,-m+2,\cdots,m,\\
l&=-n+1,-n+2,\cdots,n. 
\end{aligned}
\end{equation}
The product $\prod'_{k,l}$ in (\ref{eq:Rmn}) means that we exclude $(k,l)=(0,0)$ and $(m,n)$.
We also defined
\begin{equation}
\begin{aligned}
c&=1+\pa{b+\fr{1}{b}}^2,\\
h_i&=\fr{c-1}{24}-\l_i^2.
\end{aligned}
\end{equation}
In this paper, we consider a series expansion of the function $H^{21}_{34}(h_p|q)$ as
\be\label{eq:ckdef}
H^{21}_{34}(h_p|q)=1+\sum_{k=1}^\infty c_k(h_p) q^{k},
\ee
and focus on the series coefficients $c_k$. In the same way as (\ref{eq:Hdef}), we can also calculate the coefficients $c_k(h_p)$ recursively by the following relation,
\begin{equation}\label{eq:ck}
	c_k(h_p) = \sum_{i=1}^k \sum_{\substack{m=1, n=1\\mn=i}} \frac{R_{m,n}}{h_p - h_{m,n}} c_{k-i}(h_{m,n}+mn),
\end{equation}
where the sum is taken over $m,n=1,2,3,\cdots$ with $mn$ held fixed, i.e. the sum $\sum_{\substack{m=1, n=1\\mn=4}}$ means taking sum over $(m,n)=(1,4), (2,2) \text{ and } (4,1)$. The coefficient $c_{k}(h_{m,n}+mn)$ can be also calculated recursively by
\begin{equation}\label{eq:ckrec}
	c_k(h_{m,n}+mn) = \sum_{i=1}^k \sum_{\substack{\m=1, \n=1\\ \m\n=i}} \frac{R_{\m,\n}}{h_{m,n}+mn - h_{\m,\n}} c_{k-i}(h_{\m,\n}+\m\n),
\end{equation}
where the starting values of this recursion formula are $c_0(h_{m,n}+mn)=1$. 
\begin{quote}
{\it Examples:}
\begin{equation*}
\begin{aligned}
c_1(h_p)&=\fr{R_{1,1}}{h_p-h_{1,1}},\\
c_2(h_p)&=\fr{R_{1,1}^2}{h_p-h_{1,1}}+\fr{R_{2,1}}{h_p-h_{2,1}}+\fr{R_{1,2}}{h_p-h_{1,2}},\\
c_3(h_p)&=\fr{R_{1,1}}{h_p-h_{1,1}}\pa{R_{1,1}^2+\fr{R_{1,2}}{1+h_{1,1}-h_{1,2}}+\fr{R_{2,1}}{1+h_{1,1}-h_{2,1}}}\\
&+\fr{R_{1,2}}{h_p-h_{1,2}}\fr{R_{1,1}}{2-h_{1,1}+h_{1,2}}+\fr{R_{2,1}}{h_p-h_{2,1}}\fr{R_{1,1}}{2-h_{1,1}+h_{2,1}}+\fr{R_{1,3}}{h_p-h_{1,3}}+\fr{R_{3,1}}{h_p-h_{3,1}}.
\end{aligned}
\end{equation*}
\end{quote}

In the end of this section, we would like to comment on an important property of the function $H(h_p|q)$ and its coefficients $c_k(h_p)$. From the expression of $R_{m,n}$, we find that $R_{m,n}$ is symmetric under the exchange $h_1 h_2 \leftrightarrow h_3 h_4$, which leads to the symmetry of $H(h_p|z)$ and its coefficients $c_k(h_p)$ under the exchange $h_1 h_2 \leftrightarrow h_3 h_4$, that is,
\be\label{eq:symhAhB}
H^{21}_{34}(h_p|q)=H^{34}_{21}(h_p|q).
\ee
And also we can show from (\ref{eq:Rmn}),
\be
H^{21}_{34}(h_p|q)=H^{21}_{43}(h_p|-q),
\ee
or equivalently,
\be\label{eq:symq}
(c^{21}_{34})_n=(-1)^n (c^{21}_{43})_n.
\ee

Now that we have conformal blocks written by elliptic nome $q$, which appears in a torus partition function,
\footnote{One can understand why the elliptic nome appears in the monodromy method  \cite{Harlow2011} and in the pillow metric quantization  \cite{Maldacena2017}.}
we can reinterpret the exchanging symmetry and the crossing symmetry of the correlators by using the language of modular invariance (see (\ref{eq:bootstrap})).
\begin{center}
Exchanging Symmetry $\to$ Modular T Invariance\\
Crossing Symmetry $\to$ Modular S Invariance\\
\end{center}
And also note that this elliptic nome $q(z)$ maps the universal cover of the sphere with punctures at $z=0,1,\infty$ to the interior of the unit $q$-disk, in that, $|q|$ is always bounded by
\begin{equation}
|q|<1.
\end{equation}
This fact can be understood in terms of the relation between the modulus $\tau$ in the pillow metric and $z$  \cite{Maldacena2017}. Thus the series expansion  (\ref{eq:ckdef}) in $q$ is well-behaved and converges except for OPE singularities.

%%%%%%%%%%%%%%%%%%%%%%%%%%%%%%%%%%%%%%%%%%%%%%%%%%%%%%%%%%%%%%%%%%%%%%%%%%%%%%%%%%%%%%%%%%%%%%
%%%%%%%%%%%%%%%%%%%%%%%%%%%%%%%%%%%%%%%%%%%%%%%%%%%%%%%%%%%%%%%%%%%%%%%%%%%%%%%%%%%%%%%%%%%%%%
\section{AABB Blocks}\label{sec:AABB}
%%%%%%%%%%%%%%%%%%%%%%%%%%%%%%%%%%%%%%%%%%%%%%%%%%%%%%%%%%%%%%%%%%%%%%%%%%%%%%%%%%%%%%%%%%%%%%
%%%%%%%%%%%%%%%%%%%%%%%%%%%%%%%%%%%%%%%%%%%%%%%%%%%%%%%%%%%%%%%%%%%%%%%%%%%%%%%%%%%%%%%%%%%%%%

In this section, we focus on the AABB blocks for the correlator $\braket{O_B(\infty)O_B(1)O_A(z)O_A(0)}$ in the  $O_A O_A \to O_B O_B$ channel. First, we review our previous results on vacuum blocks from the recursion relation  \cite{Kusuki2018}  and next, we generalize this analysis to blocks with  non-zero intermediate dimensions.

Note that if setting $h_1=h_2$ and $h_3=h_4$ in (\ref{eq:Rmn}), $R_{m,n}$ with odd $mn$ always vanish and therefore we can obtain from (\ref{eq:ck}) (see \cite{Perlmutter2015,Chen2017,Kusuki2018}),
\begin{equation}\label{eq:ck=0}
c_k(h_p)=0 \ \ \ \ \ \text{if $k$ is odd.}
\end{equation}
This fact can shorten the processing time to calculate $H(h_p|q)$. That is why it is easier to study AABB blocks than ABBA blocks. In the following of this section, we implicitly assume $c_n$ with odd $n$ to be zero in all expressions.

%%%%%%%%%%%%%%%%%%%%%%%%%%%%%%%%%%%%%%%%%%%
%%%%%%%%%%%%%%%%%%%%%%%%%%%%%%%%%%%%%%%%%%%
\subsection{AABB Vacuum Blocks}\label{subsec:vacAABB}
\begin{figure}[H]
  \begin{center}
   \includegraphics[width=130mm]{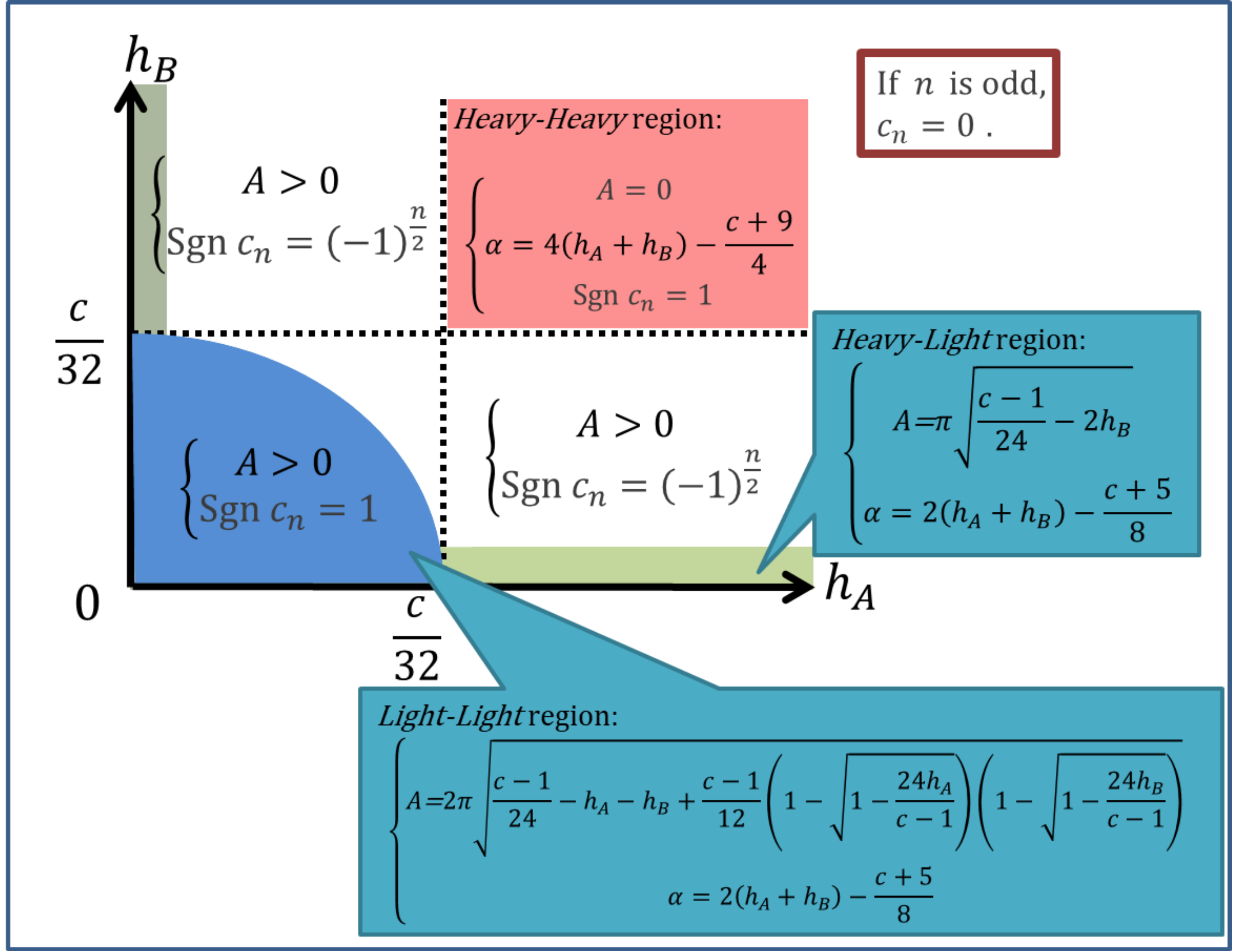}
   \end{center}
\caption{The sketch of behaviors of $c_n= \mathrm{sgn}(c_n) n^\a \ex{A \s{n}} $ for various values of $(h_A,h_B)$.}
\label{fig:HHLL-hAhBdep}
\end{figure}

%%%%%%%%%%%%%%%%%%%%%%%%%%%%%%%%%%%%%%%%%%%
%%%%%%%%%%%%%%%%%%%%%%%%%%%%%%%%%%%%%%%%%%%
First, we review the properties of AABB vacuum blocks. By using the Zamolodchikov recursion relation, we suggest in  \cite{Kusuki2018} that the coefficients $c_n$ for even $n$ have the simple asymptotic form as \footnote{Somehow, (\ref{eq:|cn|}) and (\ref{eq:AdepHL}) suggest that the coefficients $c_n$ behave like Cardy's formula. This might be the key to access large $c$ conformal blocks analytically.}

\begin{equation}\label{eq:|cn|}
\abs{c_n}
\sim \left\{
    \begin{array}{ll}
      n^\a   ,& \text{if } h_A, h_B > \fr{c}{32}  ,\\
     n^\a \ex{A \s{n}} \ \ \ \ (A>0)    ,& \text{otherwise }   ,\\
    \end{array}
  \right.\\
\ \ \ \ \ \text{for } n \gg c ,
\end{equation}
where $A$,$\a$ and the signature of $c_n$ are given in Figure \ref{fig:HHLL-hAhBdep}. This result implies that the behaviors of AABB blocks drastically change at $h_{A,B}=\fr{c}{32}$.
\footnote{The value $\fr{c}{32}$ also appears in the analytic expression of  $c_n$ for small $n$ as
	 \begin{equation}\label{eq:cnlim}
		c_{2m}\ar{c \to \infty}\fr{1}{m!} \BR{\fr{c}{2} \pa{1-\fr{32}{c}h_A}\pa{1-\fr{32}{c}h_B}}^m \ \ \ \ \text{for } 2m\ll c,
	\end{equation}
which suggests that the sign pattern of $c_{2m}$ changes at $h_{A,B}=\fr{c}{32}$ as in Figure \ref{fig:HHLL-hAhBdep}.
 }
In the heavy-light limit (the green regions in Figure \ref{fig:HHLL-hAhBdep}), we can estimate the values of $A$ and $\a$ from our analysis based on the recursion relation as 
\footnote{In  \cite{Kusuki2018}, we define $c_n$ as the coefficients of $q^{2n}$ instead of $q^n$ because of (\ref{eq:ck=0}), but now we use the definition (\ref{eq:ckdef}). Therefore, if one wants to convert the previous results in  \cite{Kusuki2018} to those in our new convention, one has to divide  the previous value of $A$ by $\s{2}$. }
\begin{equation}\label{eq:AdepHL}
\begin{aligned}
A=\pi \s{\fr{c-1}{24}-2h_B}, \ \ &\a=2(h_A+h_B)-\fr{c+5}{8} \ \ \text{if } h_A>\fr{c}{32}> h_B,\\
A=\pi \s{\fr{c-1}{24}-2h_A}, \ \  &\a=2(h_A+h_B)-\fr{c+5}{8}  \ \ \text{if } h_A<\fr{c}{32}< h_B.
\end{aligned}
\end{equation}
These values are exactly given by the Heavy-Light Virasoro blocks  \cite{Fitzpatrick2014,Fitzpatrick2015}, which supports  a validity of our statement.

In a part of  {\it light-light} region  (the blue region in Figure \ref{fig:HHLL-hAhBdep}), the values of $A$ and $\a$ are expressed by
\begin{equation}
\begin{aligned}
A&=2\pi \s{\fr{c-1}{24}-h_A-h_B+\fr{c-1}{12}\pa{1-\s{1-\fr{24h_A}{c-1}}}\pa{1-\s{1-\fr{24h_B}{c-1}}}},\\
\a&=2(h_A+h_B)-\fr{c+5}{8} .
\end{aligned}
\end{equation}
We derive this expression analytically in Appendix \ref{subsec:LLLL}.

In the {\it heavy-heavy} region (the red region in Figure \ref{fig:HHLL-hAhBdep}), the values of $A$ and $\a$ have a simple form,
\begin{equation}\label{eq:AaHHAABB}
\begin{aligned}
A&=0,\\
\a&=4(h_A+h_B)-\fr{c+9}{4}.
\end{aligned}
\end{equation}

%Note that AABB blocks are also used to calculate Renyi entropy for excited states, which receive extensive attention in the context of chaos  \cite{Caputa2014,He2014,Numasawa2016,Caputa2017,He2017,Guo2018}. The universality of $A$ in the heavy-heavy region ($h_A, h_B > \fr{c}{32} $) leads to a surprising universality of the n-th Renyi entropy ($n\geq2$) for heavy excited states ($h_O>\fr{c_{\text{cft}}}{32}$) as
%\begin{equation}\label{eq:DS}
%\D S_A^{(n)} \sim \fr{nc_{\text{cft}}}{24(n-1)}\log \fr{t}{\e}
%\end{equation}
%where $n$ is the replica number and $c_{\text{cft}}$ is the central charge of the CFT where we want to calculate the Renyi entropy. More details are in  \cite{Kusuki2018}. AABB blocks can also be used to calculate  late times behavior of OTOC which is also interesting in the context of chaos  \cite{Roberts2015,Gu2016,Caputa2016,Caputa2017a}.

%%%%%%%%%%%%%%%%%%%%%%%%%%%%%%%%%%%%%%%%%%%
%%%%%%%%%%%%%%%%%%%%%%%%%%%%%%%%%%%%%%%%%%%
\subsection{AABB Non-Vacuum Blocks}\label{subsec:AABBnon}
%%%%%%%%%%%%%%%%%%%%%%%%%%%%%%%%%%%%%%%%%%%
%%%%%%%%%%%%%%%%%%%%%%%%%%%%%%%%%%%%%%%%%%%

\begin{figure}[h]
  \begin{center}
   \includegraphics[width=70mm]{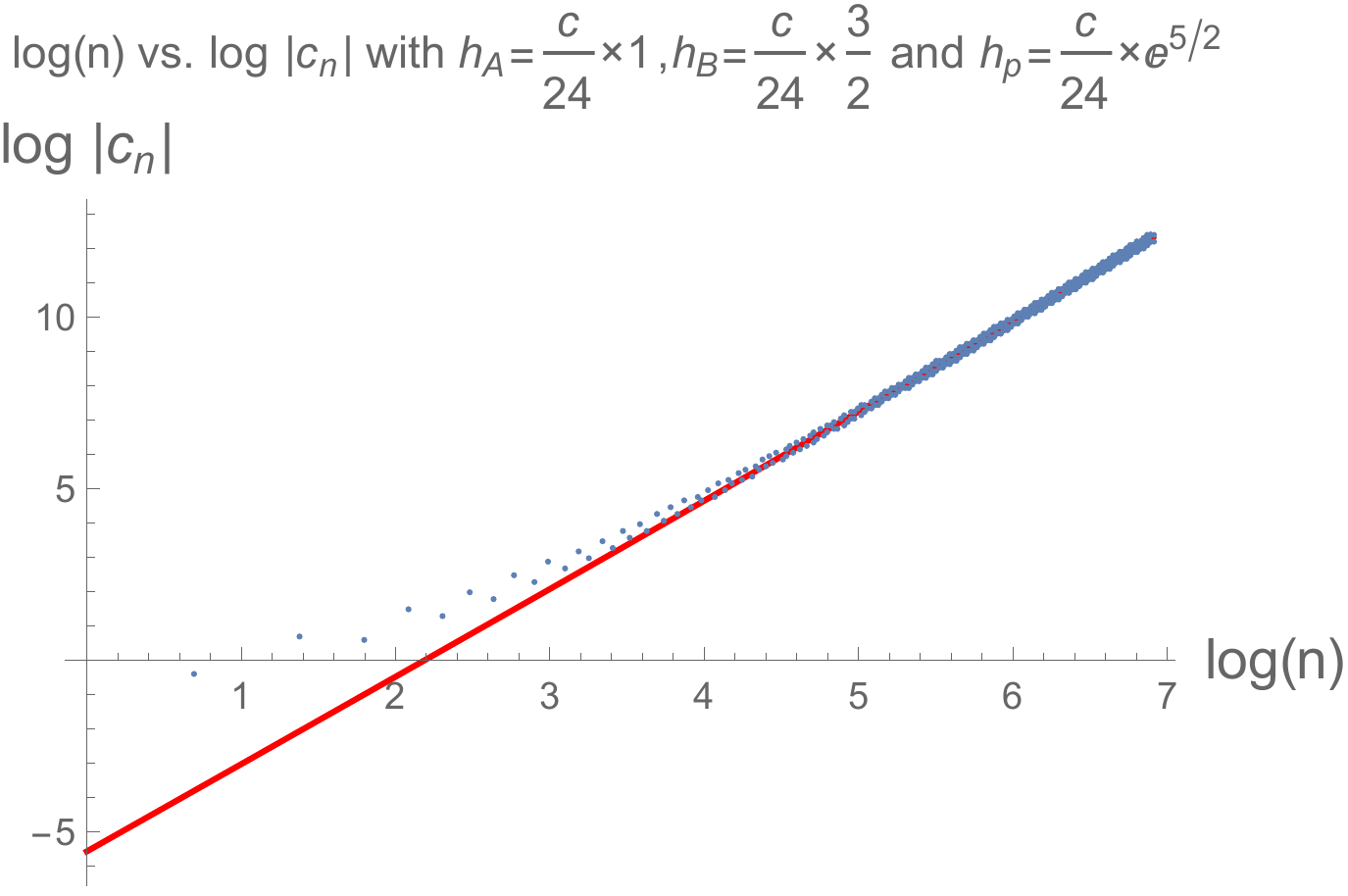}
   \includegraphics[width=70mm]{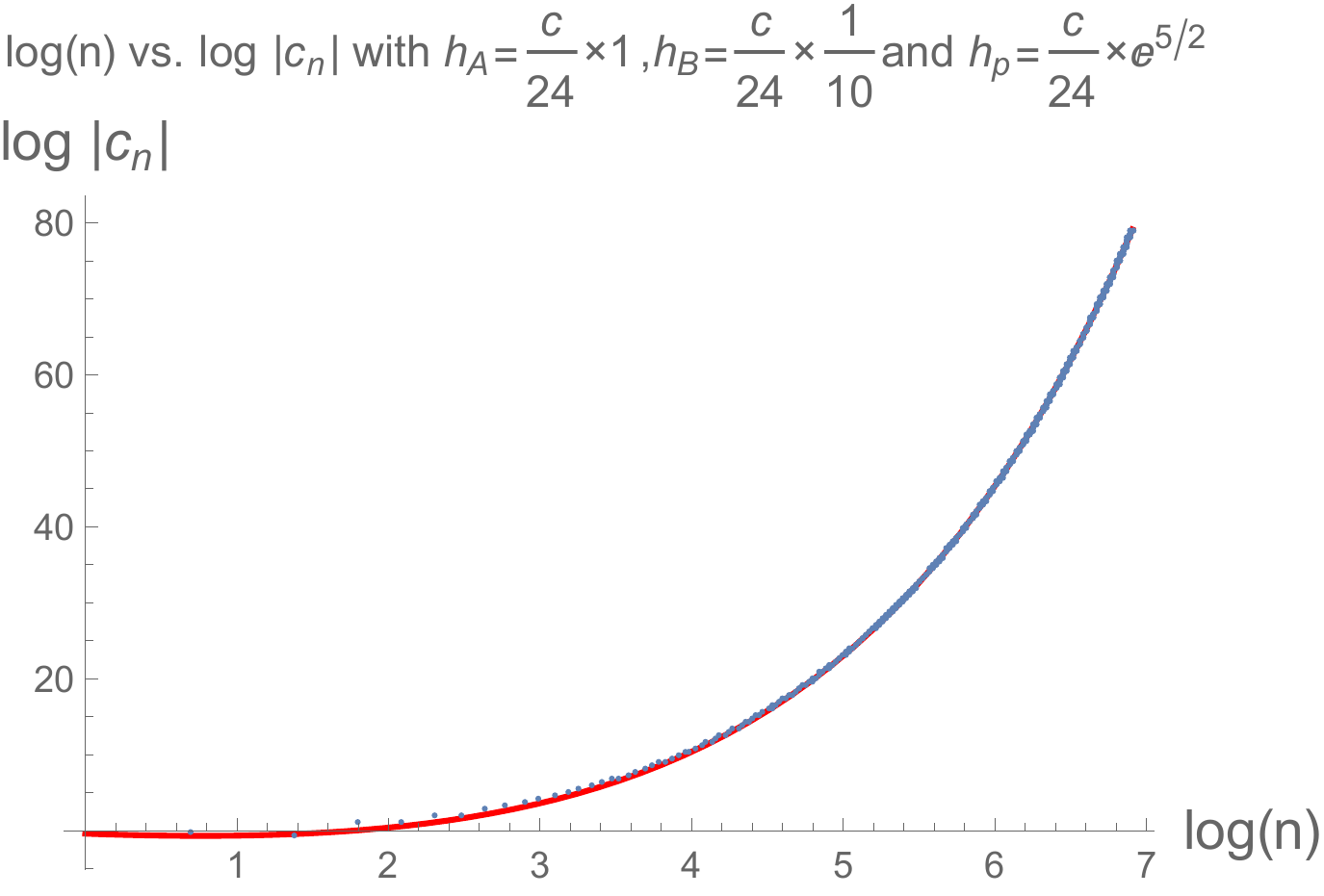}
   \end{center}
\caption{The behaviors of the coefficients $c_n$ of AABB blocks with $h_A=\fr{c}{24}$. The left is for $(h_B,h_p)=(\fr{c}{16},\fr{c}{24} \times \ex{\fr{5}{2}})$ and the right is for $(h_B,h_p)=(\fr{c}{240},\fr{c}{24}\times  \ex{\fr{5}{2}})$. The blue dots are the numerical values of $\log c_n$. The red lines are $B n^\a \ex{A\s{n}}$ with the constant $B$ determined by the fit. We now set $c=30.01$ and, to fit $A$ and $\a$, we use the numerical values of $c_n$ at $n=500\sim1000$.}
\label{fig:AABBcn}
\end{figure}

\begin{figure}[h]
  \begin{center}
   \includegraphics[width=70mm]{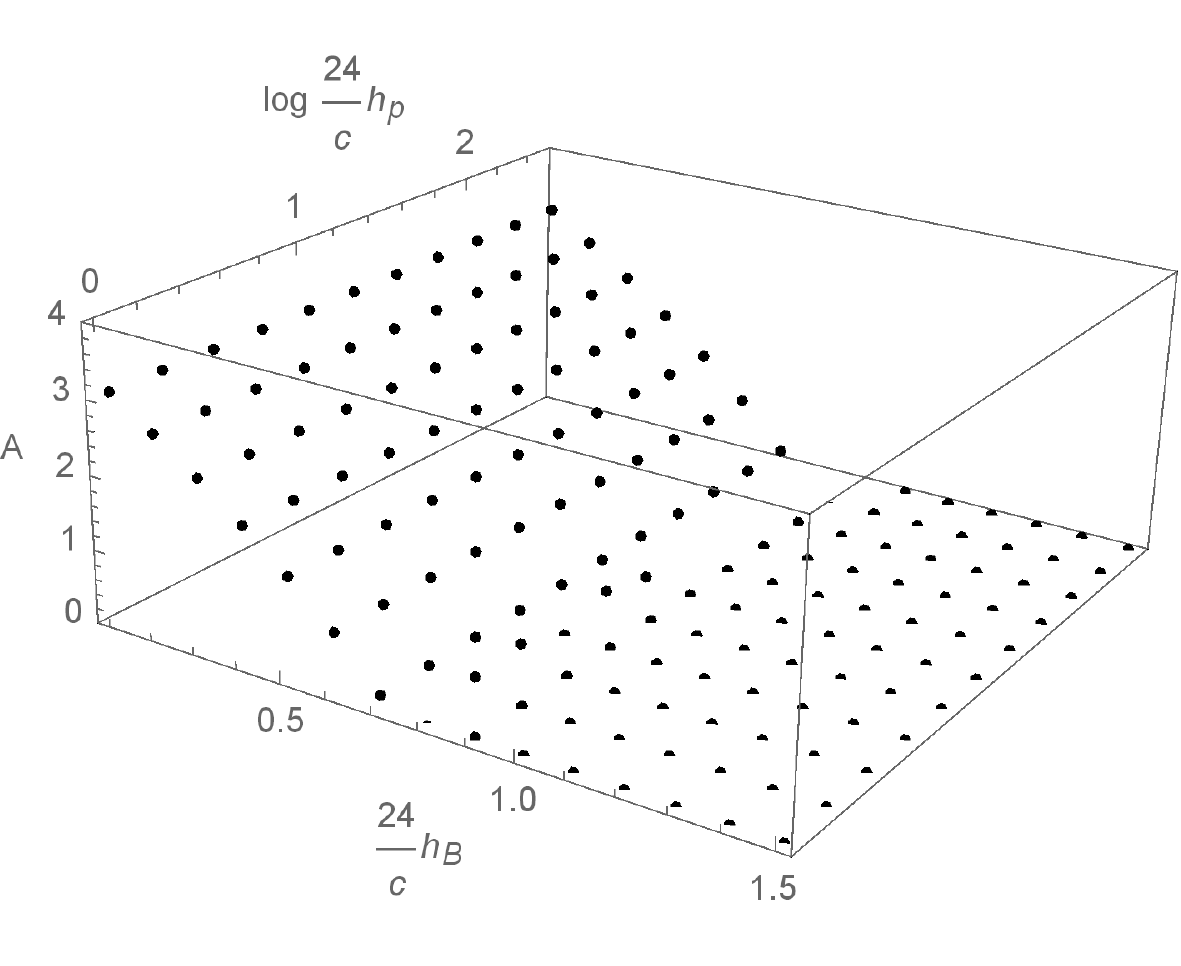}
   \includegraphics[width=70mm]{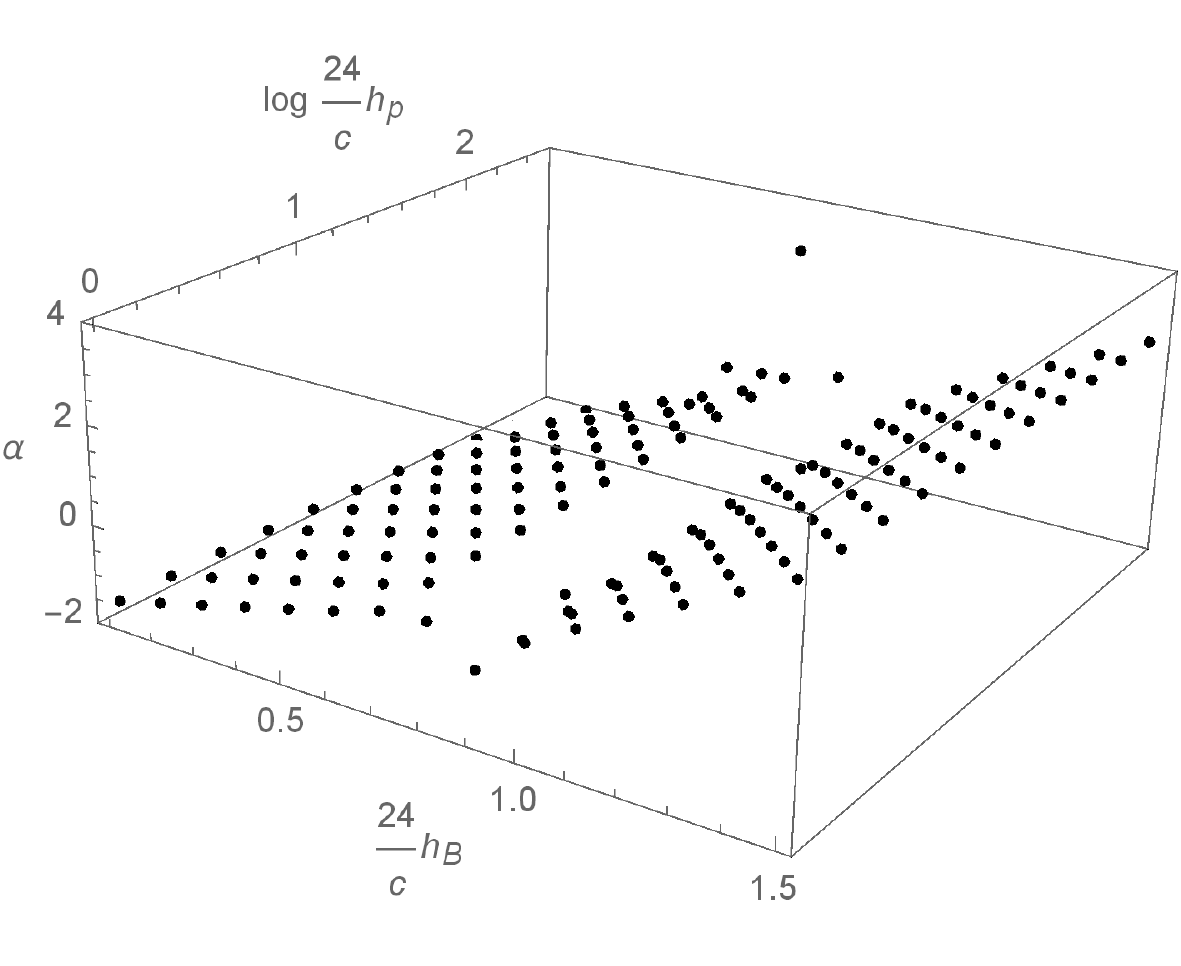}
   \end{center}
 \caption{The plots of the values of $A$ (left) and $\a$ (right) for various values of  ($h_B,h_p$) with $h_A=\fr{c}{24}$. Some strange behaviors near the line $h_B=\fr{c}{32}$ could be resolved by using the values $c_n$ for higher $n$ to fit $A$ and $\a$ (see Appendix \ref{subsec:app1}). Here we set $c=30.01$ and to fit $A$ and $\a$, we use the numerical values of $c_n$ at $n=500\sim1000$.}
\label{fig:HHLLhBhpdep}
\end{figure}

Let us move on to AABB non-vacuum blocks. In fact, the non-vacuum blocks show a similar behavior to that of the vacuum blocks, that is, the coefficients $c_n$ are well-fitted by $n^\a \ex{A \s{n}}$ for large $n$ as in Figure \ref{fig:AABBcn}. In particular, the values of $A$ and $\a$ are independent of $h_p$. What we would like to emphasize here is that there is the transition at $h_{A,B}=\fr{c}{32}$ in the behavior of AABB blocks with not only vacuum but also non-vacuum intermediate states.
These features can be seen in Figure \ref{fig:HHLLhBhpdep}.

From the values of $A$ and $\a$ for various values of  ($h_B,h_p$) in Figure \ref{fig:HHLLhBhpdep}, one can see the transition at $h_B=\fr{c}{32}$ obviously. 
This fact enhances our previous result for Renyi entropy  \cite{Kusuki2018} because the calculation of Renyi entropy in  \cite{Kusuki2018} relies on the vacuum block approximation of the correlator corresponding to the Renyi entropy. Thus the transition of the Renyi entropy can be seen at earlier times than the late time when we can approximate the correlator by the vacuum block.

Actually, we can also see the value $\fr{c}{32}$ analytically in the same way as (\ref{eq:cnlim}). By using the Zamolodchikov recursion relation, one can see that the coefficients $c_n(h_p)$ for general $h_p$ can be given by
	 \begin{equation}\label{eq:lowercn}
		c_{2m}(h_p) \ar{c \to \infty}\fr{1}{m!} \BR{\fr{c}{2} \fr{\pa{1-\fr{32}{c}h_A}\pa{1-\fr{32}{c}h_B}}{\pa{1+\fr{8}{c}h_p}}}^m \ \ \ \ \text{for } 2m\ll c.
	\end{equation}
From this observation as well as numerical supports, we can expect that the transition at $\fr{c}{32}$ of AABB blocks can be generalized to non-zero intermediate dimensions.

\begin{figure}[h]
  \begin{center}
   \includegraphics[width=70mm]{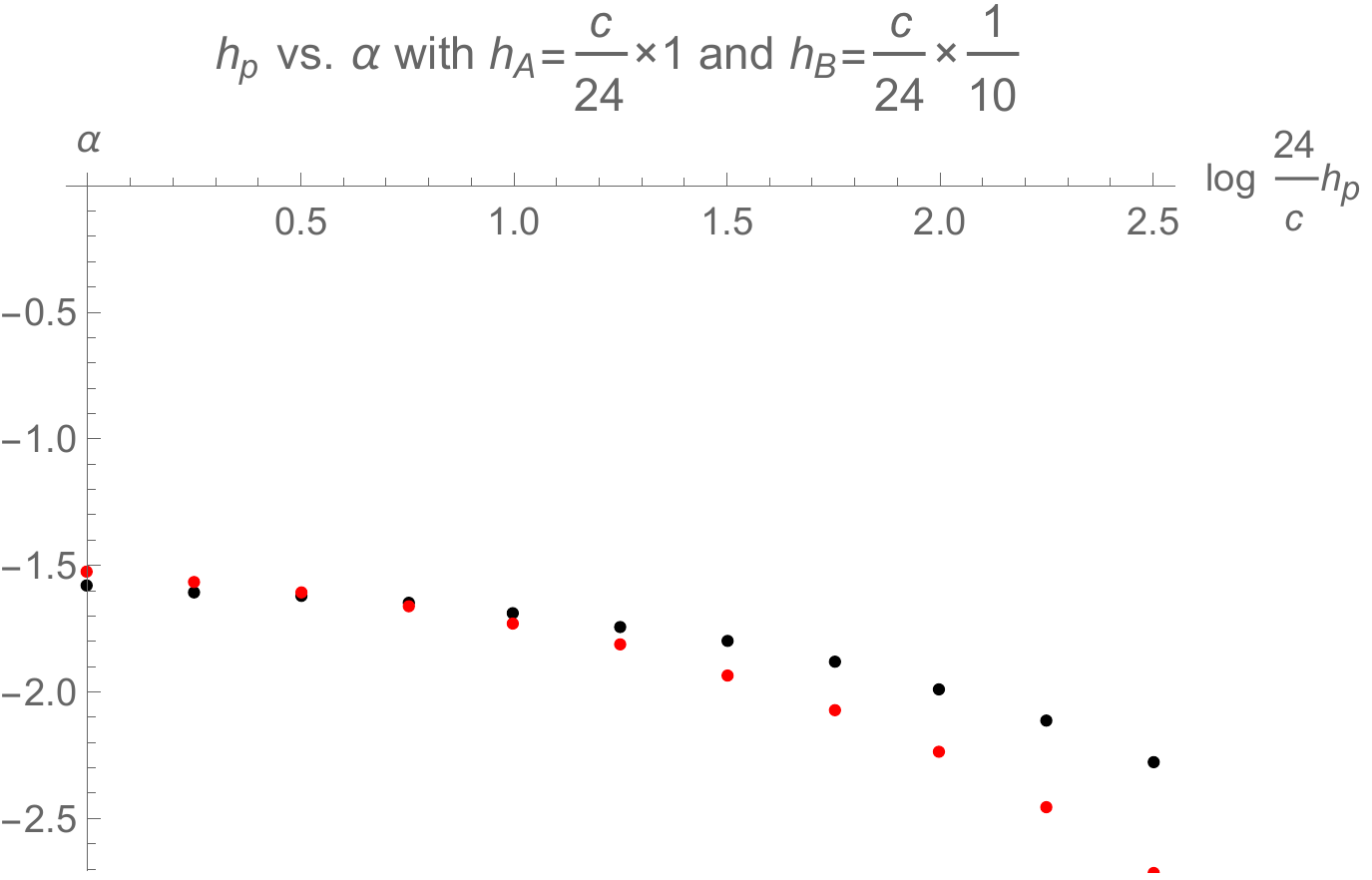}
   \includegraphics[width=70mm]{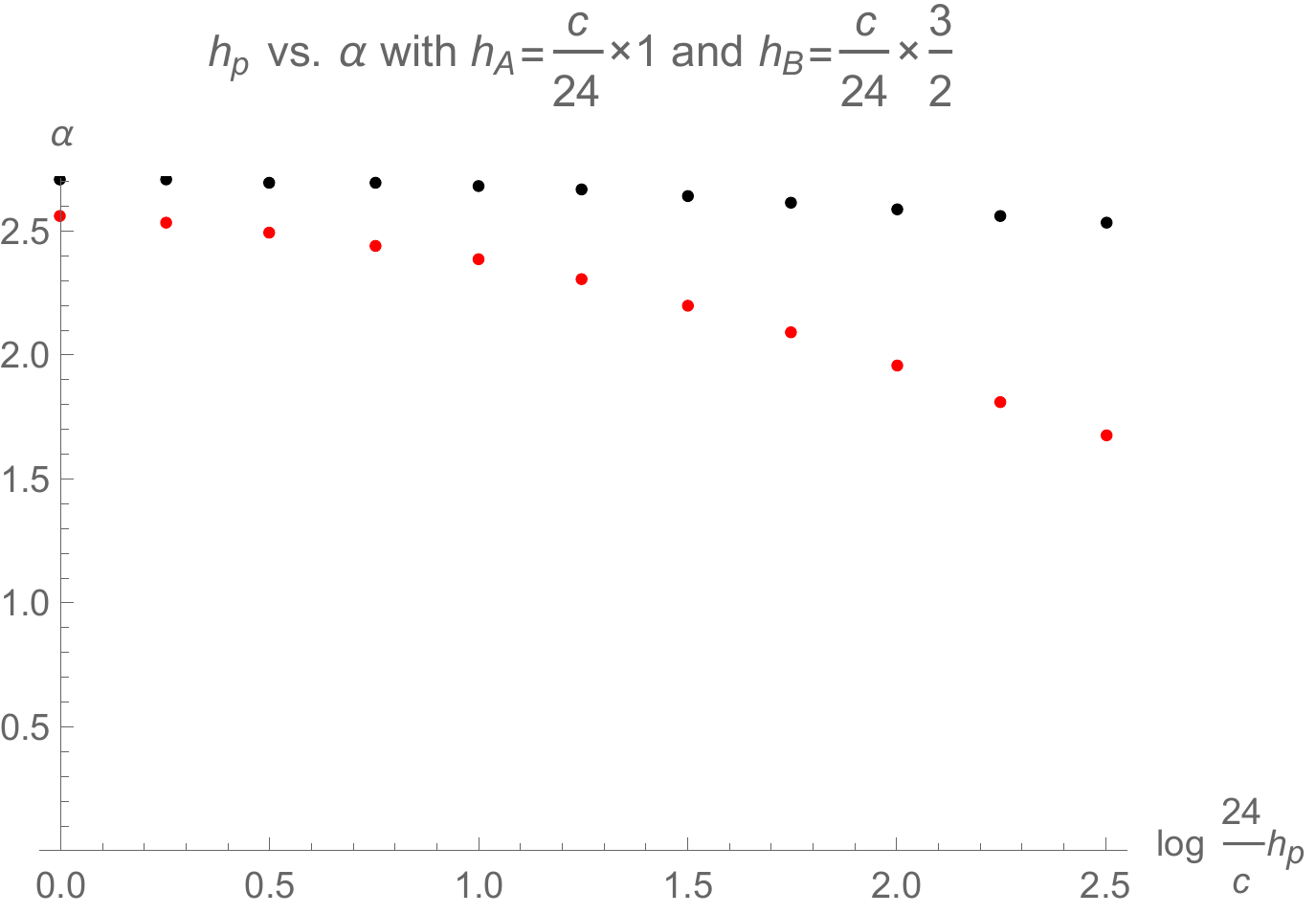}
   \end{center}
\caption{The $h_p$ dependence of $\a$. The left is for $(h_A,h_B)=(\fr{c}{24},\fr{c}{240})$ and the right is for $(h_A,h_B)=(\fr{c}{24},\fr{c}{16})$. Red dots are fitted by $c_n$ for $n=100\sim 200$ and black dots are fitted by $c_n$ for $n=500\sim 1000$. One can find that the $h_p$ dependence of $\a$ approaches to constant as we use higher $n$ to fit the values of $\a$.}
\label{fig:HHLLNdep}
\end{figure}

If one carefully observes the behaviors of $c_n$ in the right of Figure \ref{fig:HHLLhBhpdep}, one could find the values of $\a$ decrease with $h_p$. It can be seen more obviously in Figure \ref{fig:HHLLNdep}, which is one $h_B$ slice of  Figure \ref{fig:HHLLhBhpdep}.
Nevertheless, we do not think the values of $\a$ depend on $h_p$. In Figure \ref{fig:HHLLNdep}, blue dots are fitted by using $c_n$ at more higher $n$ than red dots. It suggests that the $h_p$ dependence of $\a$ approaches a constant if one uses enough large $n$ to fit $\a$. This is the reason why we think the coefficients $c_n$ for large $n$ are independent of $h_p$.
In other words, for any $h_p$, one can find one $N$ such that
\begin{equation}\label{eq:indepependent}
\abs{c_n}
\sim \left\{
    \begin{array}{ll}
      n^\a   ,& \text{if } h_A, h_B > \fr{c}{32}  ,\\
     n^\a \ex{A \s{n}} \ \ \ \ (A>0)    ,& \text{otherwise }   ,\\
    \end{array}
  \right.\\
\ \ \ \ \ \text{for } n \gg N, c ,
\end{equation}
where $A$ and $\a$ are independent of $h_p$ and are given by Figure \ref{fig:HHLL-hAhBdep}.

How can we identify the value of $N$? Actually, we can suggest $N \sim h_p$  since one can see the coefficients $c_n(h_p)$ depend only on the difference $h_p-h_{m,n}$ from the expression of the coefficients $c_n$ (\ref{eq:ck}). As a result, our Cardy-like formula 
(\ref{eq:|cn|}) for $c_n$ could break down if $n \sim h_p$. It's also interesting to find the simple form of the coefficients $c_n$ with $n \sim h_p$, however in this case, the coefficients $c_n$ depend complicatedly on many parameters $c, h_A, h_B, h_p, n$ and therefore we leave it as a future work. Nevertheless, we can extract few remarkable properties of $c_n$ with $n \sim h_p$ and we will explain it later in Section \ref{sec:heavy}.

Consequently, we can argue that the AABB blocks in the limit $ z \to 1$ is independent of intermediate dimensions due to the following reason. If one wants to reconstruct conformal blocks from the coefficients $c_n$, one has to take the sum
\begin{equation}
H^{HH}_{LL}(h_p|q)=
\sum_{n=0}^{\infty} c_n q^n.
\end{equation}
If taking the limit $z \to 1$ which corresponds to the limit $q \to 1$, we can approximate the sum of $c_n$ by that of the asymptotic form $n^\a \ex{A \s{n}}$ which is valid for large $n$, because the contributions from small $n$ terms are much less than the other infinite contributions in the limit $q \to 1$. And $c_n$ for large $n$ is independent of $h_p$, which leads to the conclusion that the AABB blocks are independent of intermediate dimensions in the limit $ z \to 1$. However, this scenario can be applied only to the case where $c_n>0$ for any $n$. It happens in the { \it heavy-heavy} or {\it light-light} region ($h_A, h_B<\fr{c}{32}$ or $h_A, h_B>\fr{c}{32}$), which can be seen in Figure \ref{fig:HHLL-hAhBdep}.  Note that even though $c_n$ is not always positive, we can read off the upper bound of the singular behavior in the limit $z \to 1$ because
\begin{equation}\label{eq:bound}
\sum_{n=0}^{\infty} c_n q^n \leq \sum_{n=0}^{\infty} \abs{c_n} q^n, \ \ \ \ \ (0<q<1).
\end{equation}
If the upper bound of the singular behavior is less than the singularity of the universal prefactor $\Lambda(h_p|q)$, then we can neglect the contribution of $H(h_p|q)$ to the singularity of the conformal block.

If one can find the limit of $z$ corresponding to $q \to i$, the above scenario can also be applied to the blocks in the {\it heavy-light} region ($h_A<\fr{c}{32}$ and $h_B>\fr{c}{32}$, or $h_A>\fr{c}{32}$ and $h_B<\fr{c}{32}$) in such a limit. Actually we can take the limit $q \to i$ by taking the limit $z \to 0$ after picking up a monodromy around $z=1$, that is, $1-z \to \ex{- 2 \pi i}\pa{1-z}$.

%%%%%%%%%%%%%%%%%%%%%%%%%%%%%%%%%%%%%%%%%%%
%%%%%%%%%%%%%%%%%%%%%%%%%%%%%%%%%%%%%%%%%%%
\subsection{Comments on Information Loss}
%%%%%%%%%%%%%%%%%%%%%%%%%%%%%%%%%%%%%%%%%%%
%%%%%%%%%%%%%%%%%%%%%%%%%%%%%%%%%%%%%%%%%%%

AdS/CFT shows that correlators decay exponentially at large time separation in a black hole background, which is known as one of the information loss problems. And this problem can be seen directly from semiclassical Virasoro blocks  \cite{Fitzpatrick2014}. We expect that this problem can be resolved by taking account of a non-perturbative correction in central charge and summing over Virasoro blocks in the $O_A O_A$ OPE channel.

Recently, it is shown numerically that exact blocks behave like power law decay $t^{-\fr{3}{2}}$  \cite{Chen2017}. It means that the non-perturbative correction to blocks ameliorates information loss. Actually, this polynomial decay can be derived from (\ref{eq:AaHHAABB}) by setting $h_A>\fr{c}{24}>h_B \ (\ >\fr{c}{32})$. And moreover the result of Section \ref{subsec:AABBnon} explains that this polynomial decay can be seen in blocks with non-vacuum intermediate states.

Semiclassical conformal blocks also exhibit the information loss problem as {\it forbidden singularities}, which are singularities not corresponding to OPE singularities. This problem can be resolved by using exact conformal blocks as mentioned in  \cite{Chen2017}, and the result of Section \ref{subsec:AABBnon} shows this resolution can be also applied to blocks with  non-zero intermediate dimensions.
%%%%%%%%%%%%%%%%%%%%%%%%%%%%%%%%%%%%%%%%%%%%%%%%%%%%%%%%%%%%%%%%%%%%%%%%%%%%%%%%%%%%%%%%%%%%%%
%%%%%%%%%%%%%%%%%%%%%%%%%%%%%%%%%%%%%%%%%%%%%%%%%%%%%%%%%%%%%%%%%%%%%%%%%%%%%%%%%%%%%%%%%%%%%%
\section{ABBA Blocks}\label{sec:ABBA}
%%%%%%%%%%%%%%%%%%%%%%%%%%%%%%%%%%%%%%%%%%%%%%%%%%%%%%%%%%%%%%%%%%%%%%%%%%%%%%%%%%%%%%%%%%%%%%
%%%%%%%%%%%%%%%%%%%%%%%%%%%%%%%%%%%%%%%%%%%%%%%%%%%%%%%%%%%%%%%%%%%%%%%%%%%%%%%%%%%%%%%%%%%%%%

To solve the conformal bootstrap program, we have to know not only AABB blocks, but also ABBA blocks, hence we are also interested in the properties of ABBA blocks. In this section, we study ABBA blocks in the same approach as AABB blocks and reveal universal behaviors of ABBA blocks in large $c$ CFTs.

The series expansion of conformal blocks in the elliptic nome $q(z)$,
\begin{equation}
\ca{F}^{21}_{34}(h_p|z)=\Lambda(z)\sum_n c_n q^n,
\end{equation}
can be seen naturally in the quantization on the pillow metric  \cite{Maldacena2017}. In particular, from a viewpoint of the pillow metric quantization, it can be shown that the coefficients $c_n$ for ABBA blocks are all positive, which is very nontrivial from the recursion relation. Our analysis is consistent with this fact, in that,  we have checked the fact that the coefficients of all ABBA blocks satisfy
\begin{equation}
(c^{BA}_{BA})_n>0 \ \ \ \ \ \text{for all } n,
\end{equation}
by using the recursion relation numerically. Note that this fact holds only for ABBA blocks, not for AABB blocks and ABAB blocks. In fact, from (\ref{eq:symq}), the coefficients for ABAB blocks are given by
\begin{equation}
\sgn[(c^{BA}_{AB})_n] = (-1)^n \ \ \ \ \ \text{for all } n,
\end{equation}
and the sign of the coefficients $(c^{AA}_{BB})_n$ are more non-trivial and illustrated in Figure \ref{fig:HHLL-hAhBdep}.

\begin{figure}[h]
 \begin{minipage}{0.5\hsize}
  \begin{center}
   \includegraphics[width=70mm]{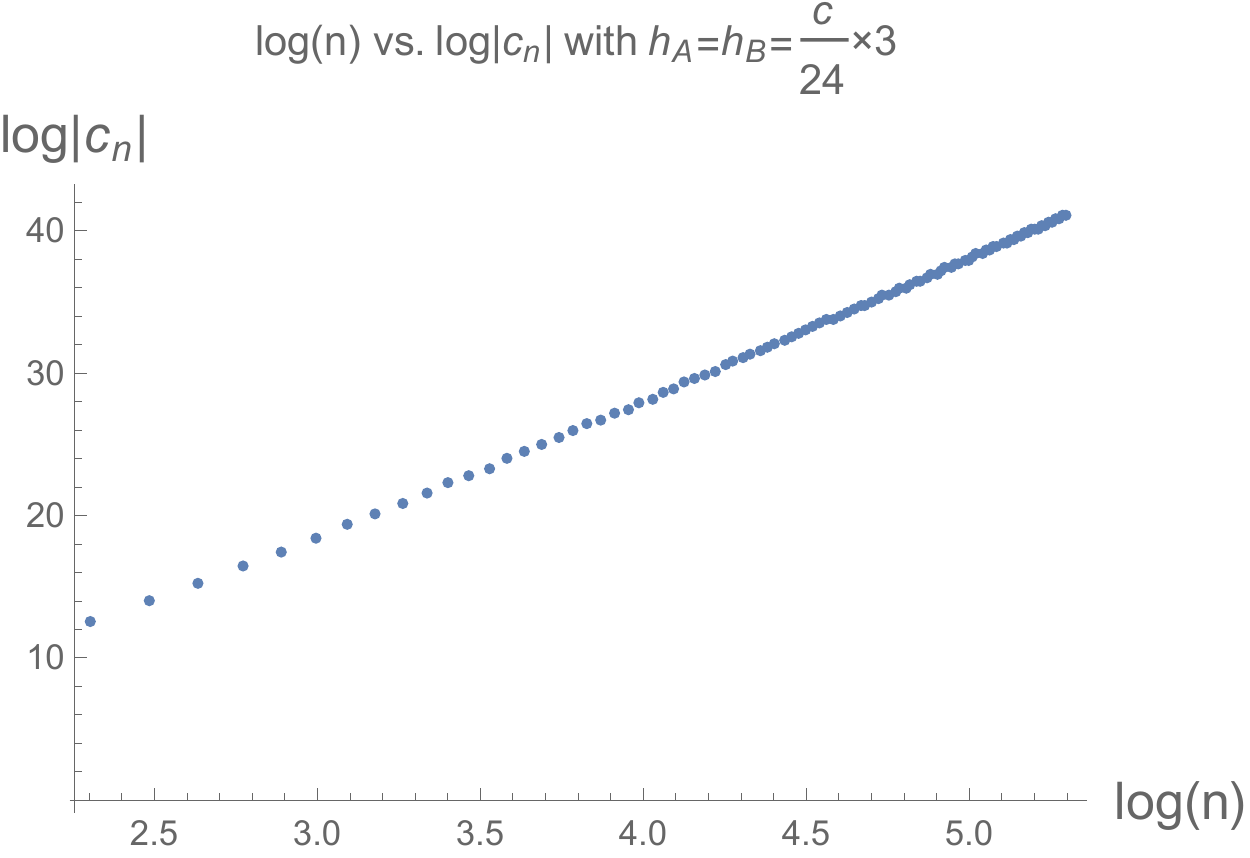}
  \end{center}
 \end{minipage}
 \begin{minipage}{0.5\hsize}
  \begin{center}
   \includegraphics[width=70mm]{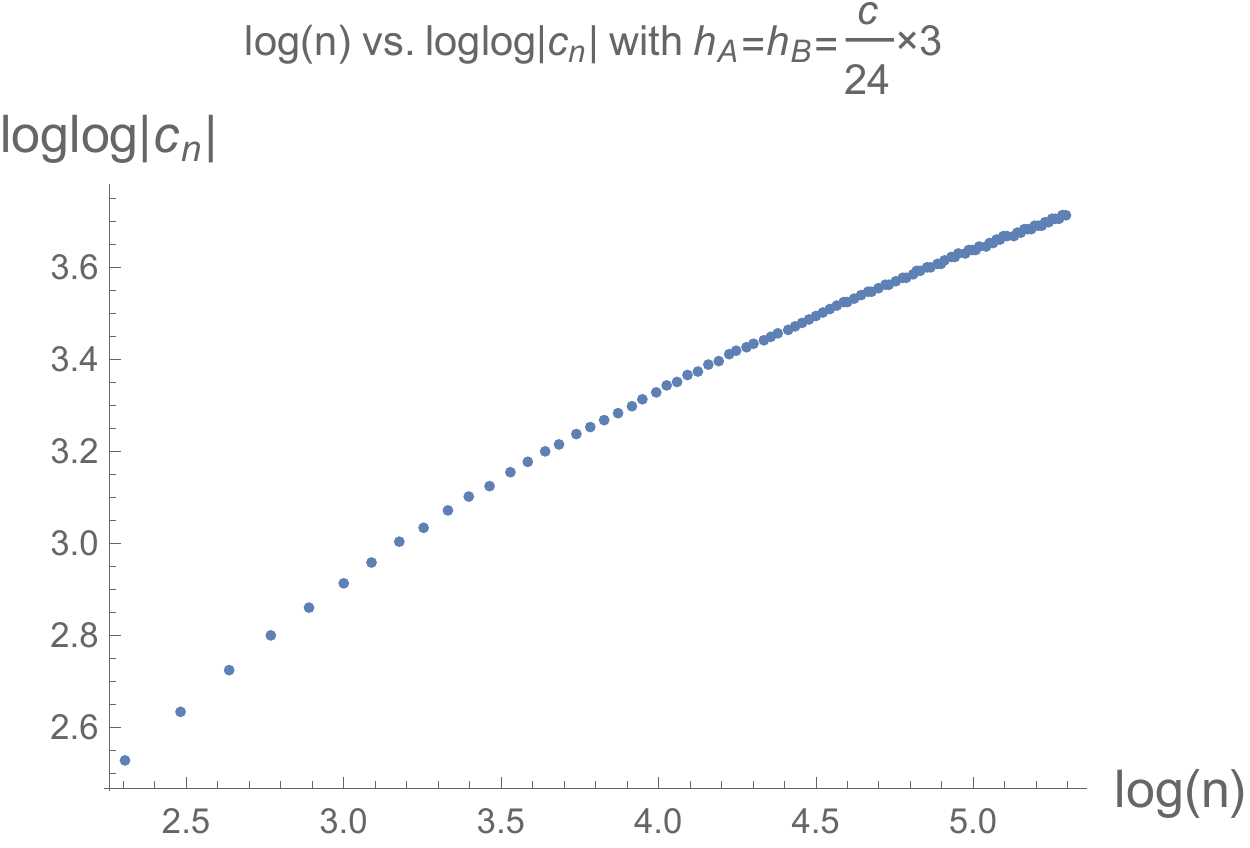}
  \end{center}
 \end{minipage}
 \begin{minipage}{0.5\hsize}
  \begin{center}
   \includegraphics[width=70mm]{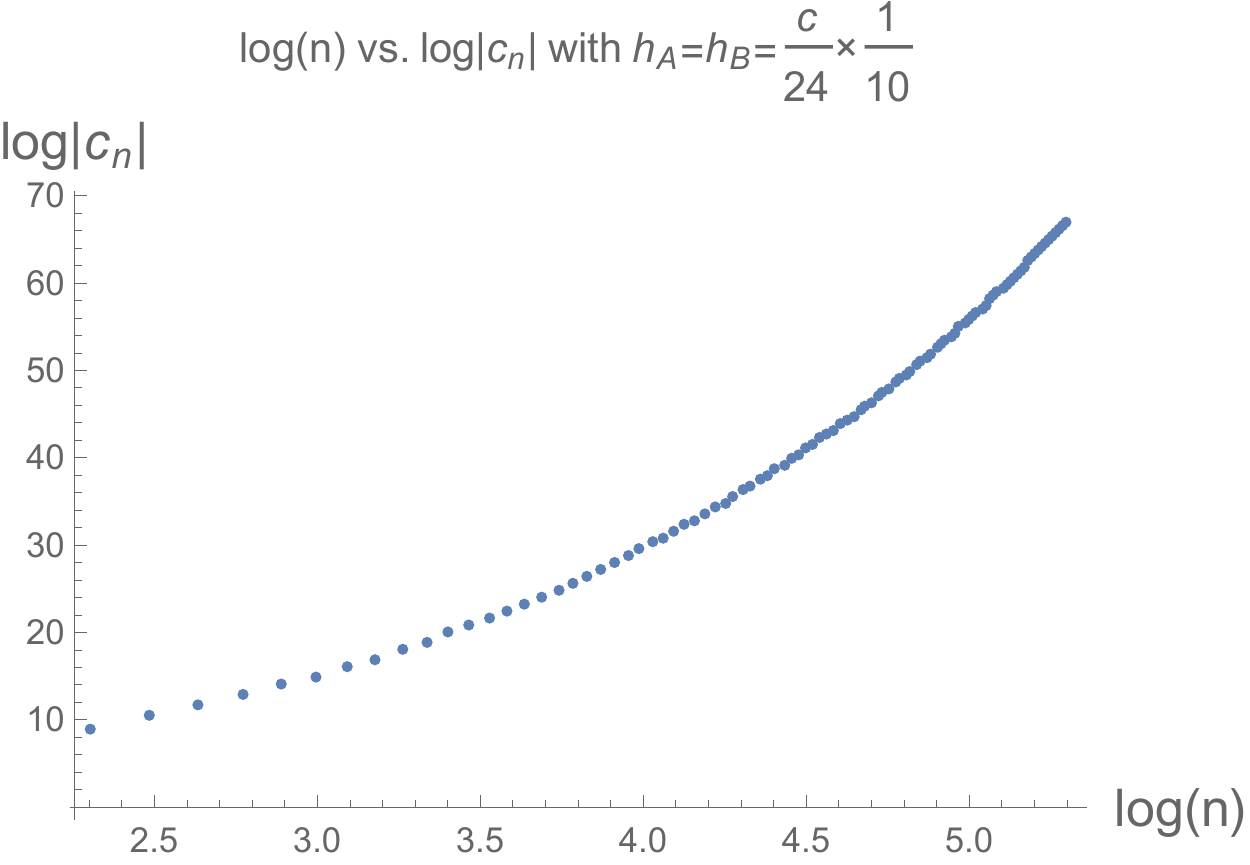}
  \end{center}
 \end{minipage}
 \begin{minipage}{0.5\hsize}
  \begin{center}
   \includegraphics[width=70mm]{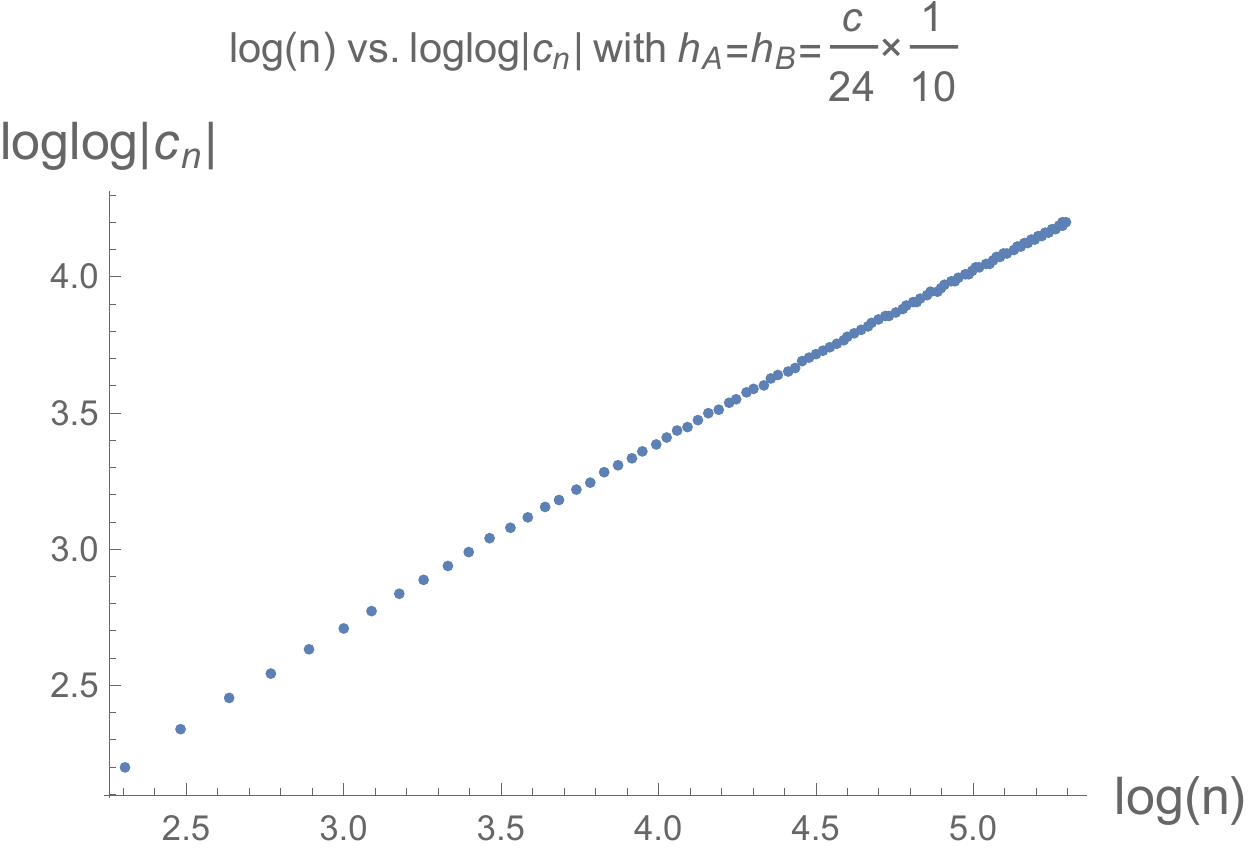}
  \end{center}
 \end{minipage}
\caption{The plots of $c_n$. The upper two plots are for $h_A=h_B=\fr{c}{8}$. We can see that the left of two shows a linear behavior, which suggest that $c_n$ grows polynomially. The lower two plots are for $h_A=h_B=\fr{c}{240}$ and we can find a linear dependence in the right,  which suggest that $c_n$ grows exponentially.}
\label{fig:cndep}
\end{figure}

Is there any similar phenomenon for ABBA blocks as AABB blocks? Surprisingly, the same asymptotic behaviors as AABB blocks can be found for ABBA blocks. In more detail, we can see that there are only two patterns in the asymptotic behaviors of ABBA blocks:
\begin{enumerate}
\item $\log c_n$ shows a linear behavior for large $n$. (The upper left in Figure \ref{fig:cndep})
\item $\log\log c_n$ shows a linear behavior for large $n$. (The lower right in Figure \ref{fig:cndep})

\end{enumerate}

~
\\
Therefore, the asymptotic form of the coefficients $c_n$ can be written by
	\begin{equation}\label{eq:cnas}
		c_n \sim n^{\a} \ex{A n^\b}
	\end{equation}
and moreover, there are universal properties for $A$ and $\beta$ as
\begin{equation}
\begin{aligned}
A&=0 \ \ \ \ \ \ \ \ \ \ \ \ \ \ \ \ \ \ \ \ \ \ \ \ \ \ \  \text{if } h_A, h_B>\fr{c}{32},\\
A&>0 \ \ \ \ \text{and} \ \ \ \ \b=\fr{1}{2} \ \ \ \ \text{otherwise}, 
\end{aligned}
\end{equation}
which are the same properties as the coefficients of AABB blocks. It is obvious that the ABBA block with $h_A=h_B$ is exactly same as the AABB block with $h_A=h_B$, which implies that one can see the same behaviors of the coefficients $c_n$ for ABBA and AABB along the line $h_A=h_B$ in Figure \ref{fig:HHLL-hAhBdep}, \ref{fig:HLHL-hAhBdep}. Therefore, it is natural that the behaviors of $c_n$  for ABBA blocks also drastically change at $h_{A,B}=\fr{c}{32}$.

%%%%%%%%%%%%%%%%%%%%%%%%%%%%%%%%%%%%%%%%%%%
%%%%%%%%%%%%%%%%%%%%%%%%%%%%%%%%%%%%%%%%%%%
\subsection{ABBA Blocks with Light Intermediate States}\label{subsec:ABBAnon}
%%%%%%%%%%%%%%%%%%%%%%%%%%%%%%%%%%%%%%%%%%%
%%%%%%%%%%%%%%%%%%%%%%%%%%%%%%%%%%%%%%%%%%%
Using our numerical results, we can estimate the values of $A$ and $\a$ in (\ref{eq:cnas}). First, we focus on the nearly vacuum blocks $\ca{F}^{BA}_{BA}(h_p \ll c|z)$. In such a case, the asymptotic coefficients (\ref{eq:cnas}) are insensitive to intermediate dimensions due to the following reason. The intermediate dimension $h_p$ appears only in the denominator of the recursion relation (\ref{eq:ck}) as the difference $h_p-h_{m,n}$, and these  zeros of the Kac determinant $h_{m,n}$ are of order $c$. It means that $h_p\ll h_{m,n}$ for all ($m,n$),  (except for $n=1$). As a result, we expect that the effect of $h_p \ll c$  to the blocks can be negligible.

\begin{figure}
  \begin{center}
   \includegraphics[width=100mm]{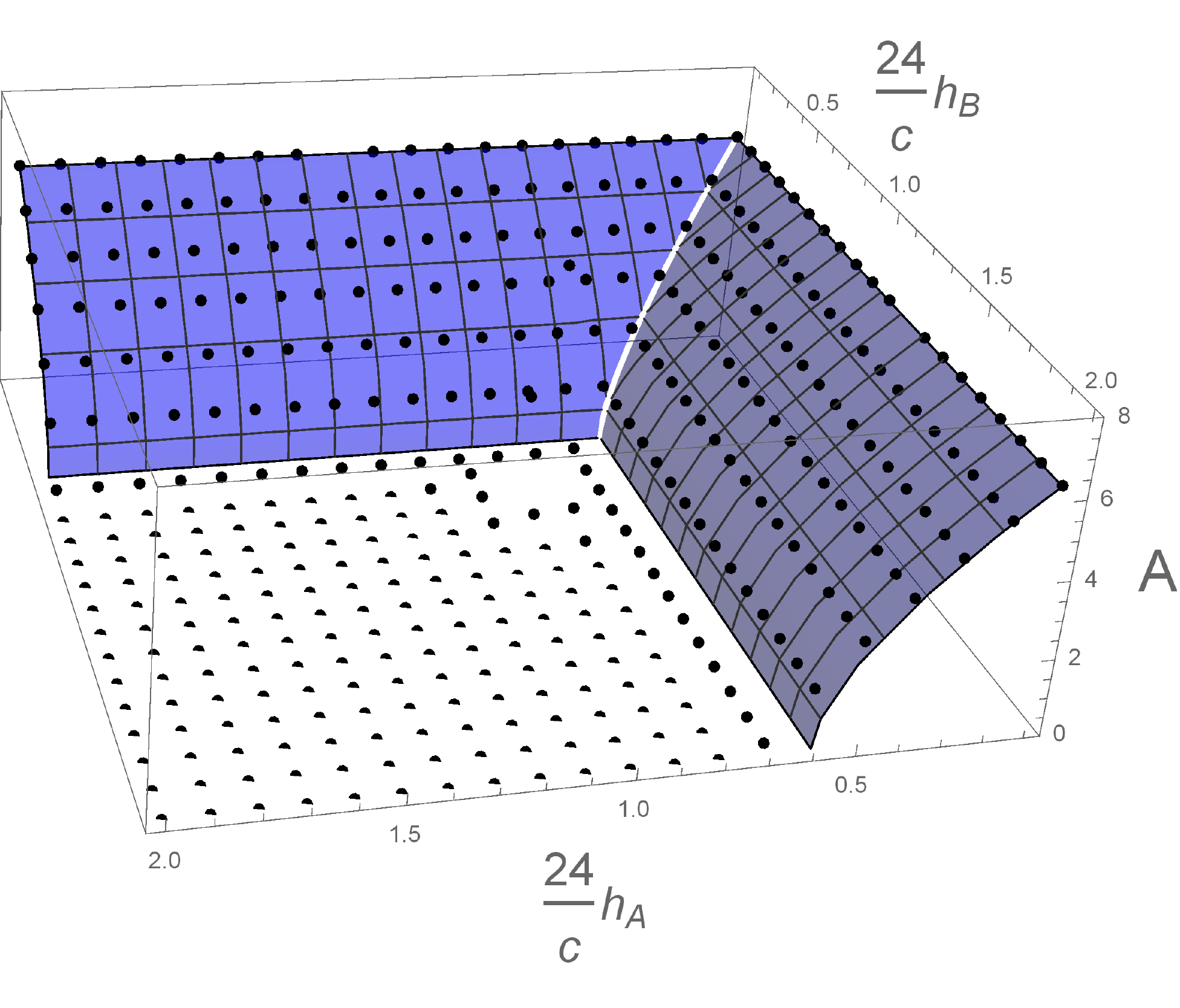}
   \includegraphics[width=100mm]{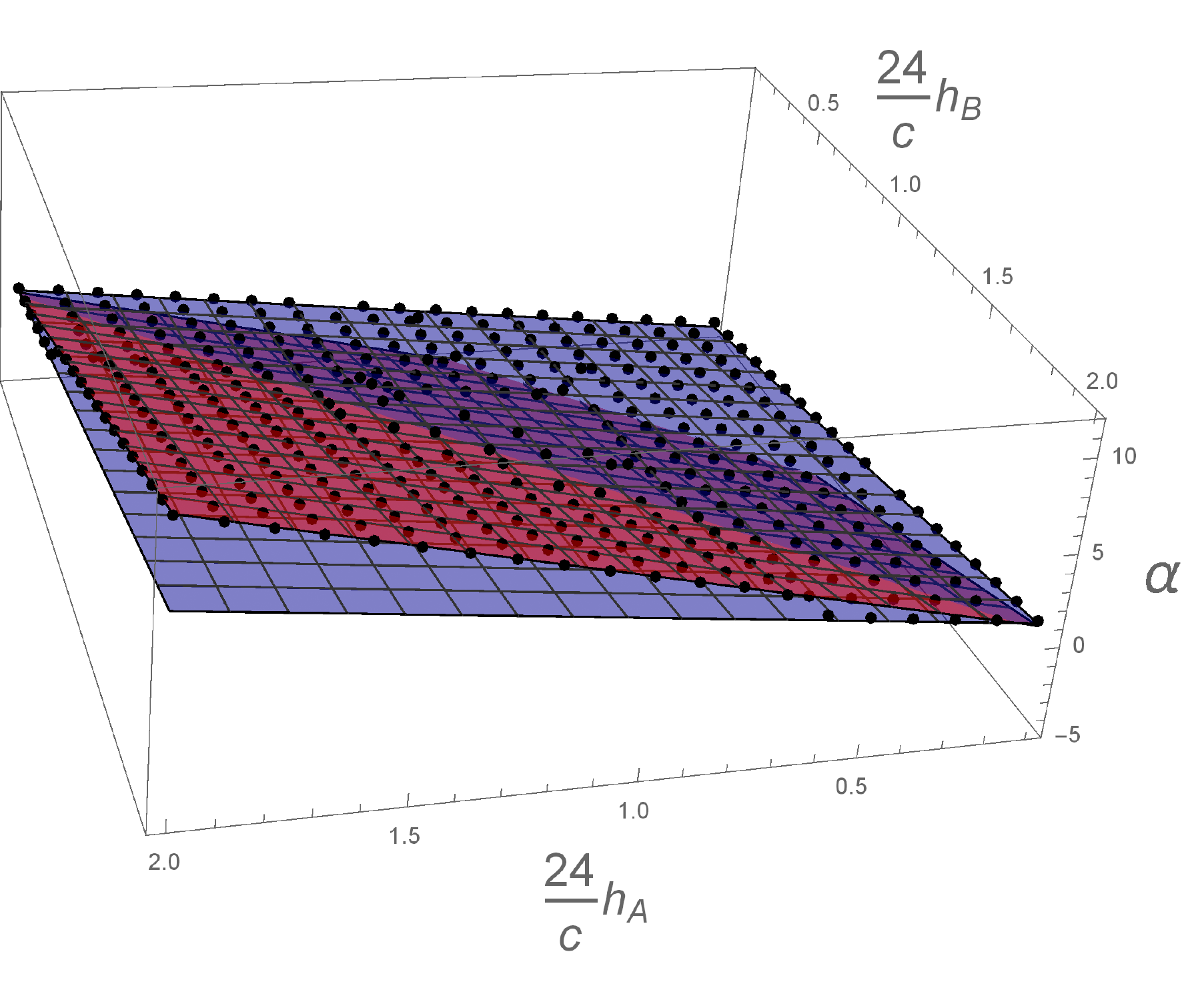}
   \end{center}
 \caption{The plots of the values of $A$ (upper) and $\a$ (lower) for various values of
 ($h_A,h_B$). The ranges are $0<h_A,h_B<\frac{c}{8}$. The black dots are the numerical values of $A$ and $\a$. The red surface describes (\ref{eq:AaHHABBA}) and the blue surfaces describes (\ref{eq:AaHLABBA}).  Some strange behaviors near the lines $h_{A,B}=\fr{c}{32}$ could be resolved by using the values $c_n$ for higher $n$ to fit $A$ and $\a$ (see Appendix \ref{subsec:app1}).
 Here we set $c=30.01$ and use the values of $c_n$ up to of $n=1000$ to fit $A$ and $\a$.}
\label{fig:Adep}
\end{figure}

Fitting the coefficients $c_n$ into the asymptotic form $n^{\a} \ex{A \s{n}}$ leads to Figure \ref{fig:Adep}.
From this observation, we can give the following expectations:
\begin{enumerate}
\item In the { \it heavy-heavy} region ($h_A>\fr{c}{32} $ and $h_B>\fr{c}{32}$), the coefficients $c_n$ have the simple asymptotic form  ($n\gg c$) described by
\begin{equation}\label{eq:AaHHABBA}
\begin{aligned}
A&=0,\\
\a&=4(h_A+h_B)-\fr{c+9}{4}.
\end{aligned}
\end{equation}

\item In the region where any $h_A$ and $h_B \ll c$, the asymptotic behavior of the coefficients $c_n$ ($n\gg c$) are determined by
\begin{equation}\label{eq:AaHLABBA}
\begin{aligned}
A&=2\pi \sqrt{ \frac{c-1}{24}-4h_B+\fr{c-1}{6}\pa{1-\s{1-\fr{24}{c-1}h_B}}},\\
\a&=2(h_A+h_B)-\fr{c+5}{8}.
\end{aligned}
\end{equation}
\end{enumerate}
These properties of ABBA blocks are very similar to AABB blocks. The difference from AABB blocks is that the sign of the coefficients $c_n$ and the value of $A$ in this region.

\begin{figure}[H]
  \begin{center}
   \includegraphics[width=120mm]{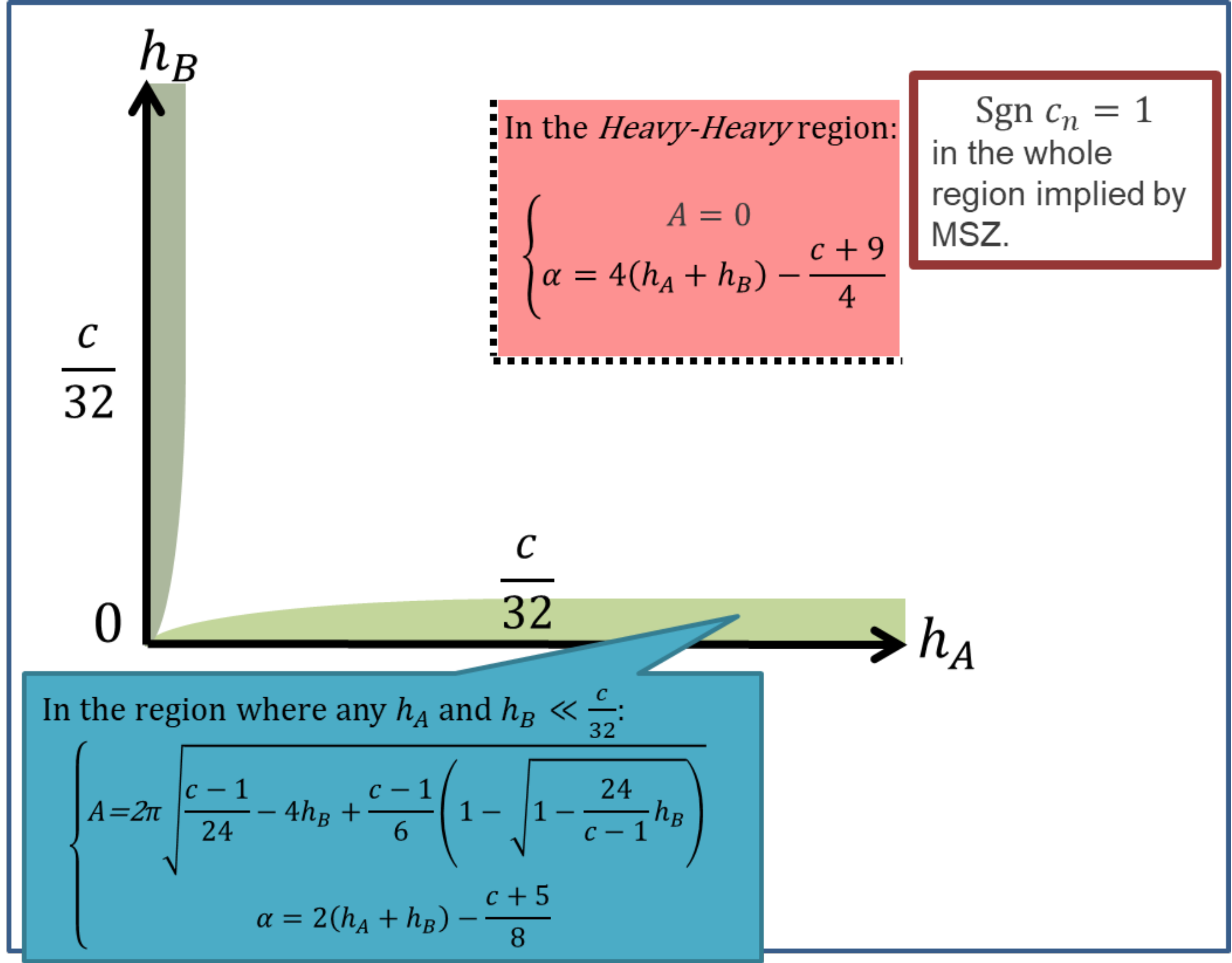}
   \end{center}
\caption{The sketch of behaviors of $c_n\sim  \mathrm{sgn}(c_n) n^\a \ex{A \s{n}} $ for various values of $(h_A,h_B)$.}
\label{fig:HLHL-hAhBdep}
\end{figure}

 In the end of this section, we summarize our results as in Figure \ref{fig:HLHL-hAhBdep}

%%%%%%%%%%%%%%%%%%%%%%%%%%%%%%%%%%%%%%%%%%%
%%%%%%%%%%%%%%%%%%%%%%%%%%%%%%%%%%%%%%%%%%%
\subsection{ABBA Blocks with General Intermediate States}
%%%%%%%%%%%%%%%%%%%%%%%%%%%%%%%%%%%%%%%%%%%
%%%%%%%%%%%%%%%%%%%%%%%%%%%%%%%%%%%%%%%%%%%

\begin{figure}[H]
  \begin{center}
   \includegraphics[width=70mm]{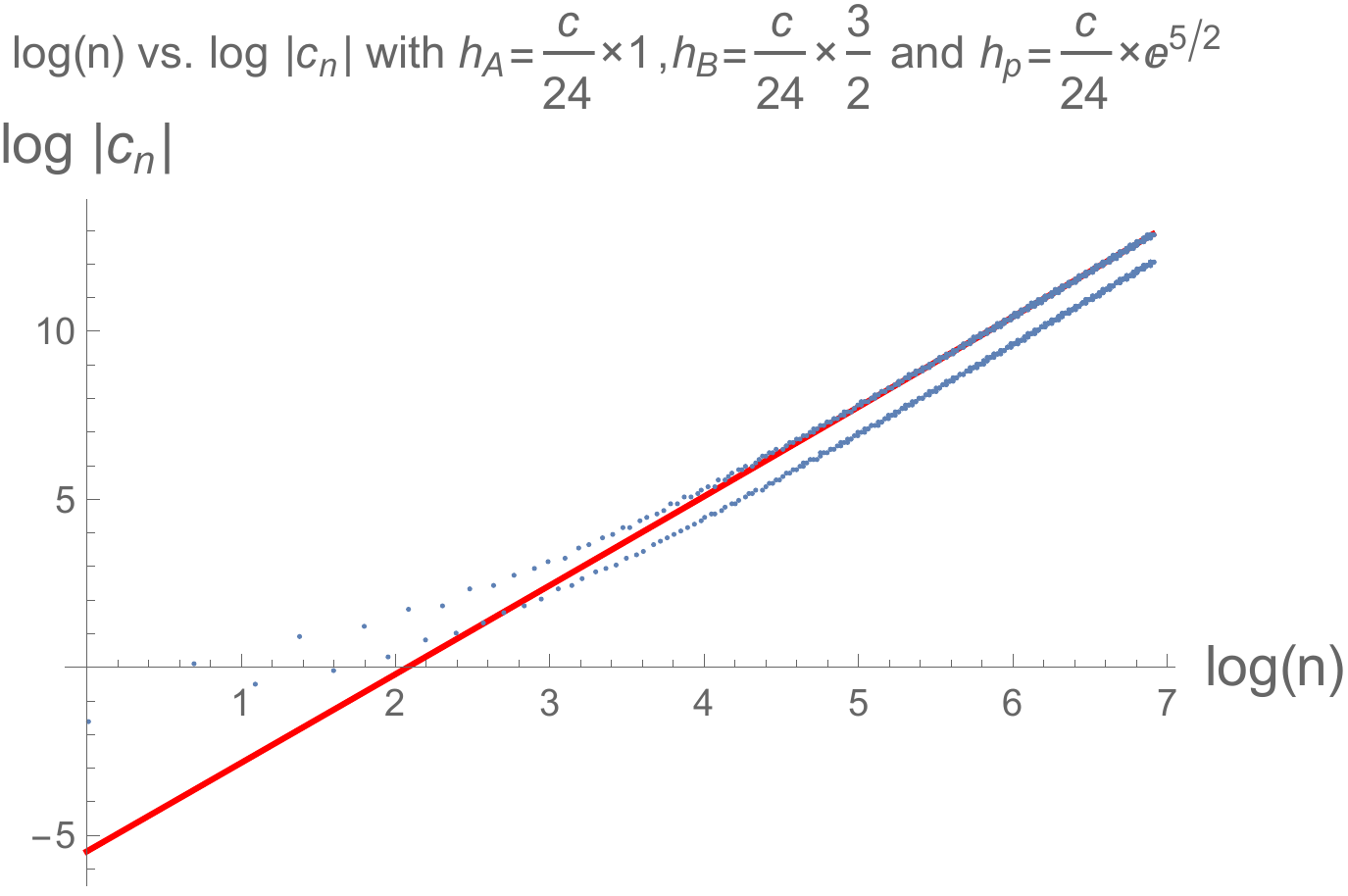}
   \includegraphics[width=70mm]{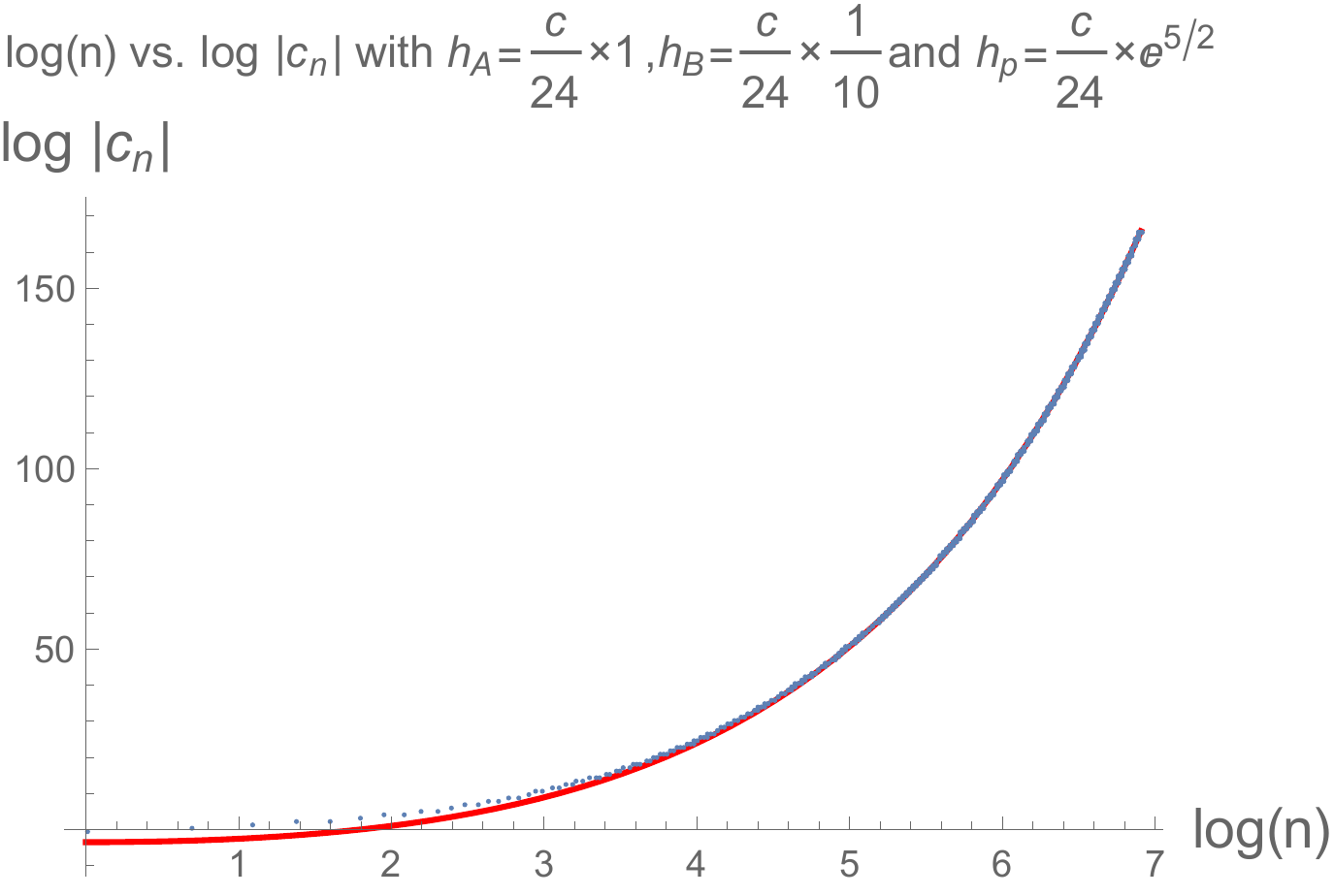}
   \end{center}
\caption{The behaviors of the coefficients $c_n$ of ABBA blocks with $h_A=\fr{c}{24}$. The left is for $(h_B,h_p)=(\fr{c}{16},\fr{c}{24} \times \ex{\fr{5}{2}})$ and the right is for $(h_B,h_p)=(\fr{c}{240},\fr{c}{24}\times  \ex{\fr{5}{2}})$. The blue dots are the numerical values of $\log c_n$. The red lines are $B n^\a \ex{A\s{n}}$ with the constant $B$ determined by the fit. We now set $c=30.01$ and to fit $A$ and $\a$, we use the numerical values of $c_n$ at $n=500\sim1000$.}
\label{fig:ABBAcn}
\end{figure}

\begin{figure}[H]
  \begin{center}
   \includegraphics[width=70mm]{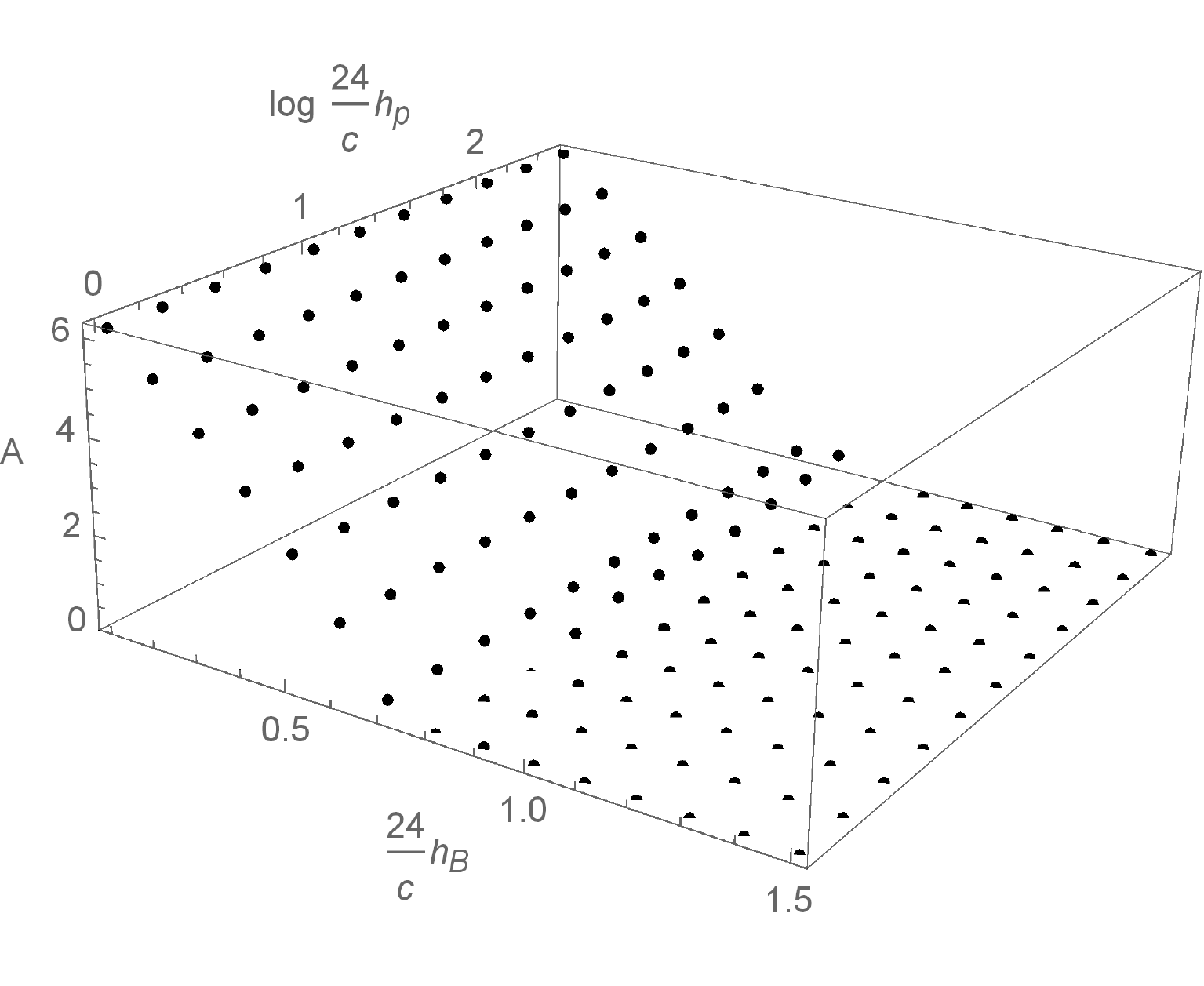}
   \includegraphics[width=70mm]{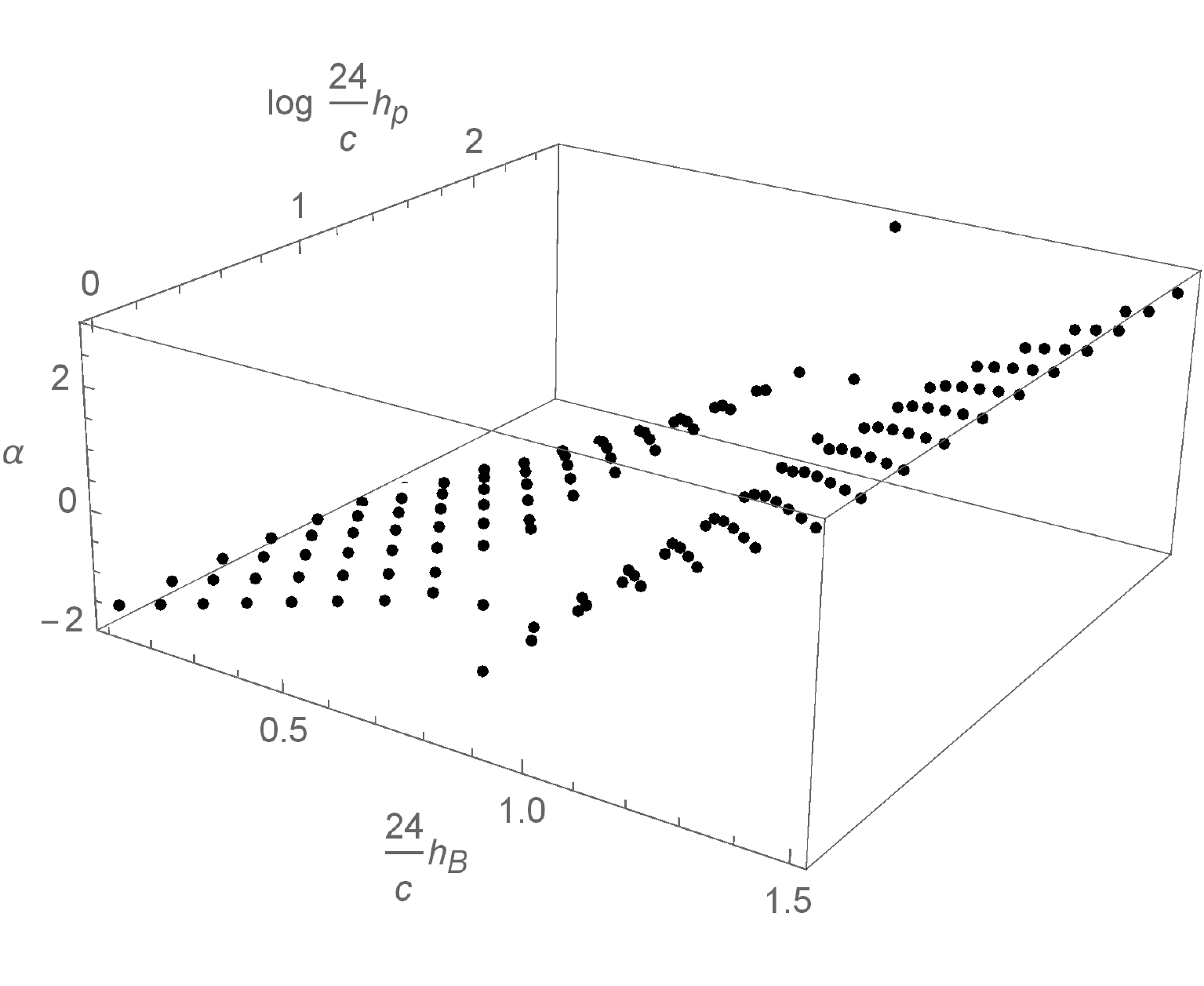}
   \end{center}
 \caption{The plots of the values of $A$ (left) and $\a$ (right) for various values of  ($h_B,h_p$) with $h_A=\fr{c}{24}$.  Some strange behaviors near the line $h_B=\fr{c}{32}$ could be resolved by using the values $c_n$ for higher $n$ to fit $A$ and $\a$. Here we set $c=30.01$ and to fit $A$ and $\a$, we use the numerical values of $c_n$ at $n=500\sim1000$.}
\label{fig:HLHLhBhpdep}
\end{figure}

The $h_p$ dependence of ABBA blocks with general intermediate dimensions has the similar feature to that of AABB. First of all, the coefficients $c_n$ are well-fitted by $n^\a \ex{A \s{n}}$ for large $n$ as in Figure \ref{fig:ABBAcn}.  We find that the behavior of the coefficients $c_n$ of general ABBA blocks also exhibits the asymptotic form (\ref{eq:cnas}). Figure \ref{fig:HLHLhBhpdep} shows the values of $A$ and $\a$ fitted by  (\ref{eq:cnas}) for various values of  ($h_B,h_p$) with $h_A=\fr{c}{24}$. One can see that the behaviors of $A$ and $\a$ drastically change at $h_{A,B}=\fr{c}{32}$ also for higher intermediate dimensions.

\begin{figure}[H]
  \begin{center}
   \includegraphics[width=70mm]{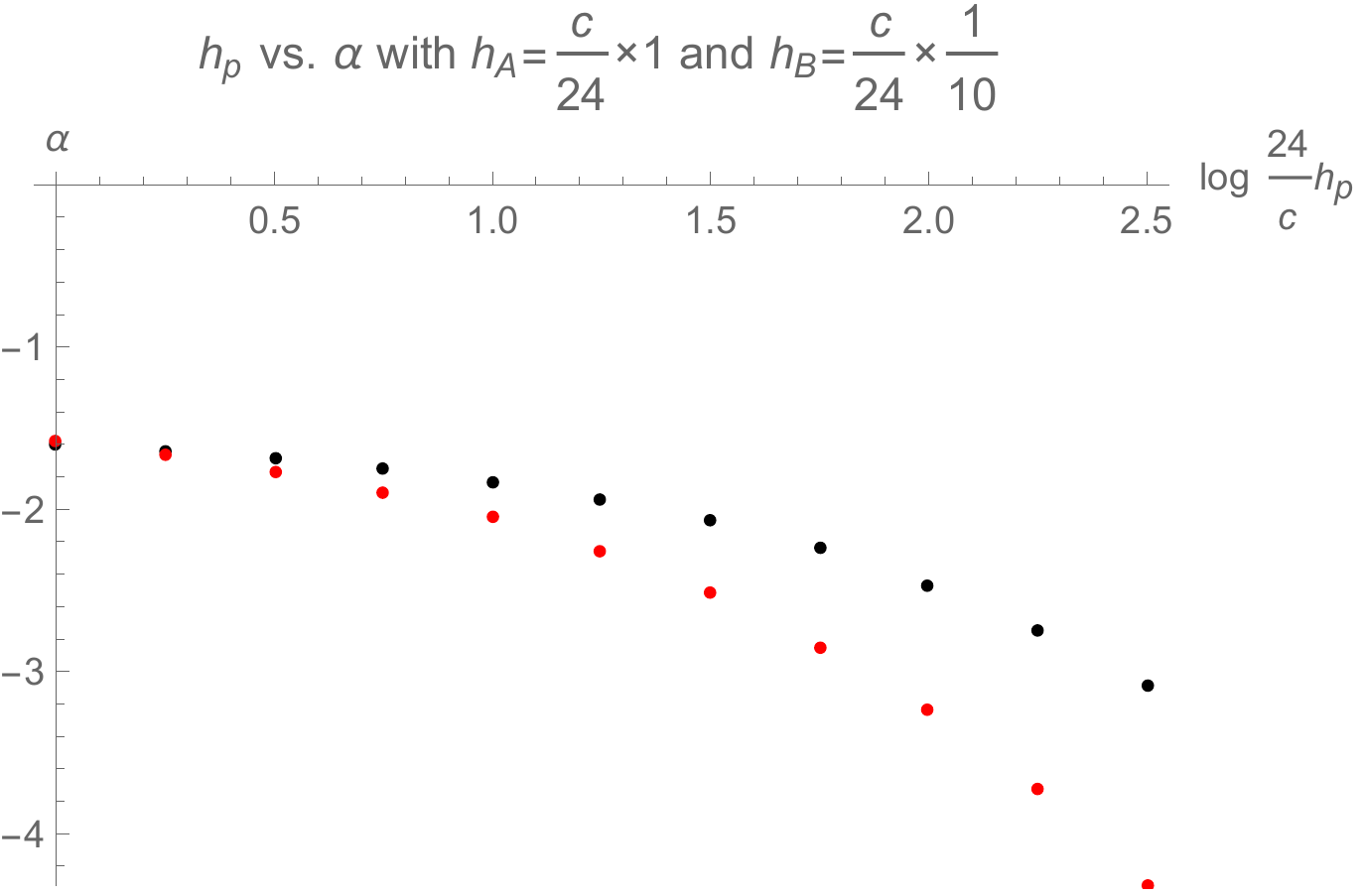}
   \includegraphics[width=70mm]{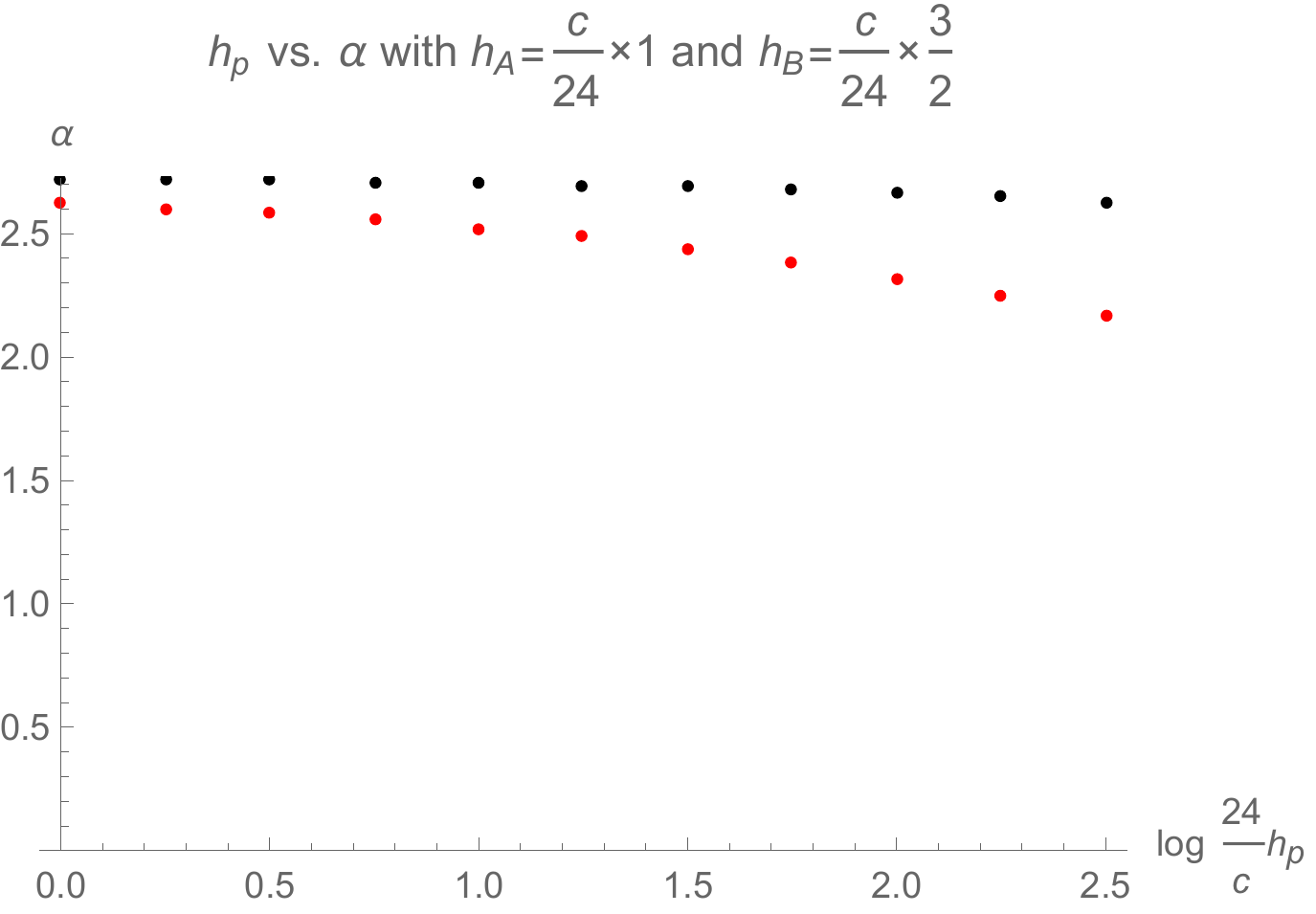}
   \end{center}
\caption{The $h_p$ dependence of $\a$. The left is for $(h_A,h_B)=(\fr{c}{24},\fr{c}{240})$ and the right is for $(h_A,h_B)=(\fr{c}{24},\fr{c}{16})$. Red dots are fitted by $c_n$ for $n=100\sim 200$ and Black dots are fitted by $c_n$ for $n=500\sim 1000$. One can find that the $h_p$ dependence of $\a$ approaches to constant as we use higher $n$ to fit the values of $\a$.}
\label{fig:HLHLNdep}
\end{figure}

From the left of Figure \ref{fig:HLHLhBhpdep}, we can see that the values of $A$ are independent of $h_p$ and thus the transition at $h_B=\fr{c}{32}$ continues to general $h_p$. And also we can see that the values of $\a$ do not depend on $h_p$ in the right of 
Figure \ref{fig:HLHLhBhpdep}. When looking at one $h_B$ slice more carefully in Figure \ref{fig:HLHLNdep}, the values of $\a$ seem to be decreasing with $h_p$. However, we think that this is due to the same reason as in Section \ref{subsec:AABBnon} because we can also see in Figure \ref{fig:HLHLNdep} that the less decreasing of $\a$ with $h_p$ we can see, the higher $n$ we use to fit $\a$.
Therefore, we conclude that the ABBA blocks are also insensitive to the intermediate dimensions $h_p$ in the limit $z \to 1$ in a similar way as the AABB block. 

Note that, as we explained in Section \ref{subsec:AABBnon}, our Cardy-like formura for the coefficients $c_n(h_p)$ could break down also for ABBA blocks if $n\sim h_p$. We can not find some simple formula for $c_n$ with $n \sim h_p$  in this paper. We will, though, exhibit some features of $c_n$ with $n \sim h_p$ in the next section \ref{sec:heavy}.


%%%%%%%%%%%%%%%%%%%%%%%%%%%%%%%%%%%%%%%%%%%%%%%%%%%%%%%%%%%%%%%%%%%%%%%%%%%%%%%%%%%%%%%%%%%%%%
%%%%%%%%%%%%%%%%%%%%%%%%%%%%%%%%%%%%%%%%%%%%%%%%%%%%%%%%%%%%%%%%%%%%%%%%%%%%%%%%%%%%%%%%%%%%%%
\section{Conformal Blocks with Very Heavy Intermediate States}\label{sec:heavy}
%%%%%%%%%%%%%%%%%%%%%%%%%%%%%%%%%%%%%%%%%%%%%%%%%%%%%%%%%%%%%%%%%%%%%%%%%%%%%%%%%%%%%%%%%%%%%%
%%%%%%%%%%%%%%%%%%%%%%%%%%%%%%%%%%%%%%%%%%%%%%%%%%%%%%%%%%%%%%%%%%%%%%%%%%%%%%%%%%%%%%%%%%%%%%

In this section, we study the $h_p$ dependence of the coefficients $c_n(h_p)$ with $n\sim h_p$. Before stating our results, we explain the motivation for it. It is known that the large $c$ conformal blocks with very heavy intermediate states ($h_p \gg  h_i, c$) can be derived by the monodromy method \cite{Zamolodchikov1984,Zamolodchikov1987,Harlow2011} and will be briefly explained in Section \ref{subsec:validity}. This block is given by
\begin{equation}\label{eq:largeh}
\ca{F}^{21}_{34}(h_p|z)=\Lambda^{21}_{34}(h_p|q),\ \ \ \ \ \ q(z)=\ex{-\pi \fr{K(1-z)}{K(z)}},
\end{equation}
where the function $\Lambda^{21}_{34}(h_p|q)$ is
\begin{equation}
 \Lambda^{21}_{34}(h_p|q)=(16q)^{h_p-\frac{c-1}{24}}z^{\frac{c-1}{24}-h_1-h_2}(1-z)^{\frac{c-1}{24}-h_2-h_3}
(\theta_3(q))^{\frac{c-1}{2}-4(h_1+h_2+h_3+h_4)}.
\end{equation}
This means that the function $H(h_p|q)$ has the following asymptotic form,
\begin{equation}\label{eq:H1}
H(h_p|q) \ar{h_p \to \infty} 1.
\end{equation}
Here, attention should be given to the fact that in the process of this approximation, the kinematic configuration is held fixed. If one would try to estimate correlators by using the saddle point approximation, the dependence on the kinematic configuration is important since the saddle points $h_{p^*}$ of the sum over intermediate states relate the kinematic configuration.

\begin{quote}
{\it Example:}

If the correlator in the limit $z,\bar{z} \to 1$ is dominated by only one saddle point, we can approximate the sum as
\begin{equation}
\sum_p C_{12p}C_{34p} \ca{F}^{21}_{34}(h_p|z)\overline{\ca{F}^{21}_{34}}(\bar{h}_p|\bar{z}) \ar{z,\bar{z} \to 1} 
C_{12p^*}C_{34p^*} \ca{F}^{21}_{34}(h_{p^*}|z)\overline{\ca{F}^{21}_{34}}(\bar{h}_{p^*}|\bar{z}) ,
\end{equation}
where $h_{p^*}$ depends on the kinematic configuration, $h_{p^*}=h_{p^*}(z)$.
\end{quote}
Therefore, the approximation (\ref{eq:H1}) under the fixed kinematics might be invalid. To be more explicit, we need to know the $h_p$ dependence of $c_n(h_p)$ not only for large $n \gg h_p$ but also for $n \sim h_p$ (see also Section \ref{subsec:TOC}). In fact, if one wants to estimate the asymptotic behavior of correlators from the knowledge of conformal blocks, one needs to know how $c_n(h_p)$ depends on $h_p$.
That's the motivation.

%%%%%%%%%%%%%%%%%%%%%%%%%%%%%%%%%%%%%%%%%%%
%%%%%%%%%%%%%%%%%%%%%%%%%%%%%%%%%%%%%%%%%%%
\subsection{Numerical Results for Very Heavy Intermediate States}\label{subsec:very heavy}
%%%%%%%%%%%%%%%%%%%%%%%%%%%%%%%%%%%%%%%%%%%
%%%%%%%%%%%%%%%%%%%%%%%%%%%%%%%%%%%%%%%%%%%

\begin{figure}[h]
 \begin{minipage}{0.5\hsize}
  \begin{center}
   \includegraphics[width=65mm]{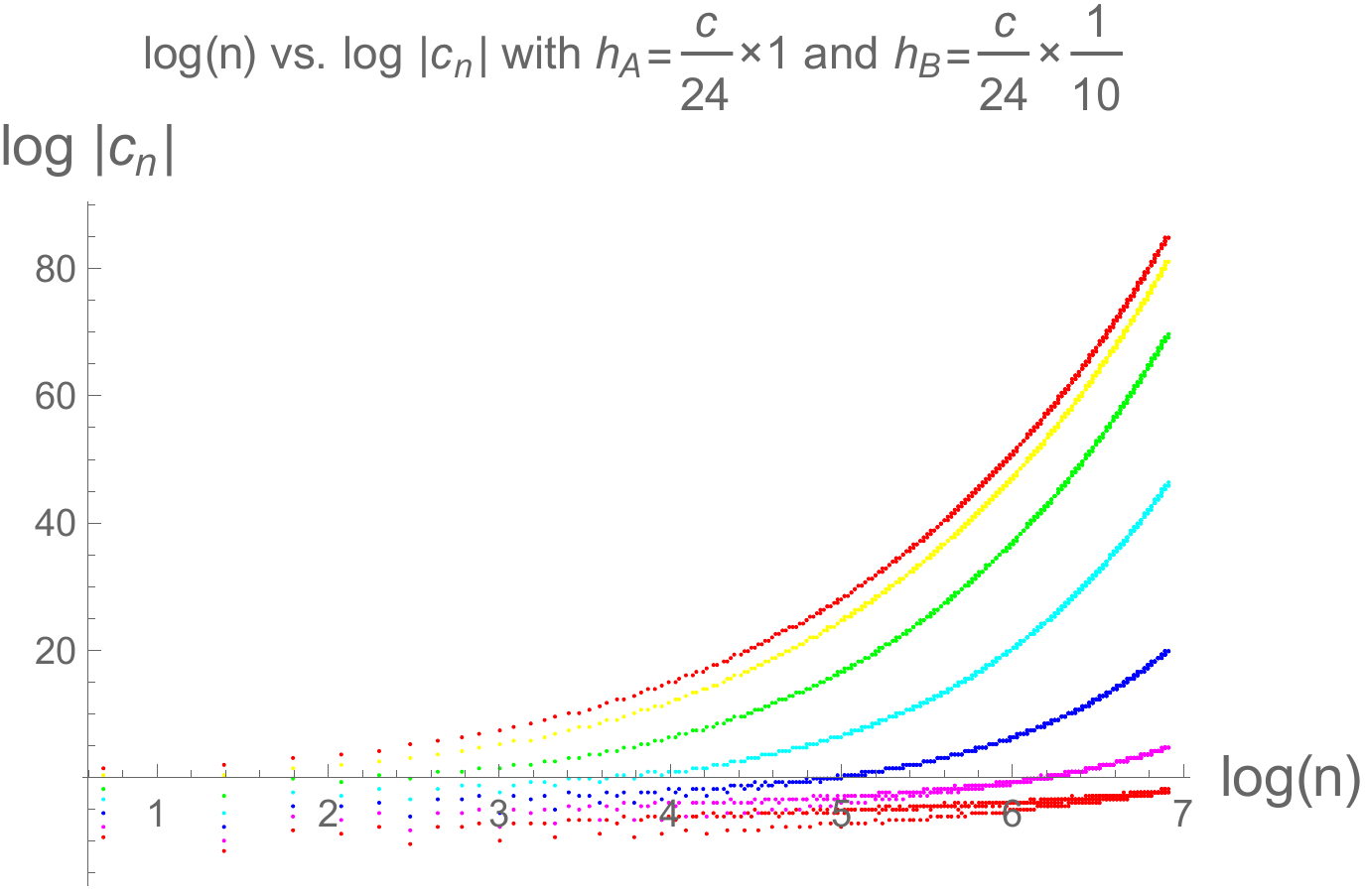}
  \end{center}
 \end{minipage}
 \begin{minipage}{0.5\hsize}
  \begin{center}
   \includegraphics[width=85mm]{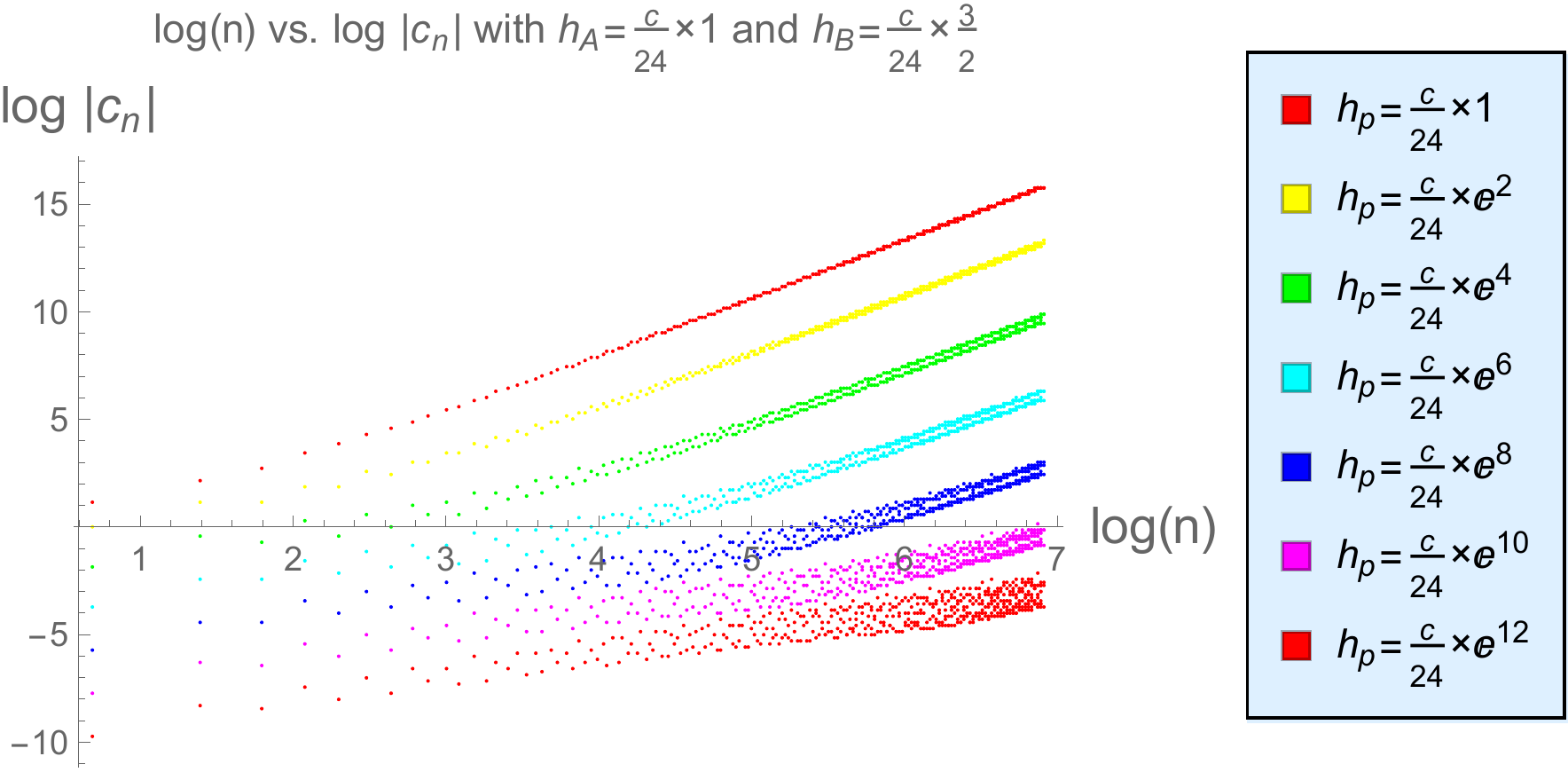}
  \end{center}
 \end{minipage}
 \begin{minipage}{0.5\hsize}
  \begin{center}
   \includegraphics[width=65mm]{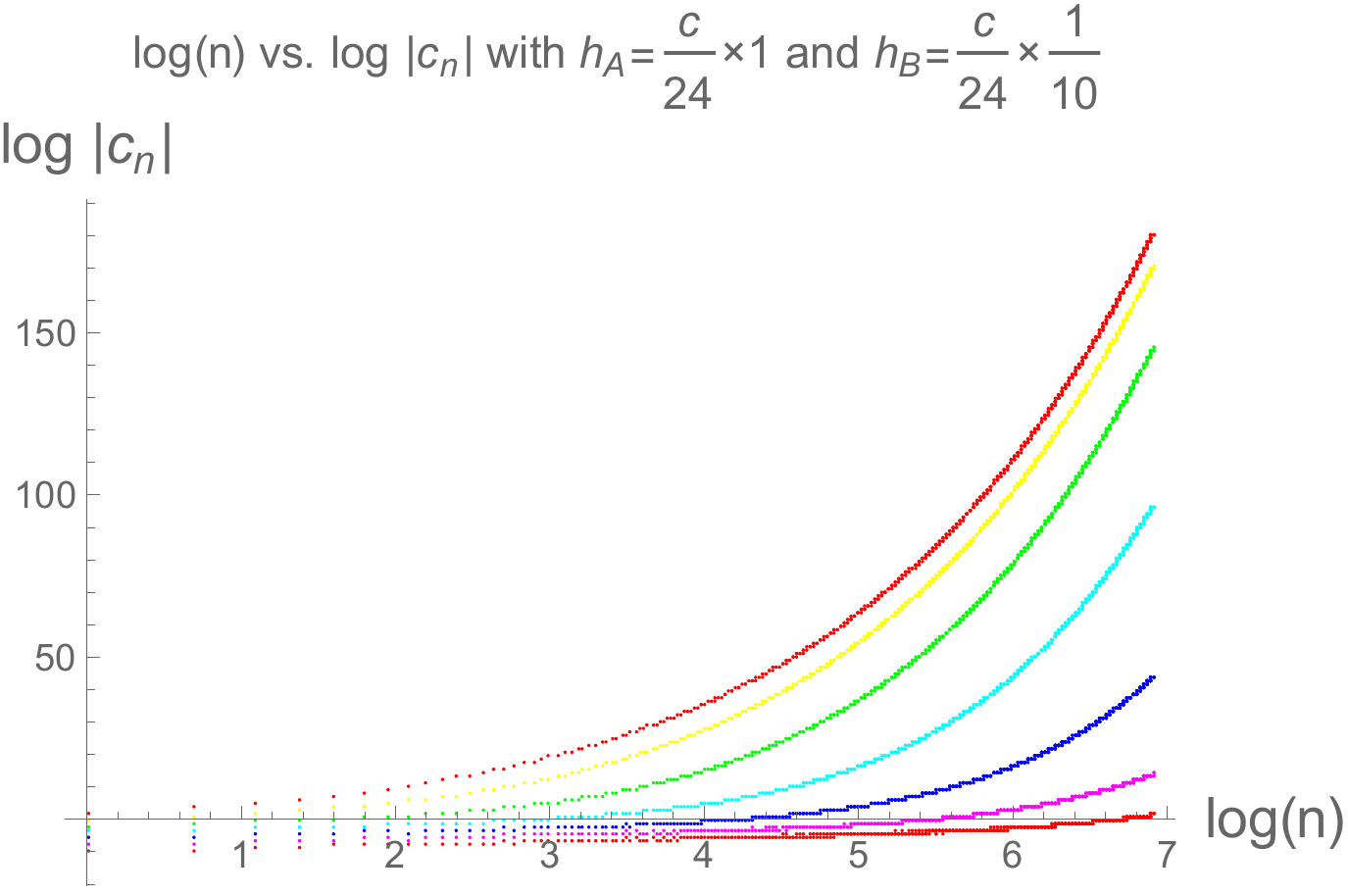}
  \end{center}
 \end{minipage}
 \begin{minipage}{0.0\hsize}
  \begin{center}
   \includegraphics[width=65mm]{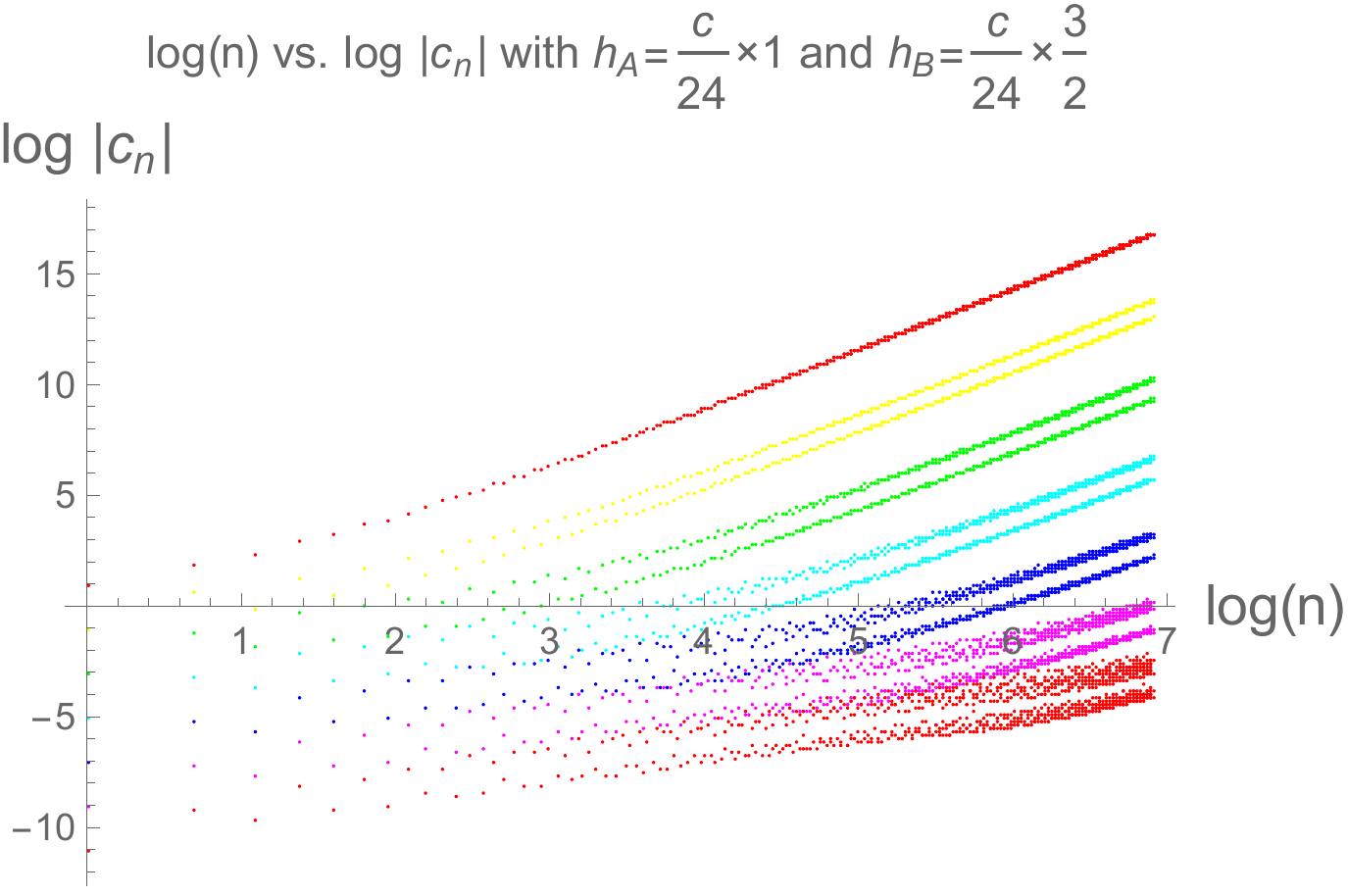}
  \end{center}
 \end{minipage}
 \caption{These figures show the behaviors of the coefficients $c_n$ of AABB (upper) and ABBA (lower) blocks simultaneously for various intermediate dimensions $h_p$. The left is for $(h_A,h_B)=(\fr{c}{24},\fr{c}{240})$ and the right is for $(h_A,h_B)=(\fr{c}{24},\fr{c}{16})$.}
\label{fig:simul}
\end{figure}

As mentioned in Section \ref{sec:AABB} and \ref{sec:ABBA}, for very heavy intermediate dimensions $h_p \sim n \gg c$, we can't describe the coefficients $c_n$ as a simple form. Nevertheless, we can find out few qualitative features for $c_n$ by directly observing the dependence of $c_n(h_p)$ with fixed $n$.
Before the observation, we compare the $n$ dependence of $c_n(h_p)$ with various $h_p$. Figure \ref{fig:simul} shows the behaviors of the coefficients $c_n$ of AABB and ABBA blocks for various intermediate dimensions $h_p$. From this figure, we can expect that the coefficients $\abs{c_n(h_p)}$ are monotonically increasing with $h_p$ in some sense. Moreover, we can expect that
\begin{equation}\label{eq:monoto}
\abs{c_n(h_p)}\geq \abs{c_n(h_p')} \ \ \ \ \text{if} \ \ h_p \leq h_p',
\end{equation}
for at least higher $n$. In fact, we find out the counterexample to (\ref{eq:monoto}) in a special case. If one considers AABB blocks with $h_A \sim h_B \sim \fr{c}{32}$, one can see the counterexample. Nevertheless, when the external dimensions are apart from the vicinity of $(h_A,h_B)=(\fr{c}{32},\fr{c}{32})$, the block satisfies the inequality (\ref{eq:monoto})  for any integer $n$. And at least, we can observe in general
\begin{equation}\label{eq:monoto2}
\abs{c_n(h_p \ll c)}\geq \abs{c_n(h_p' \nll c)}.
\end{equation}

To read off the $h_p$ dependence of the coefficients $c_n(h_p)$, we calculate $c_n(h_p)$ for various $h_p$ with fixed $n$. Here we showed only few our numerical plots and we tried not to disturb readers by too many figures. However in fact most of our plots exhibit similar properties and therefore one can see our conclusion from them. If one wants to confirm our conclusion by more examples, one can see other examples in Appendix \ref{subsec:extra}. Figure \ref{fig:AABBcnhp} shows the $h_p$ dependence of $c_n(h_p)$ with fixed $n=10 \text{ and } 1000$ for AABB blocks. The upper two figures are for $(h_A, h_B) = (\fr{c}{24}, \fr{c}{240})$, which is in the {\it heavy-light} region. These figures suggest that the $h_p$ dependence of $\log |c_n(h_p)|$ shows the steep slope from $h_p \sim n$. In other words, the coefficients $\abs{c_n(h_p)}$ behave like 
\begin{quote}
in the {\it heavy-light} region,
\begin{equation}\label{eq:veryH}
\begin{aligned}
\abs{c_n(h_p)}& \sim \left\{
    \begin{array}{ll}
     const. \ \   ,& \text{if } h_p \lesssim n  ,\\
     \pa{\fr{1}{h_p}}^{const.}  ,& \text{if } h_p \gtrsim n  .\\
    \end{array}
  \right.\\
\end{aligned}
\end{equation}
\end{quote}
This is just a rough estimate, but in fact, in the upper left of Figure \ref{fig:AABBcnhp}, $n=10$ is very small, thus most of $h_p$ satisfies $h_p \gtrsim n$ and therefore the $h_p$ dependence of $c_n$ is dominated by $  (h_p)^{const.}$ for most values of $h_p$. And in the upper right of Figure \ref{fig:AABBcnhp}, the $h_p$ dependence of $c_n$ with $n=1000$ shows gentle slope for $h_p \lesssim 1000$ and steep slope for $h_p \gtrsim 1000$. From these observation, we expect that the coefficients of $c_n(h_p)$ show the behavior as (\ref{eq:veryH}). We can see more clearly from Figure \ref{fig:AABBcnhp2} in Appendix \ref{subsec:extra}  that the transition point from a gentle slope to a steep slope is controlled by $n$. On the other hand, in the {\it heavy-heavy} region, we find more simple properties of $c_n(h_p)$. The lower two figures are for $(h_A, h_B) = (\fr{c}{24}, \fr{c}{16})$, which is in the {\it heavy-heavy} region. In this case, we can't see the transition from a gentle slope to a steep slope at $h_p \sim n$ and moreover the $h_p$ dependence of $\log |c_n|$ is clearly linear. Therefore, the statement is more probable in the {\it heavy-heavy} region than in the {\it heavy-light} region. From the above observation, we can suggest that for any $h_p$,
\begin{quote}
in the {\it heavy-heavy} region,
\begin{equation}\label{eq:veryH2}
\abs{c_n(h_p)} \sim \pa{\fr{1}{h_p}}^{\g(n)} ,
\end{equation}
\end{quote}
where $\g(n)$ is some constant for $h_p$.
Actually the same relations as (\ref{eq:veryH}) and  (\ref{eq:veryH2}) are satisfied for ABBA blocks (see Appendix \ref{subsec:extra}). 
This is one of main results in this paper. We must be able to extract this properties from the recursion relation analytically, which we leave for future work. And also it's important future work to explicitly identify  $c_n(h_p)$ as the function of $c, h_A, h_B, h_p, n$ with the aim of the motivation mentioned at the beginning of this section.

\begin{figure}[H]
  \begin{center}
   \includegraphics[width=80mm]{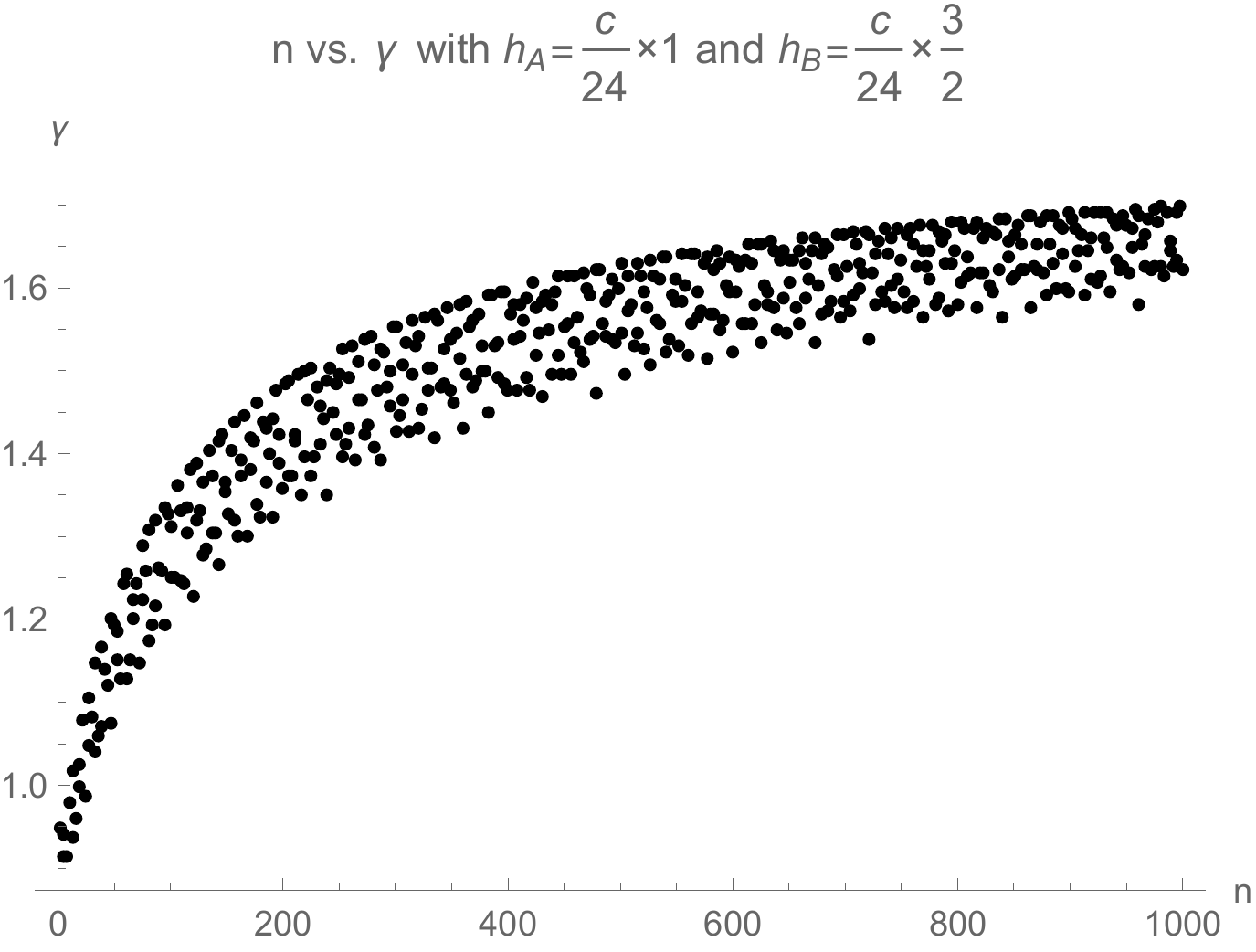}
   \end{center}
\caption{The $n$ dependence of $\g(n)$ for AABB blocks, which is the power of (\ref{eq:veryH2}). We can see that the growth of $\g(n)$ with $n$ is slower and slower as $n$ approaches infinity.}
\label{fig:gamma}
\end{figure}

Note that we can find that the power $\g(n)$ depends on $n$, however the growth of $\g(n)$ with $n$ is slower and slower as $n$ approaches infinity as in Figure \ref{fig:gamma}. This means that
\begin{equation}\label{eq:constgamma}
\abs{c_n(h_p)} \sim  \pa{\fr{1}{h_p}}^{\g}  \ \ \ \ \ \ \ \ \ \  \text{for large } n ,
\end{equation}
where $\g$ is some constant for $h_p$ and $n$. Therefore, for large $n$, the coefficients $c_n(h_p)$ can be split into two factors as
\begin{equation}
c_n(h_p) \sim P(h_p)Q(n),
\end{equation}
where $P(h_p)$ depends only on $h_p$ and $Q(n)$ depends only on $n$. This is consistent with our conjecture (\ref{eq:indepependent}), which states that the asymptotic behavior of the coefficients $c_n$ for large $n$ is independent of $h_p$ up to a constant factor. (Recall that our definition of ``$\sim$'' is the approximation up to a constant factor.) In other words, the function $Q(n)$ can be given by our Cardy-like formula for large $n$.
\begin{figure}[H]
 \begin{minipage}{0.5\hsize}
  \begin{center}
   \includegraphics[width=80mm]{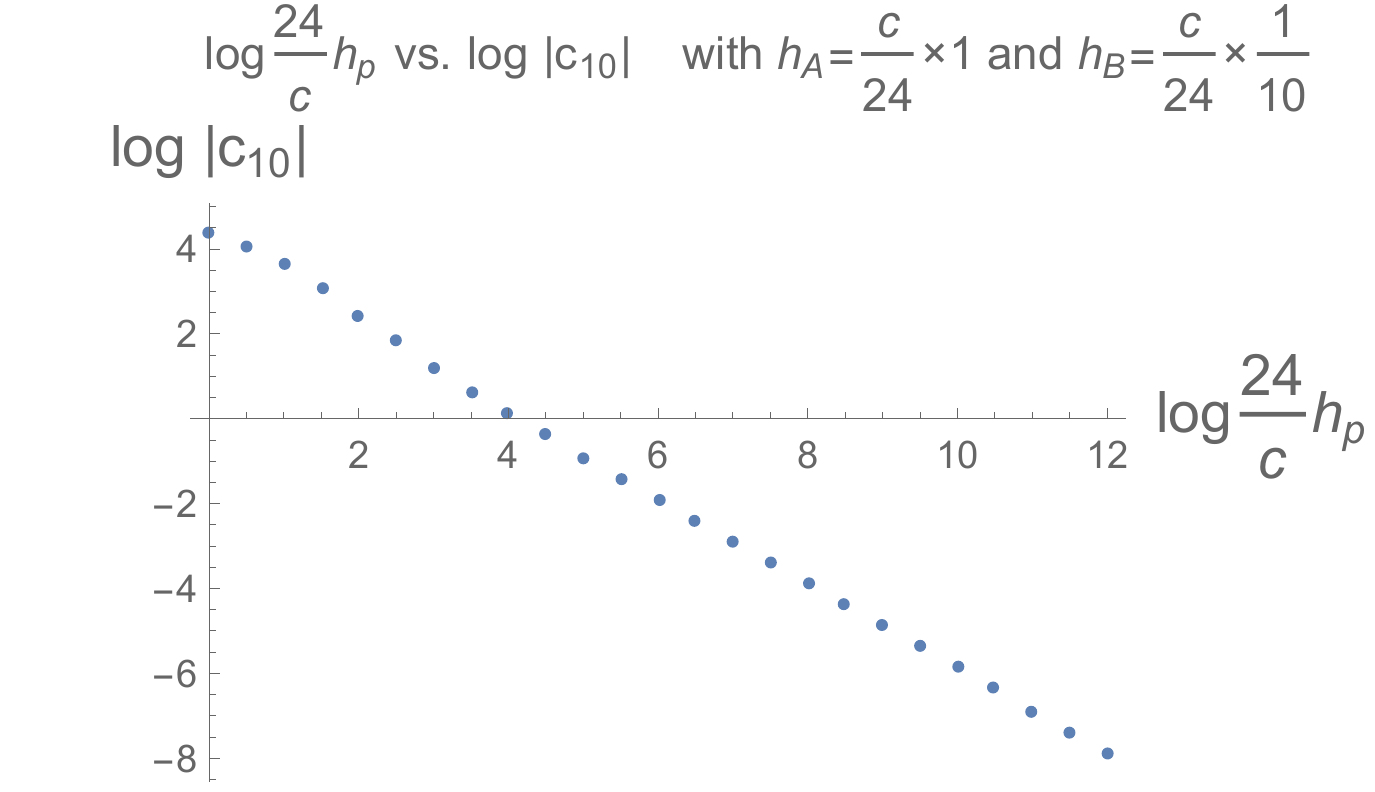}
  \end{center}
 \end{minipage}
 \begin{minipage}{0.5\hsize}
  \begin{center}
   \includegraphics[width=80mm]{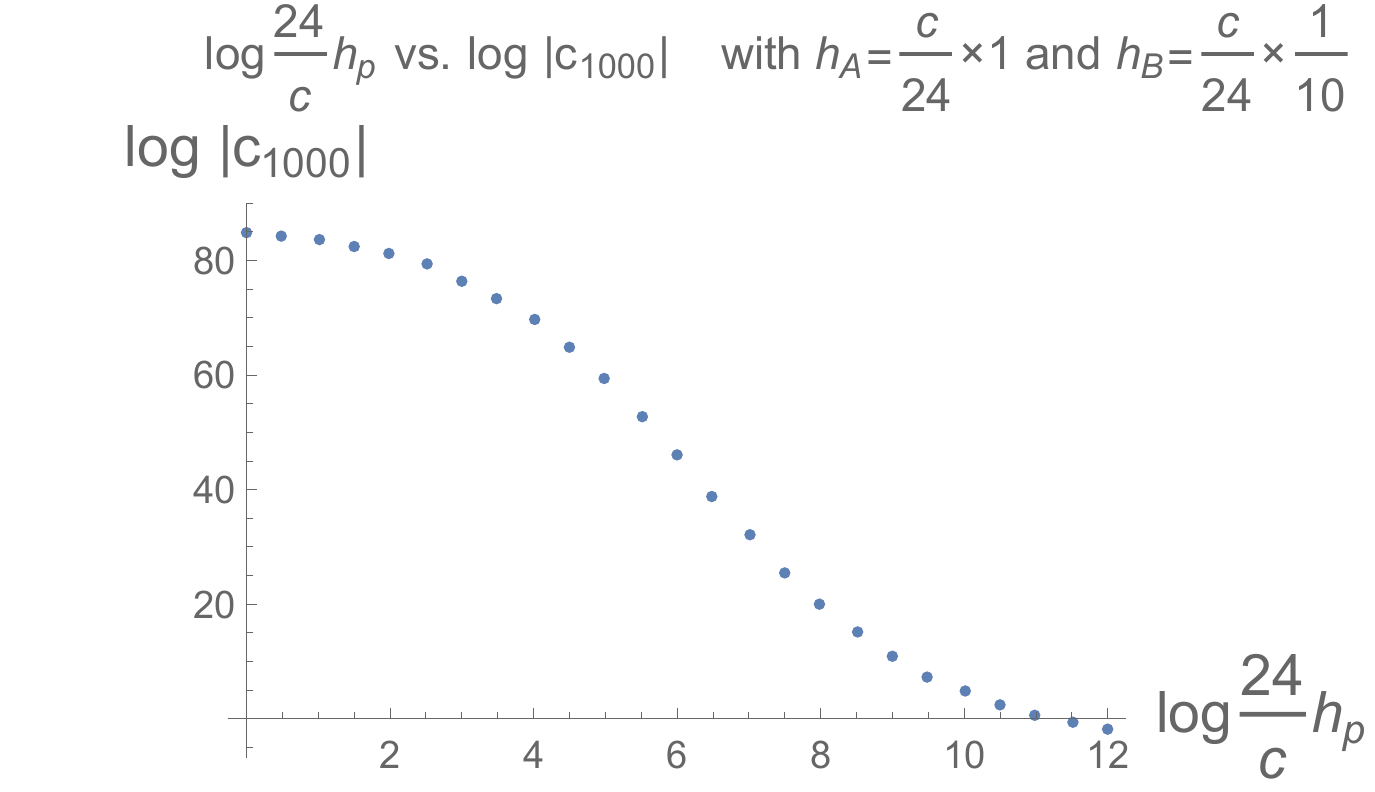}
  \end{center}
 \end{minipage}
 \begin{minipage}{0.5\hsize}
  \begin{center}
   \includegraphics[width=80mm]{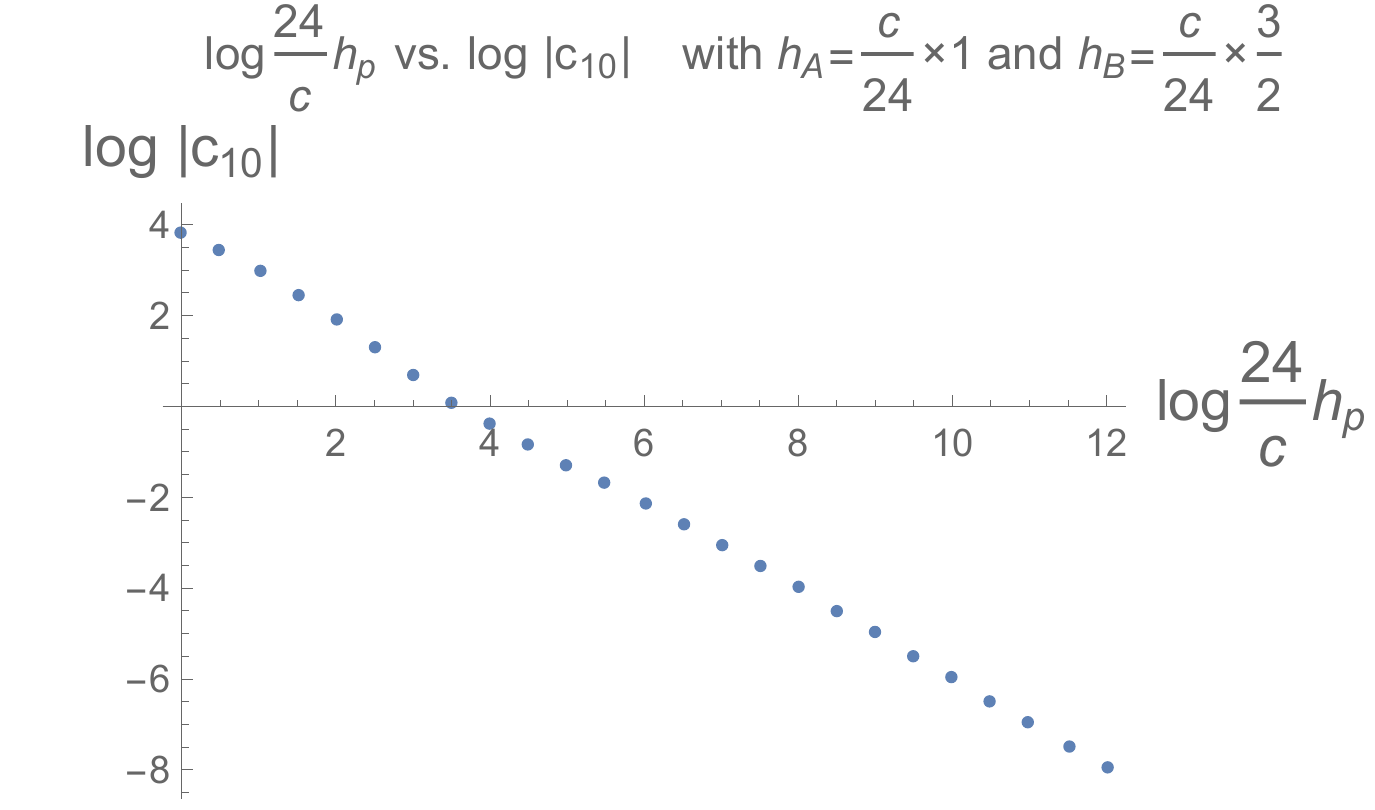}
  \end{center}
 \end{minipage}
 \begin{minipage}{0.5\hsize}
  \begin{center}
   \includegraphics[width=80mm]{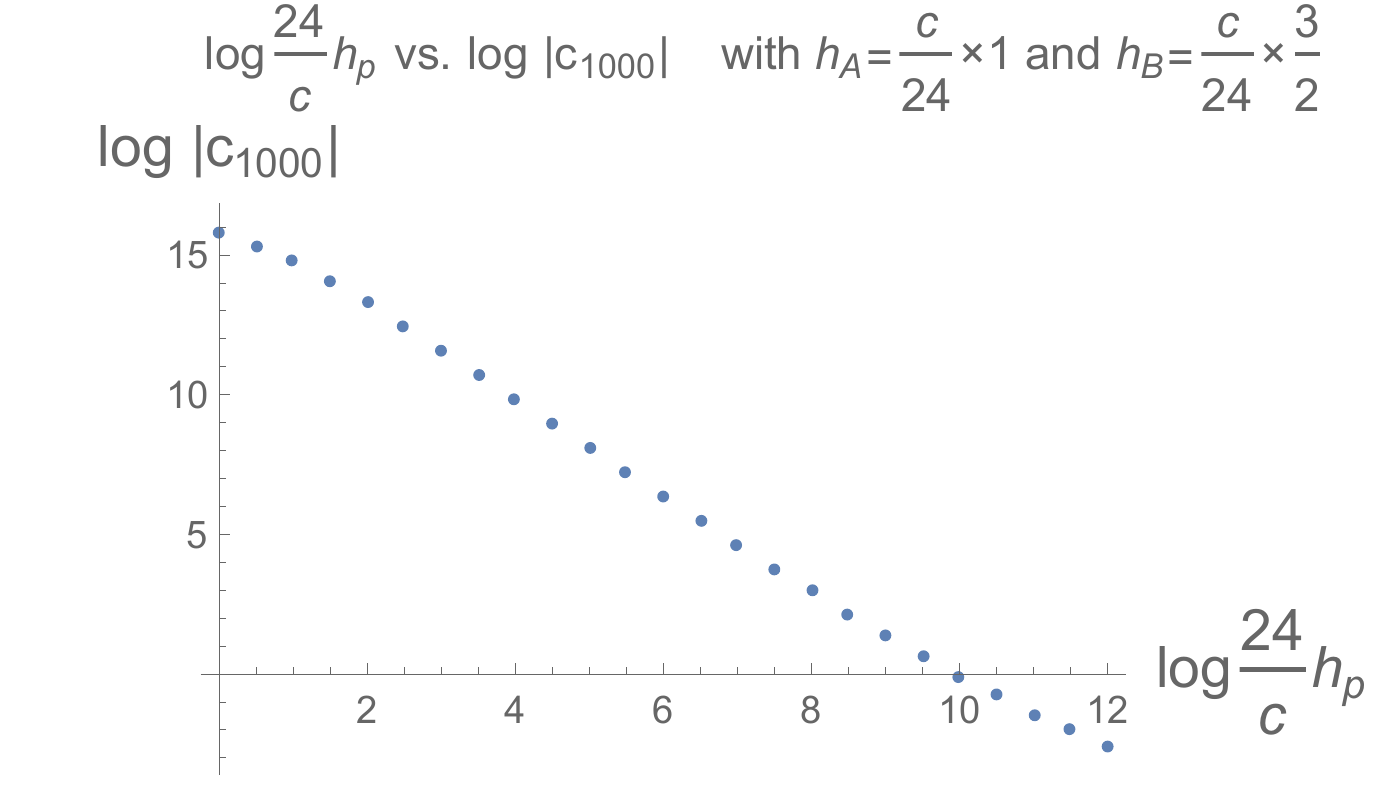}
  \end{center}
 \end{minipage}
\caption{The $h_p$ dependence of $c_n(h_p)$ with fixed $n=10$ (left) and $1000$ (right) for AABB blocks. The upper two figures are for $(h_A, h_B) = (\fr{c}{24}, \fr{c}{240})$, which is in the {\it heavy-light} region. The lower two figures are for $(h_A, h_B) = (\fr{c}{24}, \fr{c}{16})$, which is in the {\it heavy-heavy} region. }
\label{fig:AABBcnhp}
\end{figure}

%%%%%%%%%%%%%%%%%%%%%%%%%%%%%%%%%%%%%%%%%%%
%%%%%%%%%%%%%%%%%%%%%%%%%%%%%%%%%%%%%%%%%%%
\subsection{Validity of Large $h$ Asymptotics}\label{subsec:validity}
%%%%%%%%%%%%%%%%%%%%%%%%%%%%%%%%%%%%%%%%%%%
%%%%%%%%%%%%%%%%%%%%%%%%%%%%%%%%%%%%%%%%%%%
In the beginning of the section, we mentioned that if one considers the kinematics of the asymptotic blocks, one has to take care of  the regime of validity of the approximation.
Actually, we can identify the regime of validity of the approximation (\ref{eq:H1}) as
\begin{equation}\label{eq:validity}
h_p |\log q|^2 \gg c,
\end{equation}
where $h_p \to \infty$ and $q \to 1$. If the saddle point $h_p$ in the limit $q \to 1$ of the correlator satisfies (\ref{eq:validity}), then we can use the asymptotics,
\begin{equation}\label{eq:H11}
H(h_p|q) \ar{h_p \to \infty} 1.
\end{equation}
We will explain it in this subsection.

In this paper, we are interested in the holographic CFTs, therefore we restrict us to large $c$ CFTs. If the asymptotics (\ref{eq:H11}) is valid, it means that the monodromy method \cite{Zamolodchikov1984,Zamolodchikov1987,Harlow2011} can be justified.
The monodromy method is the method to derive the semiclassical conformal blocks as follows:
\begin{enumerate}

\item {\it Null ODE}\\
	The degenerate primary operator $\Psi$ with the dimension $-\fr{1}{2}-\fr{3}{4}b^2$ leads to the ODE,
	\begin{equation}\label{eq:ODE}
	\left[\frac{1}{b^2}\partial_z^2+\sum_{i=1}^4\left(\frac{h_i}{(z-z_i)^2}+\frac{1}{z-z_i}\partial_i\right)\right]
	\langle\mathcal{O}_4(z_4,\bar{z}_4)\mathcal{O}_3(z_3,\bar{z}_3)\Psi(z,\bar{z})\mathcal{O}_2(z_2,\bar{z}_2)\mathcal{O}_1(z_1,\bar{z}_1)\rangle=0.
	\end{equation}

\item {\it ODE for each intermediate states}\\
	Under some appropriate assumptions for large $c$ CFTs, the ODE (\ref{eq:ODE}) leads to a ODE for each intermediate states $\ca{O}_p$ in the OPE $\ca{O}_1 \ca{O}_2$ as
	\begin{equation}\label{eq:ODE4}
	\left[\partial_z^2+\sum_{i=1}^4\left(\frac{\delta_i}{(z-z_i)^2}-\frac{C_i}{z-z_i}\right)\right]\Psi_p=0,
	\end{equation}
	where $\delta_i=b^2h_i$ and
	\begin{equation}
	\langle\mathcal{O}_4\mathcal{O}_3\Psi\mathcal{O}_p\rangle \equiv 
	\Psi_p(z,\bar{z};z_i,\bar{z}_i)\langle\mathcal{O}_4\mathcal{O}_3\mathcal{O}_p\rangle.
	\end{equation}
	At this stage, we can not determine $C_i$, which are called as {\it accessory parameters}. This parameter is related to the conformal block as
\begin{equation}
C_2=\partial_x f_{cl},
\end{equation}
where
\begin{equation}\label{eq:semif}
\mathcal{F}^{21}_{34}(h_p|x)\sim \ex{-\frac{c}{6}f_{cl}}.
\end{equation}
\item {\it Ward-Takahashi identity}\\
	The second term of (\ref{eq:ODE4}) is can be understood as $b^2$ times the semiclassical expectation value of the stress tensor from the Ward-Takahashi identity. This fact leads to the following ODE,
	\begin{equation}\label{eq:ODE5}
	\left[\partial_z^2+\frac{\delta_1}{z^2}+\frac{\delta_2}{(z-x)^2}+\frac{\delta_3}{(1-z)^2}+\frac{\delta_1+\delta_2+\delta_3-\delta_4}{z(1-z)}-\frac{C_2 x(1-x)}{z(z-x)(1-z)}\right]\Psi_p=0.
	\end{equation}

\item {\it WKB approximation}\\
By using the WKB approximation in the limit $\d_p \to \infty$, we can solve the ODE (\ref{eq:ODE5}),
\begin{equation}\label{eq:Msolution}
\Psi_p\sim \exp\left[\pm\sqrt{x(1-x)C_2}\int_{z_0}^z\frac{dz'}{\sqrt{z'(1-z')(z'-x)}}\right].
\end{equation}

\item {\it Monodromy equation}\\
From  the usual CFT discussion for degenerate operators, we know  the OPE between $\ca{O}_p$ and $\Psi$ and therefore it is shown that the monodromy of $\Psi_p$ arond $\ca{O}_p$ can be given by
\begin{equation}
\pa{z-z_1}^{\fr{1}{2}\pa{1\pm\s{1-4b^2 h_p}}}.
\end{equation}
Hence, the solution (\ref{eq:Msolution}) needs to have the above monodromy. This fact leads to the condition,
\begin{equation}\label{eq:C2}
C_2\simeq -\frac{\pi^2 b^2 h_p}{x(1-x)K(x)^2}.
\end{equation}

\item {\it Semiclassical conformal block}\\
We have the relation
\begin{equation}
C_2=\partial_x f_{cl},
\end{equation}
and therefore we can obtain the conformal block as
\begin{equation}
\mathcal{F}^{21}_{34}(h_p|z)=\pa{16q}^{h_p},
\end{equation}
and the next order leads to the semiclassical block (\ref{eq:largeh}).
This method is called as {\it monodromy method}.
\end{enumerate}

As above, this method relies on the WKB approximation and therefore we have to take care of the regime of validity of this approximation. This regime is given by
\begin{equation}
\fr{1}{\la} \abs{\phi'^2} \gg \abs{\phi''},
\end{equation}
or
\begin{equation}
\la \abs{\fr{d}{dz}\fr{1}{\phi'}}\ll 1,
\end{equation}
where we define $\Psi_p\equiv\ex{\fr{1}{\la}\phi}$ and $h_p\equiv b^2 \fr{\eta_p}{\la}$, which are the usual convention for the WKB method. From (\ref{eq:ODE5}) and (\ref{eq:C2}), the leading order of $\phi'$ is given by
\begin{equation}
\phi' \simeq \s{-\fr{\la^2\pi^2 b^2 h_p}{K(x)^2z(z-x)(1-z)}}.
\end{equation}
As a result, we get the regime of validity as
\begin{equation}\label{eq:regime1}
\fr{h_p}{\abs{K(x)}^2} \gg c.
\end{equation}
In the limit $x\to 1$, we have the following asymptotics,
\begin{equation}
K(x)\sim \log (1-x) \sim \fr{1}{\log q(x)}.
\end{equation}
Therefore, we can reexpress (\ref{eq:regime1}) as
\begin{equation}\label{eq:regime2}
h_p \abs{\log q}^2\gg c.
\end{equation}
This is the regime of validity of the WKB approximation. In other words, the WKB solution $H(h_p|q)=1$ is valid only if $h_p \abs{\log q}^2\gg c$. Here, we don't claim that the lack of the condition (\ref{eq:regime2}) immediately leads to the breakdown of the asymptotic behavior (\ref{eq:H11}). It is just the breakdown of the WKB approximation, but it serves as a criterion of the breakdown, except for special cases.

In one of special cases, the solution from the WKB approximation is an exact solution to the ODE (\ref{eq:ODE5}) \cite{Maloney2017}. For example, if setting $\d_{1,2,3,4}=\fr{3}{16}$ (corresponding to $h_{1,2,3,4}=\fr{c}{32}$), then the ODE is solved by
\begin{equation}
\Psi_p^{(\pm)}(z)=\fr{1}{\s{t'(z)}}\ex{\pm i k t(z)}, \ \ \ \ \ \text{with } t'(z)=\fr{1}{\s{z(z-x)(z-1)}},
\end{equation}
where $C_2=\fr{1-2x+8k^2}{8x(1-x)}$. The monodromy condition leads to
\begin{equation}
C_2=\fr{1-2x}{8x(1-x)}+\fr{\pi^2\a^2}{16x(1-x)K(x)^2},
\end{equation}
where $\a=\s{1-4\d_p}$. This gives the conformal block as
\footnote{In large $c$, we can identify this conformal block with a character \cite{Cardy2017,Chang2016a},
\begin{equation}
{\ca{F}^{\fr{c}{32},\fr{c}{32}}_{\fr{c}{32},\fr{c}{32}}} (h_p|q)\sim\pa{z(1-z)}^{-\fr{c}{48}}\chi_{\fr{h_p}{2}, \fr{c}{2}}\pa{\tau}.
\end{equation}
We think that this relation relies on the fact that the value $\fr{c}{32}$ is the dimension of a twist-2 operator and a 4-pt. function of twist-2 operators is equivalent to a torus partition function.
}
\begin{equation}
\ca{F}^{\fr{c}{32},\fr{c}{32}}_{\fr{c}{32},\fr{c}{32}}(h_p|q)\sim \pa{16q}^{h_p-\fr{c}{24}}\pa{z(1-z)}^{-\fr{c}{48}}.
\end{equation}
This block is valid beyond the regime of validity of the WKB approximation.
\footnote{This might be relevant to the fact that the leading term of the coefficients $c_n$ (\ref{eq:lowercn}) vanishes when $h_{A} \text{ or } h_B=\fr{c}{32}$.} Other methods beyond the WKB approximation are discussed in \cite{Cardy2017,Fitzpatrick2017}.
%%%%%%%%%%%%%%%%%%%%%%%%%%%%%%%%%%%%%%%%%%%%%%%%%%%%%%%%%%%%%%%%%%%%%%%%%%%%%%%%%%%%%%%%%%%%%%
%%%%%%%%%%%%%%%%%%%%%%%%%%%%%%%%%%%%%%%%%%%%%%%%%%%%%%%%%%%%%%%%%%%%%%%%%%%%%%%%%%%%%%%%%%%%%%
\section{The Asymptotic Form of Conformal Blocks}\label{sec:block}
%%%%%%%%%%%%%%%%%%%%%%%%%%%%%%%%%%%%%%%%%%%%%%%%%%%%%%%%%%%%%%%%%%%%%%%%%%%%%%%%%%%%%%%%%%%%%%
%%%%%%%%%%%%%%%%%%%%%%%%%%%%%%%%%%%%%%%%%%%%%%%%%%%%%%%%%%%%%%%%%%%%%%%%%%%%%%%%%%%%%%%%%%%%%%

In this section , we estimate the simple form of the function $H(h_p|q)$ for real $q$ away from the origin $q=0$. (The function $H(h_p|q)$ in the limit $q \to 0$ is trivial and not interesting.) In order to extract the simple form of $H(h_p|q)$ , we approximate the summation
\begin{equation}\label{eq:appsum}
\sum_{n=0}^{\infty} n^\a \ex{A\s{n}}q^n
\end{equation}
by an integral, which is valid for $0 \ll q<1$.

%%%%%%%%%%%%%%%%%%%%%%%%%%%%%%%%%%%%%%%%%%%
%%%%%%%%%%%%%%%%%%%%%%%%%%%%%%%%%%%%%%%%%%%
\subsection{HLLH and LHHL Blocks}\label{subsec:HLLHblock}
%%%%%%%%%%%%%%%%%%%%%%%%%%%%%%%%%%%%%%%%%%%
%%%%%%%%%%%%%%%%%%%%%%%%%%%%%%%%%%%%%%%%%%%

Now that we have shown the simple asymptotic proprieties of the coefficients $c_n$ in the heavy-light limit, one might try to reconstruct conformal blocks. In this subsection, we focus on the ABBA block with external dimensions $h_1=h_4=h_A$ and $h_2=h_3=h_B$ because it has positive coefficients and therefore we can estimate the block easily in the following.  Note that, fortunately, especially in the heavy-light limit, it can be seen that the asymptotic form (\ref{eq:AaHLABBA}) also holds for small $n$ (see Appendix \ref{app:fitting}) and therefore the approximation by substituting our asymptotic form into (\ref{eq:appsum}) is good especially in this case. We can estimate the function $H(h|q)$ in the limit $z=1-\e$ ($\e \ll 1$) as
\begin{equation}\label{eq:A=!0}
H(h_p|q)=
\sum_{n=0}^{\infty} n^\a \ex{A\s{n}}q^n
\ar{\e \to 0} \pa{\log \e}^{2\a+\fr{3}{2}}\e^{-\fr{A^2}{4\pi^2}},
\end{equation}
where we use the following asymptotic behavior of the elliptic nome $q$,
\begin{equation}
 q(z)=\ex{-\pi \fr{K(1-z)}{K(z)}} \ar{\e \to 0} \ex{-\frac{\pi^2}{\log(16/\ep)}}.
\end{equation}
and the LHHL conformal blocks have the following asymptotic behavior,
\begin{equation}
\log \ca{F}^{HL}_{HL}(h_p|z)\ar{z \to 1} 
\pa{4h_L-2h_H-\fr{c-1}{6}\pa{1-\s{1-\fr{24}{c-1}h_L}}}\log(1-z) +o(\log\log(1-z)).
\end{equation}
On the other hand, in the same way, we can obtain the asymptotic behavior of HLLH blocks as 
\begin{equation}
\log \ca{F}^{LH}_{LH}(h_p|z)\ar{z \to 1} 
\pa{2h_L-\fr{c-1}{6}\pa{1-\s{1-\fr{24}{c-1}h_L}}}\log(1-z) +o(\log\log(1-z)).
\end{equation}
%%%%%%%%%%%%%%%%%%%%%%%%%%%%%%%%%%%%%%%%%%%
%%%%%%%%%%%%%%%%%%%%%%%%%%%%%%%%%%%%%%%%%%%
\subsection{HLHL and LHLH Blocks}
%%%%%%%%%%%%%%%%%%%%%%%%%%%%%%%%%%%%%%%%%%%
%%%%%%%%%%%%%%%%%%%%%%%%%%%%%%%%%%%%%%%%%%%

For ABAB blocks, the sign of coefficients $c_n$ oscillates and therefore we can not approximate the blocks by the same method as in Section \ref{subsec:HLLHblock} (see also Appendix \ref{subsec:HHLL}). Nevertheless, we have the inequality,
\begin{equation}
\sum_{n=0}^{\infty} c_n q^n \leq \sum_{n=0}^{\infty} \abs{c_n} q^n, \ \ \ \ \ (0<q<1).
\end{equation}
By combining this inequality, the results in Section  \ref{subsec:HLLHblock} and the equation  (\ref{eq:symq}), we get
\begin{equation}
\log \ca{F}^{LH}_{HL}(h_p|z)\lleq{z \to 1}  \pa{3h_L-h_H-\fr{c-1}{6}\pa{1-\s{1-\fr{24}{c-1}h_L}}}\log(1-z),
\end{equation}
and we can obtain the same result for LHLH block.
Here, we mean by the symbol ``$\lleq{z \to 1}$'' that an inequality holds only if $z \to 1$.

As mentioned in the last of Section \ref{subsec:AABBnon}, we can take the limit $q \to i$ by picking up the monodromy at $z =1$ and taking the limit $z \to 0$ ( in Section \ref{subsec:OTOC}, we will explain when this limit appears in more detail.).
In the similar way, we can find the limit of $z$ corresponding to the limit $q \to -1$. In fact, we can  take the limit $q \to -1$ by taking the limit $z \to \infty$ as
\begin{equation}
 q(z)=\ex{-\pi \fr{K(1-z)}{K(z)}} \ar{z=\fr{1}{\e} \to \infty} -\ex{-\frac{\pi^2}{\log(16/\ep)}}.
\end{equation}
In this limit, the block is given by
\begin{equation}
\begin{aligned}
&\log \ca{F}^{LH}_{HL}(h_p|z)
\ar{z \to \infty}
\pa{4h_L-\fr{c-1}{6}\pa{1-\s{1-\fr{24}{c-1}h_L} }}\log\pa{\fr{1}{z}} \\
& \hspace{3cm} +o\pa{\log\log\pa{\fr{1}{z}}},
\end{aligned}
\end{equation}
where we used the following property of the Jacobi theta function,
\begin{equation}
\theta_3(q(z))\ar{z=\fr{1}{\e}\to \infty}2\s{\fr{\log \fr{16}{\e}}{\pi}}\pa{\fr{\e}{16}}^{\fr{1}{4}}.
\end{equation}
In the same way, we can show that the LHLH block is given by the same expression.

%%%%%%%%%%%%%%%%%%%%%%%%%%%%%%%%%%%%%%%%%%%
%%%%%%%%%%%%%%%%%%%%%%%%%%%%%%%%%%%%%%%%%%%
\subsection{HHHH Blocks}
%%%%%%%%%%%%%%%%%%%%%%%%%%%%%%%%%%%%%%%%%%%
%%%%%%%%%%%%%%%%%%%%%%%%%%%%%%%%%%%%%%%%%%%
In this subsection, we study the asymptotics of ABBA blocks in the {\it heavy-heavy} region, which we call as {\it H$_A$H$_B$H$_B$H$_A$ block}.
The expressions (\ref{eq:AaHHABBA}) for $A$ and $\alpha$ in the {\it heavy-heavy} region lead to
\begin{equation}
H^{BA}_{BA}(h_p|q)=
\sum_{n=0}^{\infty} n^\a q^n
\ar{\e \to 0}  \pa{\log \e}^{4(h_A+h_B)-\fr{c+5}{4}},
\end{equation}
where $z=1-\e$ $(\e \ll 1)$. Here we use the asymptotic behavior of the Polylogarithm,
\begin{equation}
\mathrm{Li}_{-\a}(\ex{-\pi \tau})=\sum_{n=1}^{\infty} n^{\a}\ex{-\pi  \tau n} \ar{\tau \to 0} \fr{\Gamma(\a+1)}{\pa{\pi \tau}^{\a+1}}
\ \ \ \ \text{if } \a > -1,
\end{equation}
and we assumed $\a > -1$. In fact, almost all values $(h_A, h_B)$ in the {\it heavy-heavy} region satisfy $\a=4(h_A+h_B)-\fr{c+9}{4}>-1$, however  somehow there is a region where $\a>-1$ is not satisfied. It suggests that the exact transition point might be $h_{A,B}=\fr{c}{32}+O(c^0)$.
Note that somehow, this asymptotic form exactly matches $H^{AA}_{BB}(h_p|q)$, which is the function $H(h_p|q)$ of AABB blocks.

This $H^{BA}_{BA}$ function leads to the asymptotic form of conformal blocks in the {\it heavy-heavy} region ($h_{A,B}>\fr{c}{32}$) as
\begin{equation}\label{eq:appABBA}
\log \ca{F}^{BA}_{BA}(h_p|z)\ar{z \to 1}
\pa{\fr{c-1}{24}-2h_B}\log \pa{1-z}            -\fr{3}{2} \log \log \pa{1-z}        +    o(\log\log(1-z)),
\end{equation}
which is surprisingly simple. It means that it might be possible to derive this result analytically in some way. But we leave this problem to future work.
And for now, we do not have the clear holographic dual  description of ABBA blocks, however  this simple form also suggests that the H$_A$H$_B$H$_B$H$_A$ blocks could have some classical description in gravity side.

As we mentioned, the function $H^{AA}_{BB}(h_p|q)$ shows the same behavior as $H^{BA}_{BA}(h_p|q)$ in the {\it heavy-heavy} region. Therefore, the H$_A$H$_A$H$_B$H$_B$ block is also given by
\begin{equation}
\log \ca{F}^{AA}_{BB}(h_p|z)\ar{z \to 1}
\pa{\fr{c-1}{24}-h_A-h_B}\log \pa{1-z}            -\fr{3}{2} \log \log \pa{1-z}        +    o(\log\log(1-z)).
\end{equation}
We can see the power $\fr{3}{2}$, which is due to the same reason as that the power law $t^{-\fr{3}{2}}$ appears in the late time behavior of Virasoro blocks \cite{Chen2017}. To probe information loss, one needs to consider the analytic continuation of the correlator  \cite{Fitzpatrick2016a, Chen2017}. In more detail, we have to consider the conformal blocks undergoing a monodromy around $z=1$, whose behaviors are different from the original conformal blocks. This leads to the power law $t^{-\fr{3}{2}}$ for H$_A$H$_A$H$_B$H$_B$ blocks at late times.

For H$_A$H$_B$H$_A$H$_B$ blocks, we can obtain the bound form (\ref{eq:appABBA}) as
\begin{equation}\label{eq:appABBA2}
\log \ca{F}^{BA}_{AB}(h_p|z)\lleq{z \to 1}
\pa{\fr{c-1}{24}-h_A-h_B}\log \pa{1-z}            -\fr{3}{2} \log \log \pa{1-z}        +    o(\log\log(1-z)),
\end{equation}
and the asymptotic form as
\begin{equation}
\log \ca{F}^{BA}_{AB}(h_p|z)\ar{z \to \infty}
\fr{c-1}{24}\log \pa{\fr{1}{z}}              -\fr{3}{2} \log \log \pa{\fr{1}{z}}        +    o\pa{\log\log\pa{\fr{1}{z}}}.
\end{equation}
%%%%%%%%%%%%%%%%%%%%%%%%%%%%%%%%%%%%%%%%%%%%%%%%%%%%%%%%%%%%%%%%%%%%%%%%%%%%%%%%%%%%%%%%%%%%%%
%%%%%%%%%%%%%%%%%%%%%%%%%%%%%%%%%%%%%%%%%%%%%%%%%%%%%%%%%%%%%%%%%%%%%%%%%%%%%%%%%%%%%%%%%%%%%%
\section{Correlator, OTOC and Entanglement Entropy from Conformal Blocks}\label{sec:OTOC}
%%%%%%%%%%%%%%%%%%%%%%%%%%%%%%%%%%%%%%%%%%%%%%%%%%%%%%%%%%%%%%%%%%%%%%%%%%%%%%%%%%%%%%%%%%%%%%
%%%%%%%%%%%%%%%%%%%%%%%%%%%%%%%%%%%%%%%%%%%%%%%%%%%%%%%%%%%%%%%%%%%%%%%%%%%%%%%%%%%%%%%%%%%%%%

%%%%%%%%%%%%%%%%%%%%%%%%%%%%%%%%%%%%%%%%%%%
%%%%%%%%%%%%%%%%%%%%%%%%%%%%%%%%%%%%%%%%%%%
\subsection{Asymptotic Correlators}\label{subsec:TOC}
%%%%%%%%%%%%%%%%%%%%%%%%%%%%%%%%%%%%%%%%%%%
%%%%%%%%%%%%%%%%%%%%%%%%%%%%%%%%%%%%%%%%%%%
Now that we give the asymptotic form of conformal blocks with vacuum and non-vacuum intermediate states, one might try to extract the properties of correlators from our conformal blocks. However, one could be confronted with the following problem.
To construct correlators,  one has to take the sum of conformal blocks over intermediate dimensions as
\begin{equation}\label{eq:Fsum}
\braket{O_A(\infty)O_B(1)O_B(z)O_A(0)} =\sum_{h_p=0}^{\infty}\sum_{\bar{h}_p=0}^{\infty} \rho_{h_p,\bar{h}_p} 
\pa{C_{ABp}}^2
 \ca{F}^{BA}_{BA}(h_p|z)\overline{\ca{F}^{BA}_{BA}}(\bar{h}_p|\bar{z}),
\end{equation}
where $\rho_{h_p,\bar{h}_p} $ is the density of primary states.
Once we take $z$ near the singular point $z=1$, we can approximate it as the sum of effective contributions as
\begin{equation}
\braket{O_A(\infty)O_B(1)O_B(z)O_A(0)} \simeq \sum_{h_p=0}^{h_{p^*}}\sum_{\bar{h}_p=0}^{\bar{h}_{p^*}} \rho_{h_p,\bar{h}_p} 
\pa{C_{ABp}}^2\ca{F}^{BA}_{BA}(h_p|z)\overline{\ca{F}^{BA}_{BA}}(\bar{h}_p|\bar{z}),
\end{equation}
where $p^*$ depends on $z=1-\e \ \ \ \ (\e \ll1)$ and 
\begin{equation}
\fr{
\sum_{h_p=h_{p^*}}^{\infty}\sum_{\bar{h}_p=\bar{h}_{p^*}}^{\infty} \rho_{h_p,\bar{h}_p} 
 \pa{C_{ABp}}^2 \ca{F}^{BA}_{BA}(h_p|z)\overline{\ca{F}^{BA}_{BA}}(\bar{h}_p|\bar{z})}
{
\sum_{h_p=0}^{\infty}\sum_{\bar{h}_p=0}^{\infty} \rho_{h_p,\bar{h}_p} 
\pa{C_{ABp}}^2 \ca{F}^{BA}_{BA}(h_p|z)\overline{\ca{F}^{BA}_{BA}}(\bar{h}_p|\bar{z})}
\ll1.
\end{equation}
And in the same way, we can also define $n^*$ for the sum,
\begin{equation}\label{eq:Hsum}
H(h_p|q)=
\sum_{n=0}^{\infty} c_n(h_p) q^n,
\end{equation}
as
\begin{equation}\label{eq:nstar}
\fr{\sum_{n=0}^{n^*} c_n(h_p) q^n}
{\sum_{n=0}^{\infty} c_n(h_p) q^n}
\ll1,
\end{equation}
(or when there exists only one saddle point of the summation, one can think of $n^*$ as the saddle point of the summation (\ref{eq:Hsum}) and $h_{p^*}$ as the saddle point of the summation (\ref{eq:Fsum}). In other words, the point ($n^*,h_{p^*}$) is the saddle point of the double sum over $n$ and $h_p$.)

\begin{figure}[h]
  \begin{center}
   \includegraphics[width=120mm]{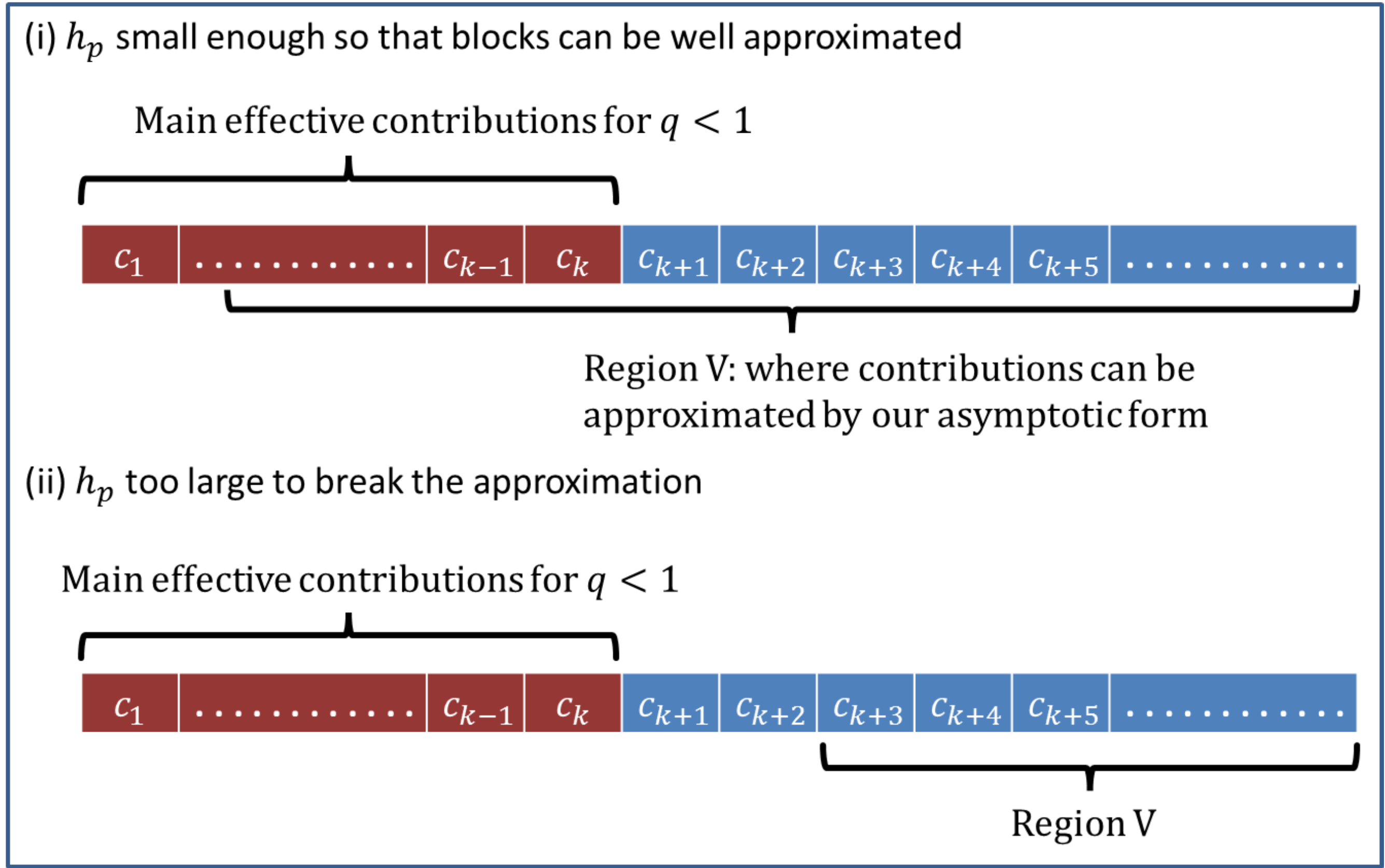}
   \end{center}
\caption{This figure explains the breakdown of the asymptotic form. Since $|q(z)|<1$ in the whole z plane except for OPE singular points, only finite number of $c_n$ \ $(n=1. . . k<\infty)$ effectively contribute to conformal blocks. On the other hand, the lower bound of the region where the coefficients $c_k$ can be well approximated by our formula depends on $h_p$ and the lower bound become large as increasing $h_p$. (The reason is explained in the main text.) Therefore, if we take $h_p$ too large, then the asymptotic form of the conformal block (\ref{eq:appABBA}) breaks down.}
\label{fig:explanation}
\end{figure}

\begin{figure}[h]
  \begin{center}
   \includegraphics[width=120mm]{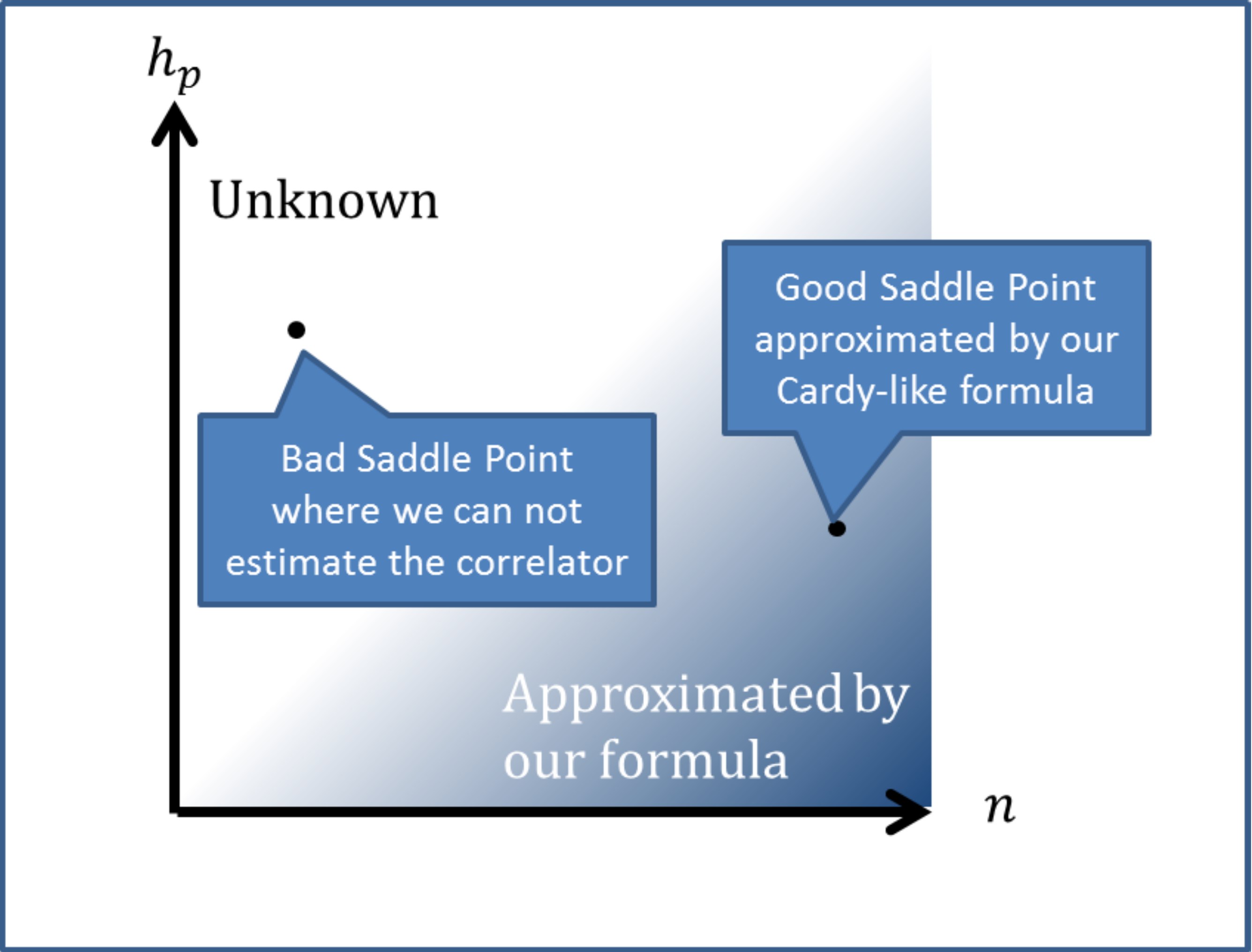}
   \end{center}
\caption{This figure also explains the breakdown of the asymptotic form in terms of the saddle point approximation. When there exists only one saddle point, the situation is simpler than Figure \ref{fig:explanation}. If the saddle point  ($n^*, h_{p^*}$) is in the lower triangle, the correlator can be estimated by our Cardy-like formula. On the other hand, if the saddle point is in the upper triangle, we can not estimate the correlator because no expression for $c_n$ is found in this region. However, if $h_{p^*}\gg n_*$ is satisfied, the contributions at $n\geq1$ are negligible, in that, $n^*=1$.}
\label{fig:saddle}
\end{figure}

Recall that the coefficients $c_n$ are given by (\ref{eq:ck}) as
 \begin{equation}
	c_k(h_p) = \sum_{i=1}^k \sum_{\substack{m=1, n=1\\mn=i}} \frac{R_{m,n}}{h_p - h_{m,n}} c_{k-i}(h_{m,n}+mn).
\end{equation}
From this expression, one can find that the asymptotic form (\ref{eq:appABBA}) breaks down as $k$ approaches to the order $h_p$. Therefore, we expect that the asymptotic form (\ref{eq:appABBA}) of $c_n$ holds only for $n \gg h_p$. In other words, if one wants to approximate the conformal block by using our asymptotic $c_n$, the condition $n^* \gg h_p$ has to be satisfied. However, it might be possible that  there are conformal blocks with $h_p \sim n^*$ in the conformal block decomposition of the correlator, that is, $h_{p^*}\sim n^*$.
As a result, the behavior of the correlator might be different from that of the conformal block. This story is illustrated in Figure \ref{fig:explanation} and it is simplified when there exists only one saddle point as explained in Figure \ref{fig:saddle}.

To derive the asymptotic behavior of the correlator from the conformal blocks, one has to understand the behavior of the coefficients $c_n$ for $n \sim h_p$. Or, possibly, one can show that $n^* \gg h_{p^*}$. If so, we can use our asymptotic conformal blocks to construct the correlator. We leave this problem to future work.

Note that one might think that it is strange in the first place that the conformal block in the limit $z\to 1$ is independent of $h_p$ because one can estimate the correlation function as
\begin{equation}
\braket{O_A(\infty)O_B(1)O_B(z)O_A(0)} \ar{z \to 1}  \abs{1-z}^{\fr{c-1}{12}-4h_B} \abs{\log\pa{1-z}}^{-3}
\end{equation}
by using our asymptotic form if the asymptotic conformal blocks are independent of $h_p$, however  we know that the correlator has a OPE singularity
\begin{equation}\label{eq:OPEsing}
\braket{O_A(\infty)O_B(1)O_B(z)O_A(0)} \ar{z \to 1}  \abs{1-z}^{-4h_B},
\end{equation}
which leads to a contradiction. But actually, in the expression of the asymptotic blocks derived in Section \ref{sec:block}, we neglect the small effect
\begin{equation}
\pa{q}^{h_p-\fr{c-1}{24}} \ar{z \to 1}   \ex{-\frac{\pi^2}{\log(16/(1-z))}\pa{h_p-\fr{c-1}{24}}},
\end{equation}
which is included in the universal prefactor (\ref{eq:pre}). If we take account of this contribution to the conformal block, we can reproduce the OPE singularity (\ref{eq:OPEsing}) by tunning the OPE coefficients. We will explain it in more detail in Section \ref{sec:bootstrap}.

%%%%%%%%%%%%%%%%%%%%%%%%%%%%%%%%%%%%%%%%%%%
%%%%%%%%%%%%%%%%%%%%%%%%%%%%%%%%%%%%%%%%%%%
\subsection{Analytic Continuation of Correlators}\label{subsec:OTOC}
%%%%%%%%%%%%%%%%%%%%%%%%%%%%%%%%%%%%%%%%%%%
%%%%%%%%%%%%%%%%%%%%%%%%%%%%%%%%%%%%%%%%%%%

From the consideration in the above subsection, one would think that there could be a significant difference between the asymptotic behavior of conformal blocks and correlators and therefore the features important in physics could not be obtained from the asymptotics of conformal blocks. But actually, we can use our asymptotic form directly in considering analytic continuations around OPE singular points, which appears in the calculation of OTOCs .  In more details, OTOCs are obtained by analytic continuation of the same Euclidean four point function. If one considers a 2d CFT on a thermal cylinder, OTOCs can be calculated by taking the map $(1-z) \to \ex{-2 \pi i}(z-1)$ while leaving $\bar{z}$ as it is and taking the limit $z, \bar{z} \to 0$  \cite{Roberts2015,Caputa2016,Caputa2017a}. Let denote the function after this operation by $f_{mono}(z)$.
\begin{quote}
{\it Example:}

If one considers $f(z)=\log (1-z)$, then
\begin{equation*}
\begin{aligned}
f_{mono}(z)&=f(z)-2 \pi i ,\\
f_{mono}(\bar{z})&=f(\bar{z}).
\end{aligned}
\end{equation*}
\end{quote}
In this notation, the OTOC for $O_A$ and $O_B$ can be obtained by calculating the correlator as
\begin{equation}
\sum_p C_{AAp}C_{BBp} \ca{F}^{AA}_{BB, mono}(h_p|z)\overline{\ca{F}^{AA}_{BB}}(\bar{h}_p|\bar{z}).
\end{equation} 
Taking the limit $z,\bar{z} \to 0$ which corresponds to increasing time $t$, we can approximate this sum by the identity block as
\begin{equation}\label{eq:id}
\sum_p C_{AAp}C_{BBp} \ca{F}^{AA}_{BB, mono}(h_p|z)\overline{\ca{F}^{AA}_{BB}}(\bar{h}_p|\bar{z}) \ar{z,\bar{z} \to 0}  \ca{F}^{AA}_{BB, mono}(0|z)\overline{\ca{F}^{AA}_{BB}}(0|\bar{z}).
\end{equation} 
As we mentioned before, the elliptic nome has the limit $q_{mono}(z) \ar{z \to 0} i$, thus we can use our asymptotic form of AABB blocks for the coefficients $c_n$. Finally, we can obtain the behavior of OTOCs at late times as 
\footnote{One might wonder we can not estimate the OTOC since we do not know the asymptotic behavior of the {\it heavy-heavy} block in the limit $q \to i$. However, we can estimate the bound of the block as mentioned in Section \ref{subsec:AABBnon} (see eq.(\ref{eq:bound})). As a result, we can find that the contribution of the function $H(h_p|q)$ to the block can be neglected compared to the universal prefactor $\Lambda(h_p|q)$. This leads to the estimation of the OTOC (\ref{eq:OTOCH}).}

\begin{equation}\label{eq:OTOCH}
\fr{\ave{O_A (t)O_B O_A(t) O_B}_\beta}{\ave{O_A O_A}_\beta \ave{O_B O_B}_\beta} \ar{t \to \infty}
\ex{-\fr{c-1}{12}\fr{\pi t}{\beta}}, \ \ \ \ \ \ \mbox{if } h_A, h_B>\fr{c}{32},
\end{equation}
where the relation between cross ratios and time is given by
\begin{equation}
z \sim -\ex{- 2\fr{\pi(t-x)}{\b}}, \ \ \ \ \  \bar{z} \sim -\ex{- 2\fr{\pi(t+x)}{\b}}.
\end{equation}
Moreover, in the heavy-light limit, we can reproduce the results in  \cite{Roberts2015,Perlmutter2016}. From our asymptotic form, we can suggest that OTOCs in the holographic CFT have the exponential decay at late times for any operators. And this exponential decay can be seen in no other CFT  \cite{Gu2016,Caputa2016,Caputa2017a} than the holographic CFT. This may suggest that this late time behavior can also be used as a criterion of chaotic nature of a given quantum field theory, in addition to the existing arguments on the Lyapunov exponent  \cite{Fitzpatrick2016b,Maldacena2016,Roberts2015}.

Actually the result of the Renyi entropy after a local quench \cite{Kusuki2018} can be also obtained in the almost same way as above, that is, all we have to do is calculate only the identity block as in the right hand side of (\ref{eq:id}). In other words, the Renyi entropy after a local quench is a kind of OTOC. The dynamics of the Renyi entropy also receive extensive attention in the context of chaos  \cite{Caputa2014,He2014,Numasawa2016,Caputa2017,He2017,Guo2018} as well as OTOC. And from our result in \cite{Kusuki2018},  in holographic CFT, this behaviors of the Renyi entropy after a local quench dramatically change when we use {\it heavy} operators ($h>\fr{c}{32}$) to excite vacuum states. We expect that this transition exhibits the interesting physics in the holographic CFT.

As these examples of OTOC and Entanglement entropy, we can extract the interesting physics directly from our asymptotic form of the coefficients $c_n$, even though we have the problem mentioned in Section \ref{subsec:TOC}.

%%%%%%%%%%%%%%%%%%%%%%%%%%%%%%%%%%%%%%%%%%%%%%%%%%%%%%%%%%%%%%%%%%%%%%%%%%%%%%%%%%%%%%%%%%%%%%
%%%%%%%%%%%%%%%%%%%%%%%%%%%%%%%%%%%%%%%%%%%%%%%%%%%%%%%%%%%%%%%%%%%%%%%%%%%%%%%%%%%%%%%%%%%%%%
\section{Towards the Conformal Bootstrap}\label{sec:bootstrap}
%%%%%%%%%%%%%%%%%%%%%%%%%%%%%%%%%%%%%%%%%%%%%%%%%%%%%%%%%%%%%%%%%%%%%%%%%%%%%%%%%%%%%%%%%%%%%%
%%%%%%%%%%%%%%%%%%%%%%%%%%%%%%%%%%%%%%%%%%%%%%%%%%%%%%%%%%%%%%%%%%%%%%%%%%%%%%%%%%%%%%%%%%%%%%

\newsavebox{\boxpb}
\sbox{\boxpb}{\includegraphics[height=130pt]{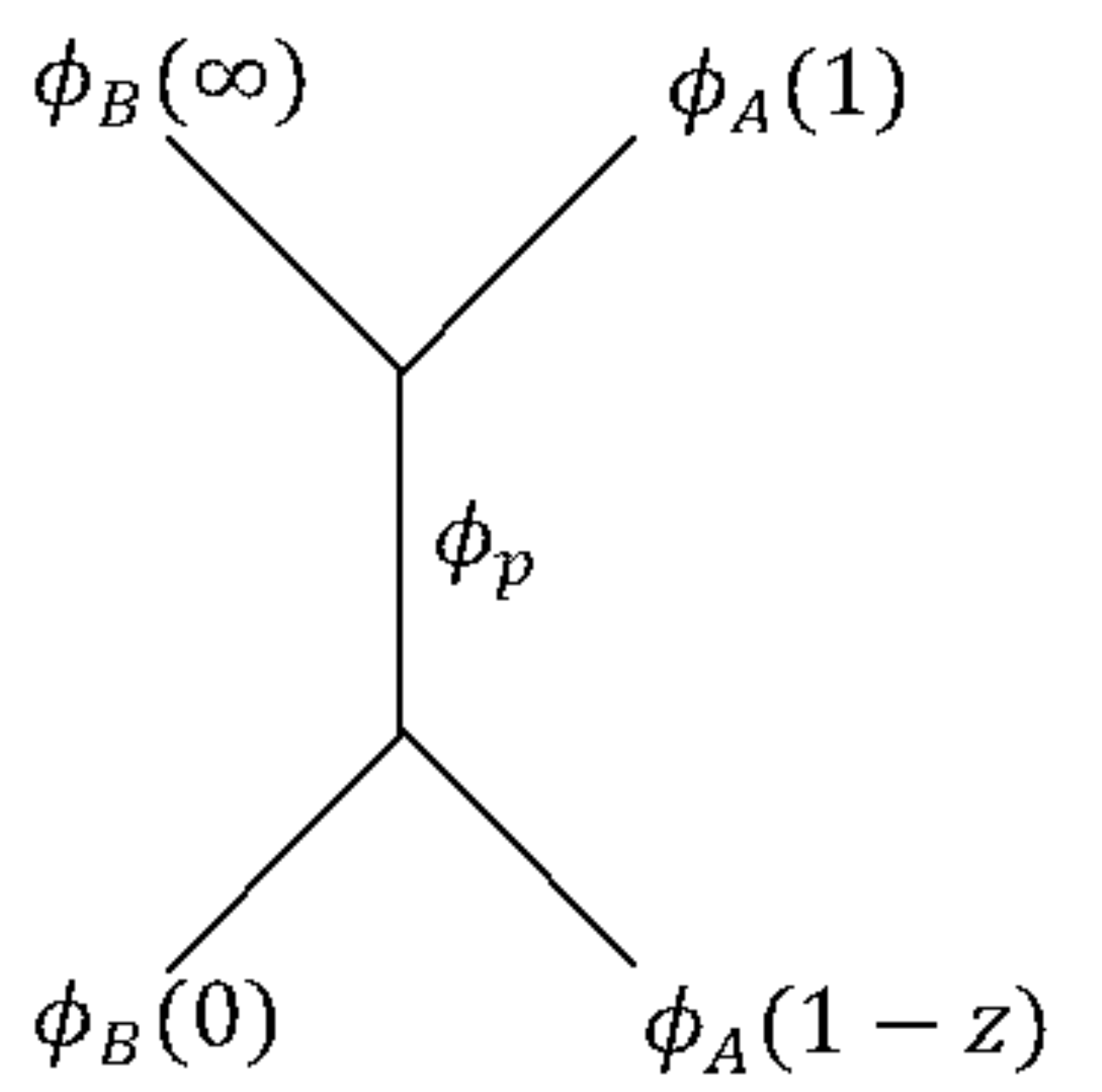}}
\newlength{\pbw}
\settowidth{\pbw}{\usebox{\boxpb}}

We would like to comment on the conformal bootstrap between AABB and ABBA (correctly, not ABBA but BAAB) blocks. First, let us consider the bootstrap equation in terms of the elliptic nome $q$ instead of $z$. By using the relations $z=\pa{\fr{\theta_2(q)}{\theta_3(q)}}^4$ and  $1-z=\pa{\fr{\theta_4(q)}{\theta_3(q)}}^4$, we can reexpress conformal blocks as 
\begin{equation}
\parbox{\paw}{\usebox{\boxpa}}=\pa{16q}^{h_p-\fr{c-1}{24}}\pa{16\eta(q)^{12}}^{\fr{c-1}{24}-h_A-h_B}\pa{\fr{\theta_3(q)^{2h_A}}{\theta_2(q)^{h_A-h_B}}}^4H^{AA}_{BB}(h_p|q)
\end{equation}
and
\begin{equation}
\begin{aligned}
\parbox{\pbw}{\usebox{\boxpb}}&=\pa{16\ti{q}}^{h_p-\fr{c-1}{24}}\pa{16\eta(\ti{q})^{12}}^{\fr{c-1}{24}-h_A-h_B}\pa{\fr{\theta_3(\ti{q})^{2h_A}}{\theta_4(\ti{q})^{h_A-h_B}}}^4 H^{AB}_{AB}(h_p|\ti{q})\\
&=\pa{-i \tau}^{\fr{c-1}{4}-2(h_A+h_B)}\pa{16\ti{q}}^{h_p-\fr{c-1}{24}}\pa{16\eta(q)^{12}}^{\fr{c-1}{24}-h_A-h_B}\pa{\fr{\theta_3(q)^{2h_A}}{\theta_2(q)^{h_A-h_B}}}^4 H^{AB}_{AB}(h_p|\ti{q})
\end{aligned}
\end{equation}
where $\eta$ is the Dedekind eta function and $\ti{q}(z)=q(1-z)$. We will set $q=\bar{q}=\ex{-\fr{\beta}{2}}$, then we have the following bootstrap equation,
\begin{equation}\label{eq:bootstrap}
\begin{aligned}
&\sum_{p}C_{AAp}C_{BBp}\pa{16}^{\D_p} 
\ex{-\fr{\beta}{2}\pa{\D_p-\fr{c-1}{12}}} 
H^{AA}_{BB}(h_p|q) \overline{H^{AA}_{BB}(h_p|q)}\\
&=
\sum_{p}C_{ABp}^2  \pa{\fr{\beta}{2\pi}}^{\fr{c-1}{2}-4(\D_A+\D_B)}\pa{16}^{\D_p}
\ex{-\fr{2\pi^2}{\beta}\pa{\D_p-\fr{c-1}{12}}} 
H^{AB}_{AB}(h_p|\ti{q})\overline{ H^{AB}_{AB}(h_p|\ti{q})}
\end{aligned}
\end{equation}
where we used the following identities,
\begin{equation}
\theta_3\pa{-\fr{1}{\tau}}=\s{-i\tau}\theta_3(\tau), \ \ \ \ \
\theta_4\pa{-\fr{1}{\tau}}=\s{-i\tau}\theta_2(\tau), \ \ \ \ \
\eta\pa{-\fr{1}{\tau}}=\s{-i\tau}\eta(\tau).
\end{equation}

In most cases, the bootstrap equation can be applied to two limits, the {\it high-low temperature limit} and the  {\it medium temperature limit}. Let us see each limit in the following:
\begin{description}
%%%%%%%%%%%%%%%%%%%%%%%%%%%%%%%%%%%%%
\item[The high-low temperature limit]~\\
%%%%%%%%%%%%%%%%%%%%%%%%%%%%%%%%%%%%%
The most famous consequence of the high-low temperature limit is Cardy's formula  \cite{Cardy1986a}, which can be derived by setting $h_A=h_B=\fr{c}{32}$ in (\ref{eq:bootstrap}). If we take the limit $\beta \to \infty$ in (\ref{eq:bootstrap}), then we have
\begin{equation}\label{eq:bootstrapsum}
\begin{aligned}
&\ex{\fr{\beta}{2}\fr{c-1}{12}}
H^{AA}_{BB}(0|q) \overline{H^{AA}_{BB}(0|q)}\\
&=
\sum_{p}C_{ABp}^2  \pa{\fr{\beta}{2\pi}}^{\fr{c-1}{2}-4(\D_A+\D_B)}\pa{16}^{\D_p}
\ex{-\fr{2\pi^2}{\beta}\pa{\D_p-\fr{c-1}{12}}} 
H^{AB}_{AB}(h_p|\ti{q})\overline{ H^{AB}_{AB}(h_p|\ti{q})}.
\end{aligned}
\end{equation}
From the expression of the $H(h_p|q)$ function (\ref{eq:Hdef}), we can see that
\begin{equation}
H^{21}_{34}(h_p|q) \ar{q \to 0} 1.
\end{equation}
Here, we assume that
\begin{equation}\label{eq:Hone}
H^{21}_{34}(h_p|q) \ar{h_p \to \infty} 1.
\end{equation}
We have to mention again that this is valid only in the regime (\ref{eq:validity}),
\footnote{
In \cite{Kraus2016}, the asymptotics for heavy-heavy-light three point coefficints is derived in the similar way.
We believe that the regime (\ref{eq:validity2}) is corresponding to (45) in \cite{Kraus2016},
\begin{equation}\label{eq:validitytorus}
h_p \abs{\log q}^2\gg 1,
\end{equation}
 which is shown by the large $h$ limit of a 1-point conformal block on a torus. In that case, we can estimate the large $h$ expansion of the block on a torus by using the Virasoro algebra. However, we can not find out the corresponding derivation for the 4-point block. Note that the r.h.s of (\ref{eq:validity2} 
) and (\ref{eq:validitytorus}) are different in spite of the Poghossian identities \cite{Hadasz2010b,Poghossian2009}. We believe that it comes from the special property for the block with $h_{A} \text{ or } h_B=\fr{c}{32}$ as explained in the last of Section \ref{subsec:validity}. Note that the regime (\ref{eq:validitytorus}) comes from the expectation that no descendants contribute to a 1-point block on a torus at large $h_p$ in large $c$ CFTs. However, it does not occur in the bootstrap for 4-point function (see Appendix \ref{sec:descendants}).}
\begin{equation}\label{eq:validity2}
h_p |\log q|^2 \gg c,
\end{equation}
except for special cases.
In the high-low temperature limit, the bootstrap equation (\ref{eq:bootstrap})  is simplified as
\begin{equation}\label{eq:highlow}
\begin{aligned}
\ex{\fr{\beta}{2}\fr{c-1}{12}}
=
\sum_{p}C_{ABp}^2  \pa{\fr{\beta}{2\pi}}^{\fr{c-1}{2}-4(\D_A+\D_B)}\pa{16}^{\D_p}
\ex{-\fr{2\pi^2}{\beta}\pa{\D_p-\fr{c-1}{12}}} .
\end{aligned}
\end{equation}
Here we assume $c>1$ and the summation in the right hand side is dominated by $h_p, \bar{h}_p \gg 1$. Thus there are many heavy primary states, and the sum in (\ref{eq:highlow}) can be approximated by an integral as
\begin{equation}\label{eq:bootstrapint}
\begin{aligned}
\ex{\fr{\beta}{2}\fr{c-1}{12}}
=
\int \dd \D_p \ \ \rho(\D_p)
\overline{C_{ABp}^2}  \pa{\fr{\beta}{2\pi}}^{\fr{c-1}{2}-4(\D_A+\D_B)}\pa{16}^{\D_p}
\ex{-\fr{2\pi^2}{\beta}\pa{\D_p-\fr{c-1}{12}}} ,
\end{aligned}
\end{equation}
where $\rho(\D_p)$ is the density of states which has the asymptotic formula called as Cardy's formula:
\begin{equation}
\rho(\D_p)\ar{\D_p \to \infty} \ex{4\pi\s{\fr{c-1}{12}\pa{\D_p-\fr{c-1}{12}}}} \ \ \ \ (\D_p \gg c),
\end{equation}
and the average is over all primary operators of fixed dimension $\D_p$.

By using the inverse Laplace transformation, we can obtain the mean-squared  OPE coefficients as
\begin{equation}\label{eq:CABp}
\overline{C_{ABp}^2}  \ar{\D_p \to \infty}
\fr{\pi}{16^{\D_p}\pa{\fr{12\D_p}{c-1}-1}^{4(\D_A+\D_B)-\fr{c+1}{2}}}\ex{-2\pi\s{\fr{c-1}{12}\pa{\D_p-\fr{c-1}{12}}}}.
\end{equation}

Here we can reexpress (\ref{eq:CABp}) by using entropy as
\begin{equation}\label{eq:HLLcoef}
\s{\overline{C_{ABp}^2}} \ar{\D_p \to \infty}
4^{-\D_p}\ex{-\fr{S(\D_p)}{4}}.
\end{equation}

Note that in  \cite{Das2017}, the mean-squared OPE coefficients $\overline{C_{AAp}^2}$ are given in much the same way as the above approach (they used the quantization on the pillow metric and they restrict attention to the case $\D_A=\D_B$) and they give the holographic dual interpretation of $\overline{C_{AAp}^2}$.

However, the saddle point of the inverse Laplace transformation for (\ref{eq:bootstrapint}) is given by
\begin{equation}
\D_p\abs{\log \ti{q}}^2\simeq \pi^2\fr{c-1}{12}.
\end{equation}
This does not satisfy the condition (\ref{eq:validity2}), therefore the above derivation of the three point function is subtle.
\footnote{Note that (\ref{eq:HLLcoef}) is analytically justified in \cite{Kusuki2018a}.}
As discussed in Section \ref{subsec:TOC}, it is possible that the saddle point $h_{p^*}$ of the r.h.s of (\ref{eq:bootstrapsum}) is smaller than $n^*$, which is  defined by (\ref{eq:nstar}). If the inequality $h_{p^*}<n^*$ is satisfied, then we should use our asymptotic formula,
\begin{equation}
\begin{aligned}
H^{AB}_{AB}(h_{p^*}|\ex{-\fr{2\pi^2}{\b}})& \lleq{\b \to \infty} \left\{
    \begin{array}{ll}
     \ex{\fr{\b}{2}\pa{\fr{c-1}{24}-4h_B+\fr{c-1}{6}\pa{1-\s{1-\fr{24}{c-1}h_B}}}}  ,& \text{if } h_B \ll c  ,\\
      \b^{4(h_A+h_B)-\fr{c+5}{4}} ,& \text{if } h_{A,B} >\fr{c}{32},  \\
    \end{array}
  \right.\\
\end{aligned}
\end{equation}
instead of (\ref{eq:Hone}). From our numerical observations (\ref{eq:monoto}) or (\ref{eq:monoto2}),
 we expect at least that
\begin{equation}
\begin{aligned}
0&\lleq{\b \to \infty}
\log \pa{H^{AB}_{AB}(h_{p^*}|\ex{-\fr{2\pi^2}{\b}})} \\
& \lleq{\b \to \infty} \left\{
    \begin{array}{ll}
     \fr{\b}{2}\pa{\fr{c-1}{24}-4h_B+\fr{c-1}{6}\pa{1-\s{1-\fr{24}{c-1}h_B}}} ,& \text{if } h_B\ll c  ,\\
      \pa{4(h_A+h_B)-\fr{c+5}{4}} \log \b ,& \text{if } h_{A,B} >\fr{c}{32},  \\
    \end{array}
  \right.\\
\end{aligned}
\end{equation}
in the limit $\b \to  \infty$.

From the above observation, it is shown that for $h_B \ll c$, the bootstrap equation can be described instead of (\ref{eq:bootstrapint}) as 
\begin{equation}
\begin{aligned}
\ex{\fr{\beta}{2}\fr{c-1}{12}}
\lleq{\b \to \infty}&
\int \dd \D_p \ \ \rho(\D_p)
\overline{C_{ABp}^2}  \pa{\fr{\beta}{2\pi}}^{\fr{c-1}{2}-4(\D_A+\D_B)}\pa{16}^{\D_p}\\
&\times
\ex{-\fr{2\pi^2}{\beta}\pa{\D_p-\fr{c-1}{12}}} \ex{\fr{\b}{2}\pa{\fr{c-1}{12}-4\D_B+\fr{c-1}{3}\pa{1-\s{1-\fr{12}{c-1}\D_B}}}},
\end{aligned}
\end{equation}
where we set $h_A=\overline{h_A}$ and $h_B=\overline{h_B}$ for simplicity. Here the above inequality is satisfied only if $\b \to \infty$. By using the inverse Laplace transformation, we can get the asymptotic three point function by
\begin{equation}
\begin{aligned}
16^{-\D_p} \ex{4\pi\s{\pa{\D_B-\fr{c-1}{12}\pa{1-\s{1-\fr{12}{c-1}\D_B}}}\pa{\D_p-\fr{c-1}{12}}} - 4\pi\s{\fr{c-1}{12}\pa{\D_p-\fr{c-1}{12}}}}
&\lleq{\D_p \to \infty} \overline{C_{ABp}^2} \\ 
&\lleq{\D_p \to \infty}  16^{-\D_p} \ex{- 2\pi\s{\fr{c-1}{12}\pa{\D_p-\fr{c-1}{12}}}},
\end{aligned}
\end{equation}
where the second inequality is led by (\ref{eq:CABp}).

For $h_{A,B} >\fr{c}{32}$, the bootstrap equation is
\begin{equation}
\begin{aligned}
\ex{\fr{\beta}{2}\fr{c-1}{12}}
\lleq{\b \to \infty} 
\int \dd \D_p \ \ \rho(\D_p)
\overline{C_{ABp}^2}  \b^{-3}
\ex{-\fr{2\pi^2}{\beta}\pa{\D_p-\fr{c-1}{12}}}16^{\D_p} ,
\end{aligned}
\end{equation}
where we set $h_A=\overline{h_A}$ and $h_B=\overline{h_B}$ again. This inequality leads to the asymptotic inequality as
\begin{equation}
16^{-\D_p} \ex{- 2\pi\s{\fr{c-1}{12}\pa{\D_p-\fr{c-1}{12}}}}
\lleq{\D_p \to \infty}  \overline{C_{ABp}^2}  
\lleq{\D_p \to \infty}  16^{-\D_p} \ex{- 2\pi\s{\fr{c-1}{12}\pa{\D_p-\fr{c-1}{12}}}},
\end{equation}
which means that the leading asymptotic behavior of the three point function $C_{ABp}$ with $h_{A,B} >\fr{c}{32}$ in the limit $h_p \to \infty$ is perfectly determined by the conformal bootstrap as

\begin{equation}
\overline{C_{ABp}^2}  \ar{\D_p \to \infty} 16^{-\D_p} \ex{- 2\pi\s{\fr{c-1}{12}\pa{\D_p-\fr{c-1}{12}}}}.
\end{equation}
Actually, it is possible that this asymptotics is true for any external dimensions. However, we have shown that the asymptotics of blocks drastically changes at $\fr{c}{32}$ (which means that the contribution from the descendants differs based on the external dimensions) and therefore it is also possible that the asymptotics of three point functions also has the transition.

%%%%%%%%%%%%%%%%%%%%%%%%%%%%%%%%%%%%%
\item[The medium temperature limit]~\\
%%%%%%%%%%%%%%%%%%%%%%%%%%%%%%%%%%%%%

One of the main contributions of the medium temperature limit is to derive the upper bound on the gap, which is called the Hellerman bound  \cite{Hellerman2009,Friedan2013,Collier2016} and which is revisited by using semiclassical conformal blocks \cite{Chang2016a}. And this limit is also used in the numerical bootstrap  \cite{Rattazzi2008,Rychkov2009,Simmons-Duffin2015,Simmons-Duffin2016}.

If one hopes to make use of the bootstrap equation in this limit, we have to resolve the problem mentioned in Section \ref{subsec:TOC}. In other words, we have to understand the behavior of the coefficients $c_n$ for $n \sim h_p$. This is an important future work.
\end{description}
It is also interesting to consider the bootstrap equation in various limits unexplored after analytic continuations.  It might be possible to solve the bootstrap equation analytically in some of them by using our asymptotic formula.
%%%%%%%%%%%%%%%%%%%%%%%%%%%%%%%%%%%%%%%%%%%%%%%%%%%%%%%%%%%%%%%%%%%%%%%%%%%%%%%%%%%%%%%%%%%%%%
%%%%%%%%%%%%%%%%%%%%%%%%%%%%%%%%%%%%%%%%%%%%%%%%%%%%%%%%%%%%%%%%%%%%%%%%%%%%%%%%%%%%%%%%%%%%%%
\section{Discussion}\label{sec:dis}
%%%%%%%%%%%%%%%%%%%%%%%%%%%%%%%%%%%%%%%%%%%%%%%%%%%%%%%%%%%%%%%%%%%%%%%%%%%%%%%%%%%%%%%%%%%%%%
%%%%%%%%%%%%%%%%%%%%%%%%%%%%%%%%%%%%%%%%%%%%%%%%%%%%%%%%%%%%%%%%%%%%%%%%%%%%%%%%%%%%%%%%%%%%%%

In this paper, we study large-c conformal blocks of 4-pt functions on a sphere and find the simple asymptotic form of the blocks and the transition of the behavior of the blocks at $h_{A,B}=\fr{c}{32}$. This strongly suggests the interesting structures or physical phenomena in CFT and gravity theory, but for now we can not answer what happens at that point. We hope to understand how to interpret this interesting phenomena as physics.

One might ask whether there are other situations where the value $\fr{c}{32}$ appears. Actually, this dimension can be seen in the twist-2 operator. As mentioned in the last of Section \ref{subsec:validity}, we can find  the following two facts:
\begin{enumerate}
\item 
For the blocks with the external operators $h_i =\fr{c}{32}$, the monodromy method can be solved exactly \cite{Maloney2017}.

\item
We can find a relation between the universal prefactor $\L(h_p|q)$ and a character as
\begin{equation}
{\L^{\fr{c'}{32},\fr{c'}{32}}_{\fr{c'}{32},\fr{c'}{32}}}_{c=c'-1} (h_p|q)=16^{h_p} \ex{S_{anomaly}} \chi_{\fr{h_p}{2}, c=\fr{c'}{2}}\pa{\tau},
\end{equation}
where the conformal anomaly factor is given by \cite{Lunin2001},
\begin{equation}
\braket{\sigma_2 \sigma_2 \sigma_2 \sigma_2 }=\ex{\fr{c'}{2}S_{anomaly}}Z_{c=\fr{c'}{2}}(\tau,\bar{\tau})
=\abs{2^8 z(1-z)}^{-\fr{c'}{24}}Z_{c=\fr{c'}{2}}(\tau,\bar{\tau}).
\end{equation}
Moreover, in the large $c$ limit, we can neglect the contribution of the function $H(h_p|q)$ to the Virasoro block \cite{Chang2016a}, in that, 
\begin{equation}
\fr{\log {H^{\fr{c'}{32},\fr{c'}{32}}_{\fr{c'}{32},\fr{c'}{32}}}_{c=c'-1} (h_p|q)}{c} \ar{c \to \infty} 0.
\end{equation}
This is consistent with that the coefficients $c_n$ vanish at $h_{A} \text{ or } h_B=\fr{c}{32}$, which can be seen from our formula (\ref{eq:lowercn}). As a result, we obtain the following relation in the large $c$ limit,
\begin{equation}
{\ca{F}^{\fr{c}{32},\fr{c}{32}}_{\fr{c}{32},\fr{c}{32}}} (h_p|q)=16^{h_p} \ex{S_{anomaly}} \chi_{\fr{h_p}{2}, \fr{c}{2}}\pa{\tau}.
\end{equation}

\end{enumerate}
We believe that these facts, our formula  (\ref{eq:lowercn}) and our conjectures illustrated in Figure \ref{fig:HHLL-hAhBdep}, \ref{fig:HLHL-hAhBdep} correlate to each other. It could be a key to understand the transition at $\fr{c}{32}$ analytically.
\footnote{Actually, one can also find the transition at $\fr{c}{32}$ in the fusion matrix \cite{Chang2016}.}

Another future work is to derive the simple asymptotic form in Figure \ref{fig:HHLL-hAhBdep}, \ref{fig:HLHL-hAhBdep} by using the Zamolodchikov recursion relation analytically. Actually it seems to be hard work because this simplification is attributed to many cancellations of terms in (\ref{eq:ck}).
\footnote{ One can easily see this cancellations when trying to observe our formula (\ref{eq:cnlim}) for small $n$.}
However, our numerical computation suggests that large $c$ conformal blocks have a simple expression,  which means that it might be possible to derive the large $c$ conformal blocks analytically in some way. We hope to give the large $c$ blocks, at least the coefficients $c_n$, in some analytic way.

It is also interesting to generalize the sparseness condition  \cite{Hartman2014} from the modular invariance to the condition to OPE coefficients from the crossing symmetry (\ref{eq:bootstrap}) as in  \cite{Chang2016a} by using our results for large $c$ conformal blocks.

In this paper, we show only the asymptotic behavior of conformal blocks in the vicinity of $z=1$, but those after picking up a monodromy at OPE singularities also interesting, for example, they appear in the calculation of entanglement entropy and OTOC as argued in Section \ref{subsec:OTOC}. And also, as mentioned in  \cite{Chen2017}, it is intriguing to  consider the limit after picking up a monodromy, which could be a new limit to solve the conformal bootstrap equation analytically.

One can generalize our analysis to more general blocks, in that, { \it ABCD blocks}, which have four different external dimensions $h_A, h_B, h_C, h_D$. We expect that one can also  see the transition at $\fr{c}{32}$ for ABCD blocks and find a simple asymptotic expression for the blocks. However, if one wants to uncover the properties of ABCD blocks, one has to investigate the blocks with  six parameters ($h_A, h_B, h_C, h_D, h_p$ and $c$). As one easily expects, it takes much more time than studying AABB or ABBA blocks, therefore, we leave it as a future work.

%%%%%%%%%%%%%%%%%%%%%%%%%%%%%%%%%%%%%%%%%%%%%%%%%%%%
\section*{Acknowledgments}
%%%%%%%%%%%%%%%%%%%%%%%%%%%%%%%%%%%%%%%%%%%%%%%%%%%%

We are grateful to Tadashi Takayanagi and Henry Maxfield for fruitful discussions and
comments. We also thank Kanato Goto, Sylvain Ribault, Diptarka Das and Shouvik Datta for useful discussions, and in particular Pawel Caputa, Yasuaki Hikida and Jared Kaplan for reading the draft of this paper and giving us valuable comments. 
YK is supported by JSPS fellowship.

\appendix
%%%%%%%%%%%%%%%%%%%%%%%%%%%%%%%%%%%%%%%%%%%%%%%%%%%%%%%%%%%%%%%%%%%%%%%%%%%%%%%%%%%%%%%%%%%%%%
%%%%%%%%%%%%%%%%%%%%%%%%%%%%%%%%%%%%%%%%%%%%%%%%%%%%%%%%%%%%%%%%%%%%%%%%%%%%%%%%%%%%%%%%%%%%%%
\section{More Details of Numerical Computations} \label{app:fitting}
%%%%%%%%%%%%%%%%%%%%%%%%%%%%%%%%%%%%%%%%%%%%%%%%%%%%%%%%%%%%%%%%%%%%%%%%%%%%%%%%%%%%%%%%%%%%%%
%%%%%%%%%%%%%%%%%%%%%%%%%%%%%%%%%%%%%%%%%%%%%%%%%%%%%%%%%%%%%%%%%%%%%%%%%%%%%%%%%%%%%%%%%%%%%%

%%%%%%%%%%%%%%%%%%%%%%%%%%%%%%%%%%%%%%%%%%%
%%%%%%%%%%%%%%%%%%%%%%%%%%%%%%%%%%%%%%%%%%%
\subsection{Well-Fitted for Any $n$ in The Heavy-Light Limit}\label{subsec:app1}
%%%%%%%%%%%%%%%%%%%%%%%%%%%%%%%%%%%%%%%%%%%
%%%%%%%%%%%%%%%%%%%%%%%%%%%%%%%%%%%%%%%%%%%
In this subsection, we display plots of the $n$ dependence of the coefficients $c_n$ of ABBA blocks and we show that  the coefficients $c_n$ are well-fitted by
\begin{equation}\label{eq:cnfit}
c_n \sim n^{\a} \ex{A \s{n}}
\end{equation}
for higher $n$. 

\begin{figure}[h]
 \begin{minipage}{0.5\hsize}
  \begin{center}
   \includegraphics[width=70mm]{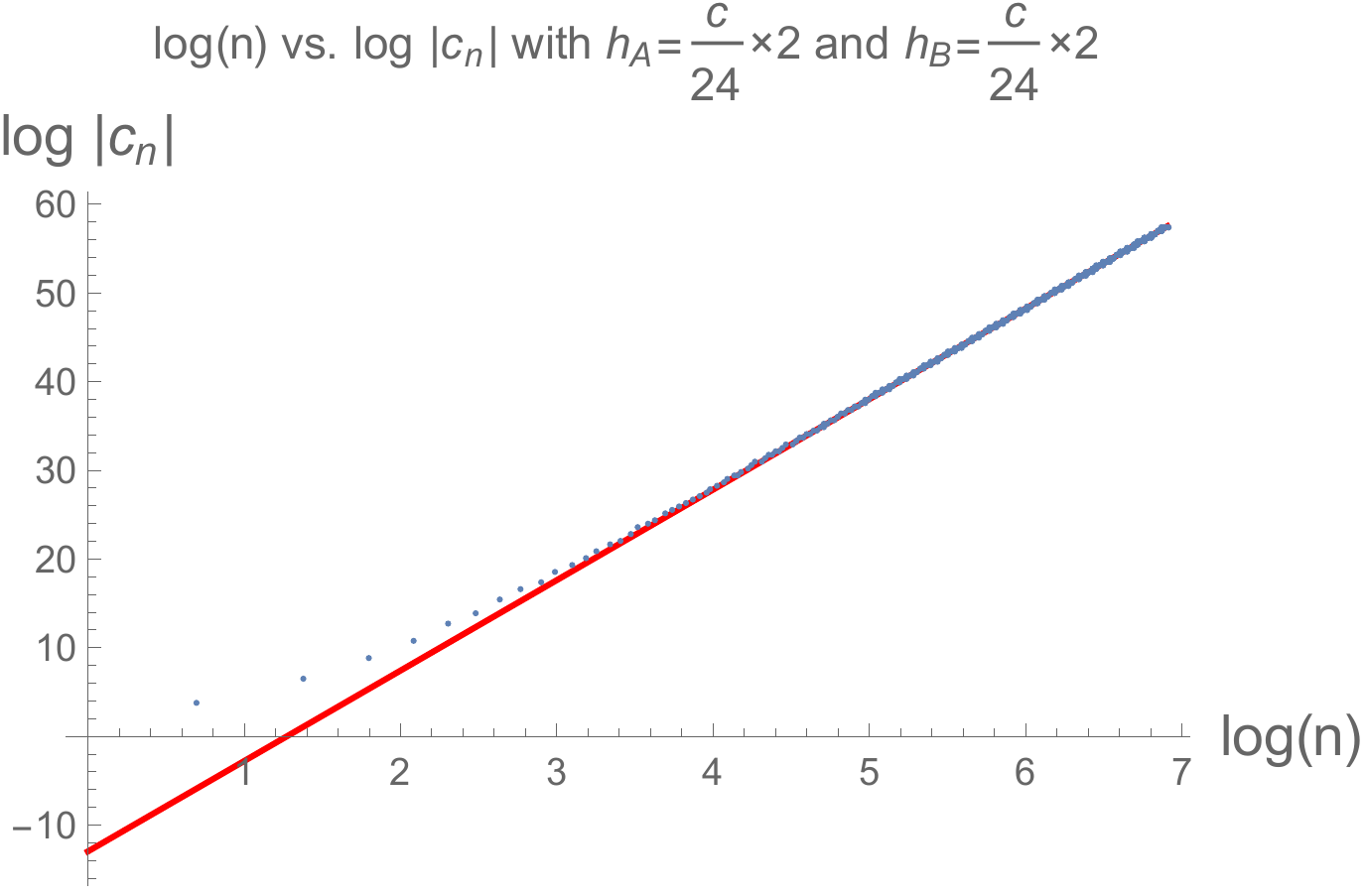}
  \end{center}
 \end{minipage}
 \begin{minipage}{0.5\hsize}
  \begin{center}
   \includegraphics[width=70mm]{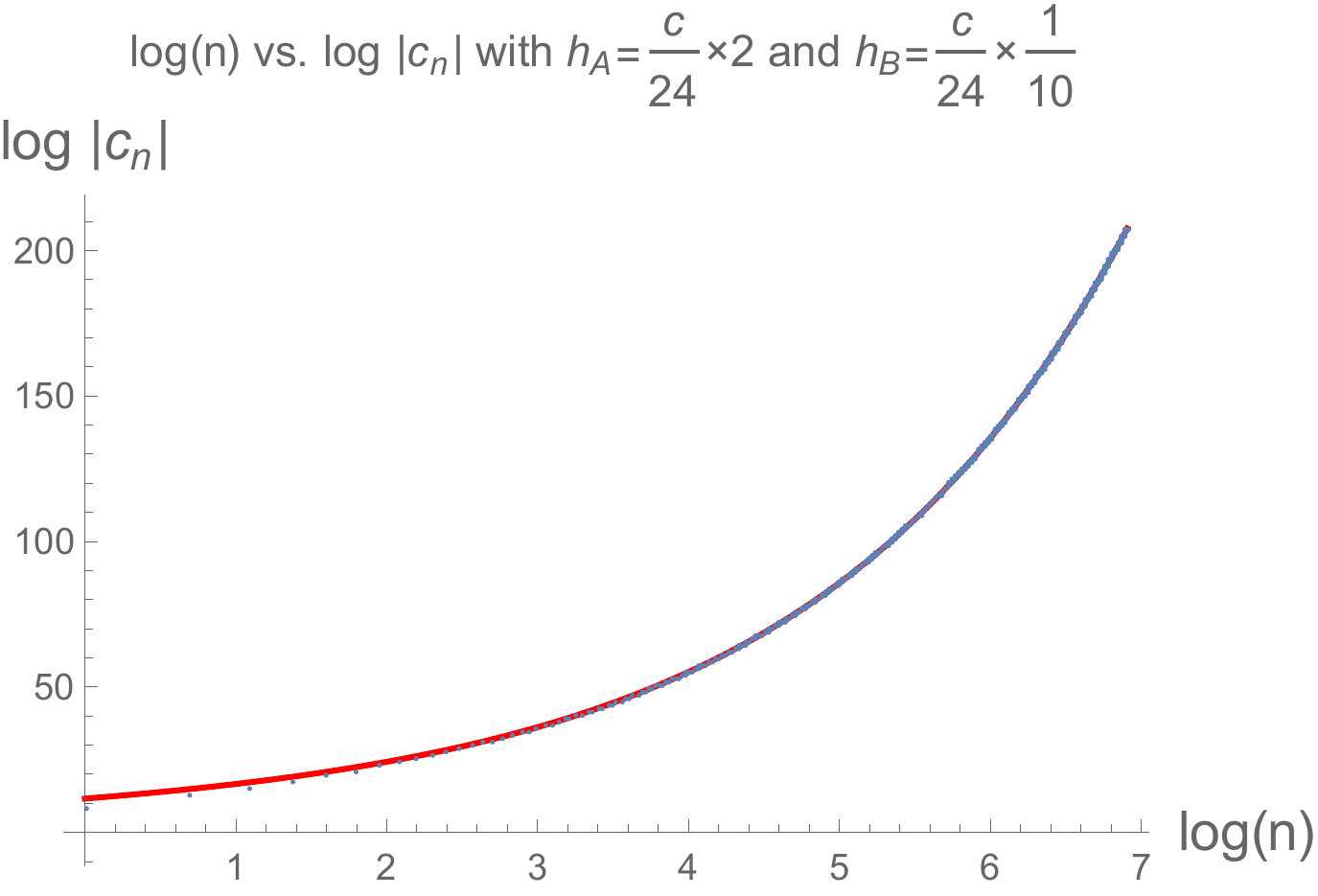}
  \end{center}
 \end{minipage}
\caption{The behaviors of the coefficients $c_n$ of ABBA blocks. The left is for $(h_A,h_B)=(\fr{c}{12},\fr{c}{12})$ and the right is for $(h_A,h_B)=(\fr{c}{12},\fr{c}{240})$. The blue dots are the numerical values of $\log c_n$. The red lines are $B n^\a \ex{A\s{n}}$ with the constant $B$ determined by the fit. We now set $c=30.01$ and $h_p=0.01$.}
\label{fig:cndep2}
\end{figure}

Figure \ref{fig:cndep2}  shows the $n$ dependence of the numerical values of $c_n$ (blue dots) and the fitted function by $n^{\a} \ex{A \s{n}}$ (red line). From these figures, one can find that the numerical values of $c_n$ are well-fitted by $n^{\a} \ex{A \s{n}}$ for higher $n$. And moreover one can find that in the heavy-light limit (the right in Figure \ref{fig:cndep2}), the coefficients $c_n$ are also well-fitted for small $n$, which implies that 
\begin{equation}
H(h_p|q)\simeq 1+\sum_{n=1}  n^{\a} \ex{A \s{n}}q^n
\end{equation}
can be applied for $z$ away from the point $z=1$.

\begin{figure}[H]
 \begin{minipage}{0.5\hsize}
  \begin{center}
   \includegraphics[width=70mm]{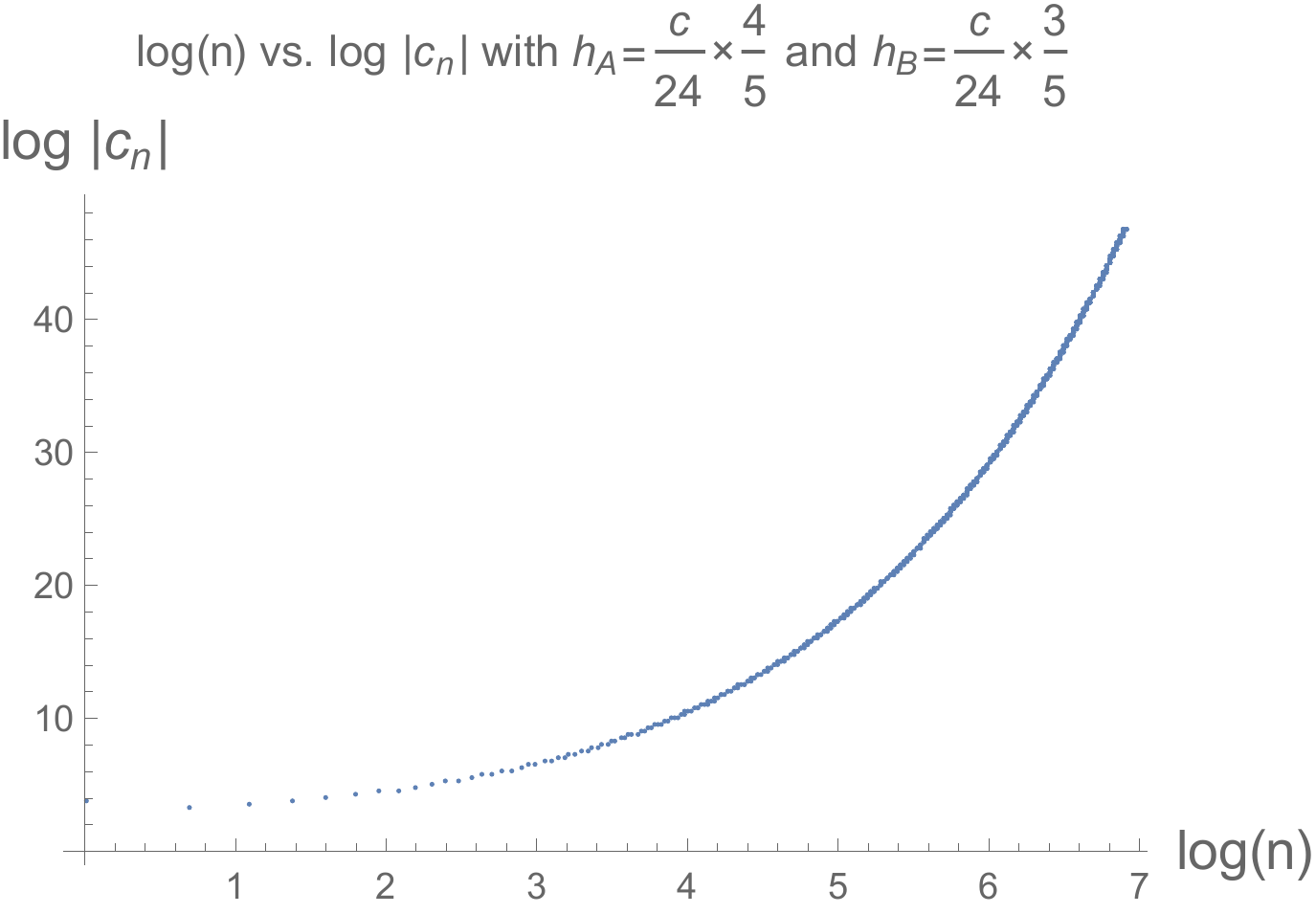}
  \end{center}
 \end{minipage}
 \begin{minipage}{0.5\hsize}
  \begin{center}
   \includegraphics[width=70mm]{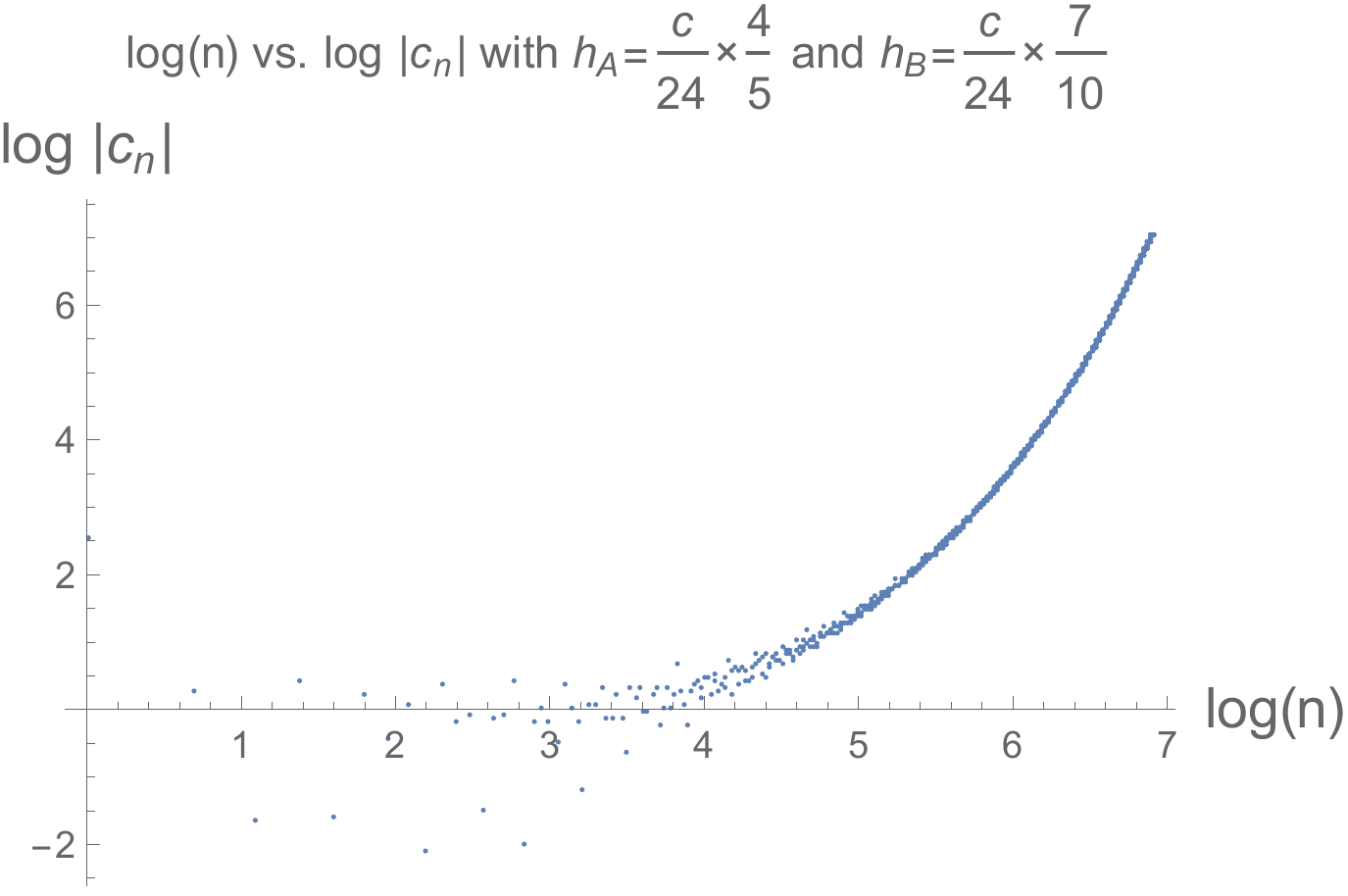}
  \end{center}
 \end{minipage}
 \begin{minipage}{0.5\hsize}
  \begin{center}
   \includegraphics[width=70mm]{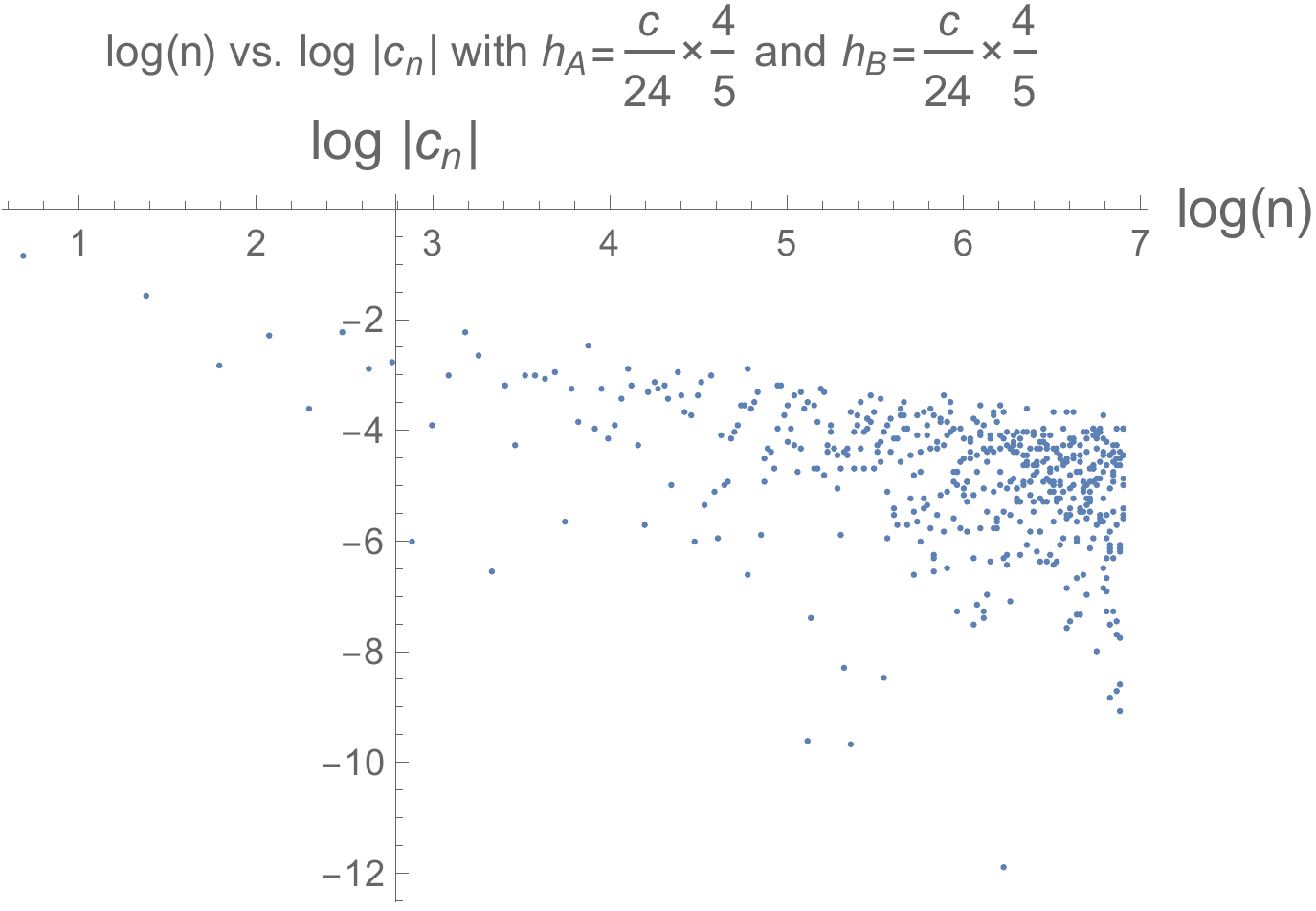}
  \end{center}
 \end{minipage}
 \begin{minipage}{0.5\hsize}
  \begin{center}
   \includegraphics[width=70mm]{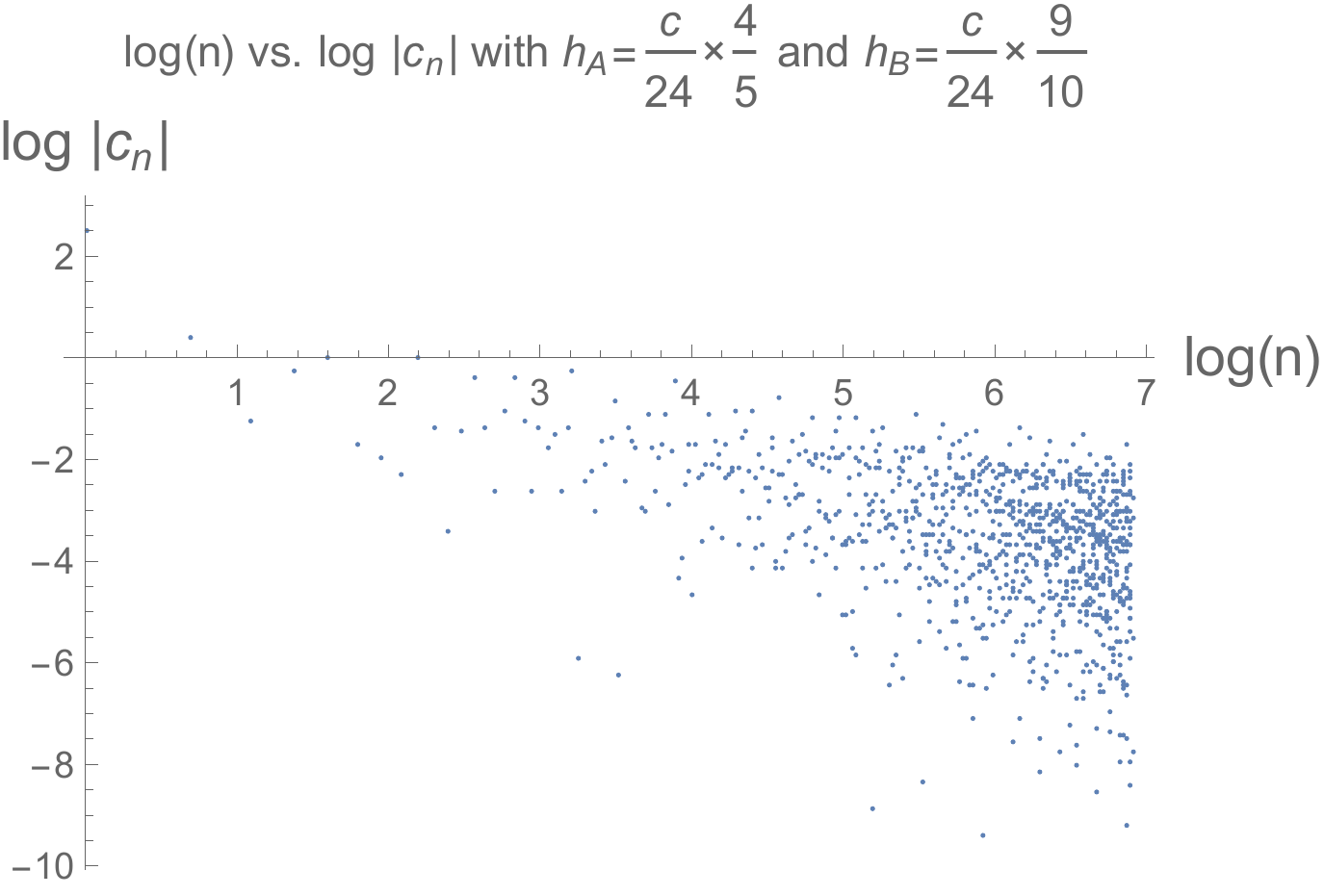}
  \end{center}
 \end{minipage}
\caption{The plots of $c_n$ for $h_A=\fr{c}{30}$ and $h_B=\fr{c}{24} \times \fr{6}{10}, \fr{7}{10}, \fr{8}{10}, \fr{9}{10}$. One can see that approaching $h_B=\fr{c}{32}$ causes the oscillation of the coefficients $c_n$.}
\label{fig:osc}
\end{figure}

Note that in the vicinity of the value $h_{A,B}=\fr{c}{32}$, the coefficients $c_n$ widely oscillate for small $n$ as in Figure \ref{fig:osc}. This is the reason why one can see some strange dots near the line $h_{A,B}=\fr{c}{32}$ in some figures (for example, Figure \ref{fig:HHLLhBhpdep}). However, if we extend the analysis of the behavior of the coefficients $c_n$ to higher $n$, we can again obtain the behavior well-fitted by (\ref{eq:cnfit}).

\begin{figure}[H]
 \begin{minipage}{0.5\hsize}
  \begin{center}
   \includegraphics[width=70mm]{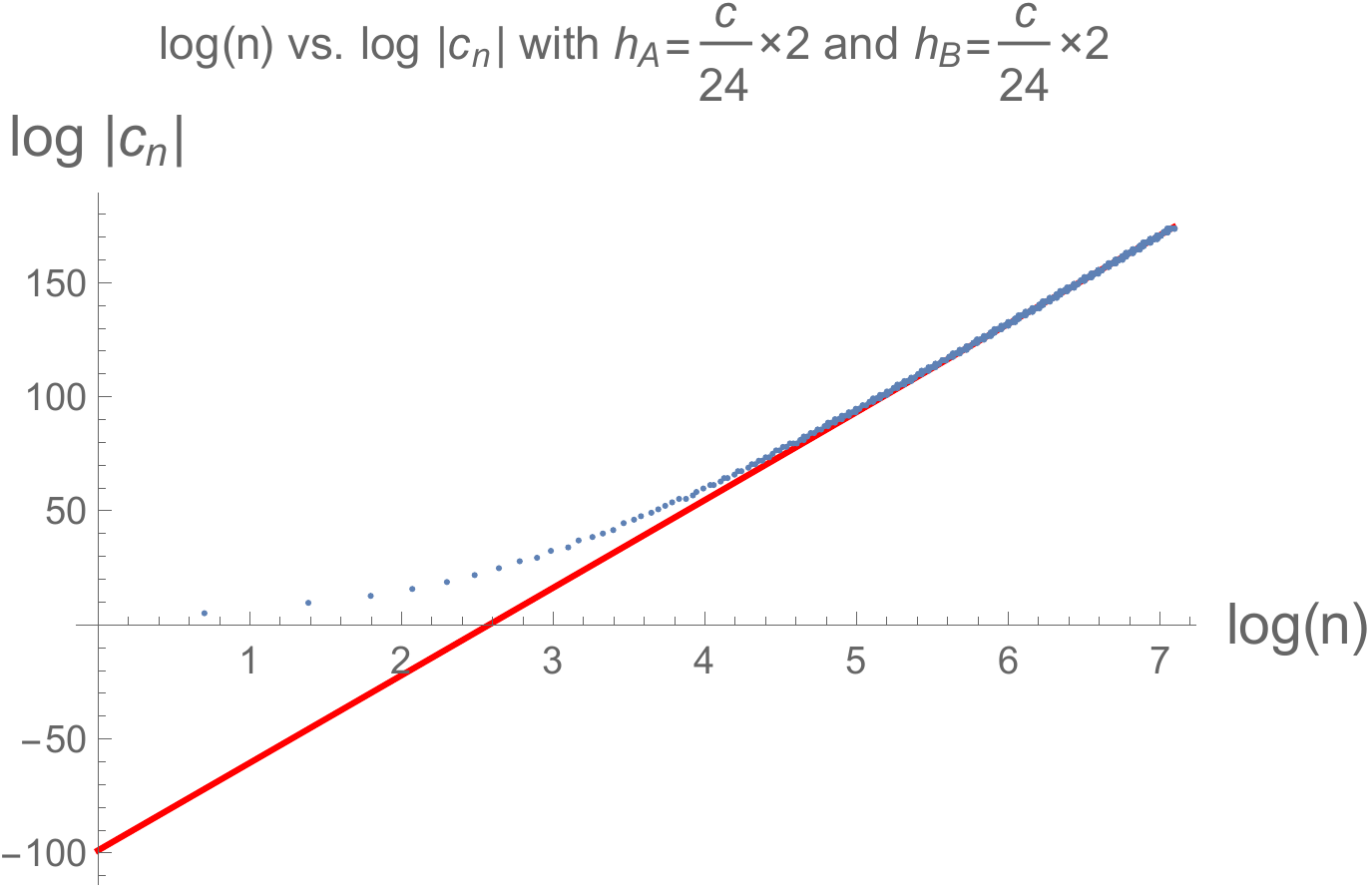}
  \end{center}
 \end{minipage}
 \begin{minipage}{0.5\hsize}
  \begin{center}
   \includegraphics[width=70mm]{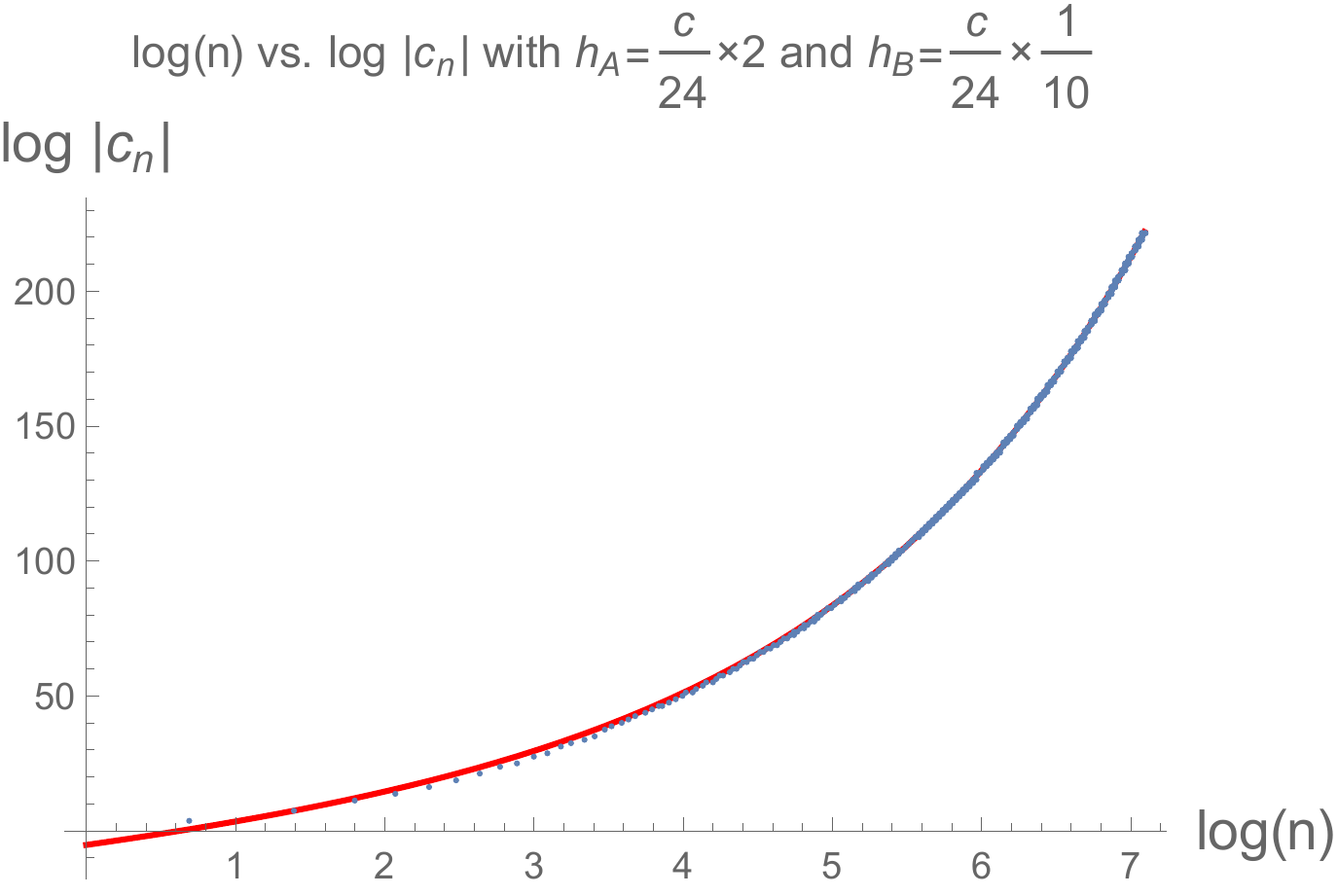}
  \end{center}
 \end{minipage}
\caption{The behaviors of the coefficients $c_n$ of AABB blocks. The left is for $(h_A,h_B)=(\fr{c}{12},\fr{c}{12})$ and the right is for $(h_A,h_B)=(\fr{c}{12},\fr{c}{240})$. The blue dots are the numerical values of $\log c_n$. The red lines are $B n^\a \ex{A\s{n}}$ with the constant $B$ determined by the fit. We now set $c=100.01$ and $h_p=0$.}
\label{fig:AABBcndep2}
\end{figure}

Note also that the phenomena that $c_n$ are well-fitted in the heavy-light limit can be also seen in the coefficients of AABB blocks as in Figure \ref{fig:AABBcndep2}.

%%%%%%%%%%%%%%%%%%%%%%%%%%%%%%%%%%%%%%%%%%%
%%%%%%%%%%%%%%%%%%%%%%%%%%%%%%%%%%%%%%%%%%%
\subsection{Some Extra Plots}\label{subsec:extra}
\begin{figure}[h]
  \begin{center}
   \includegraphics[width=70mm]{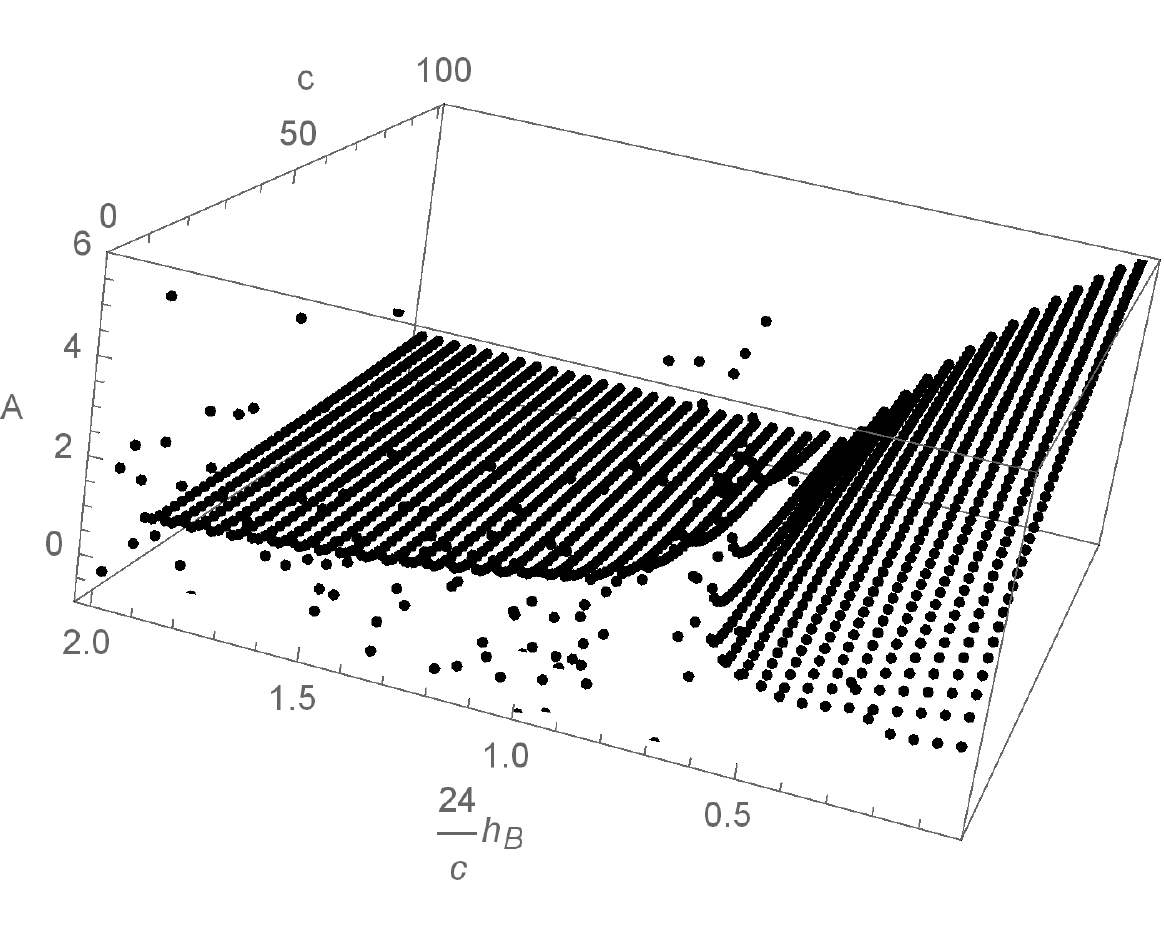}
   \includegraphics[width=70mm]{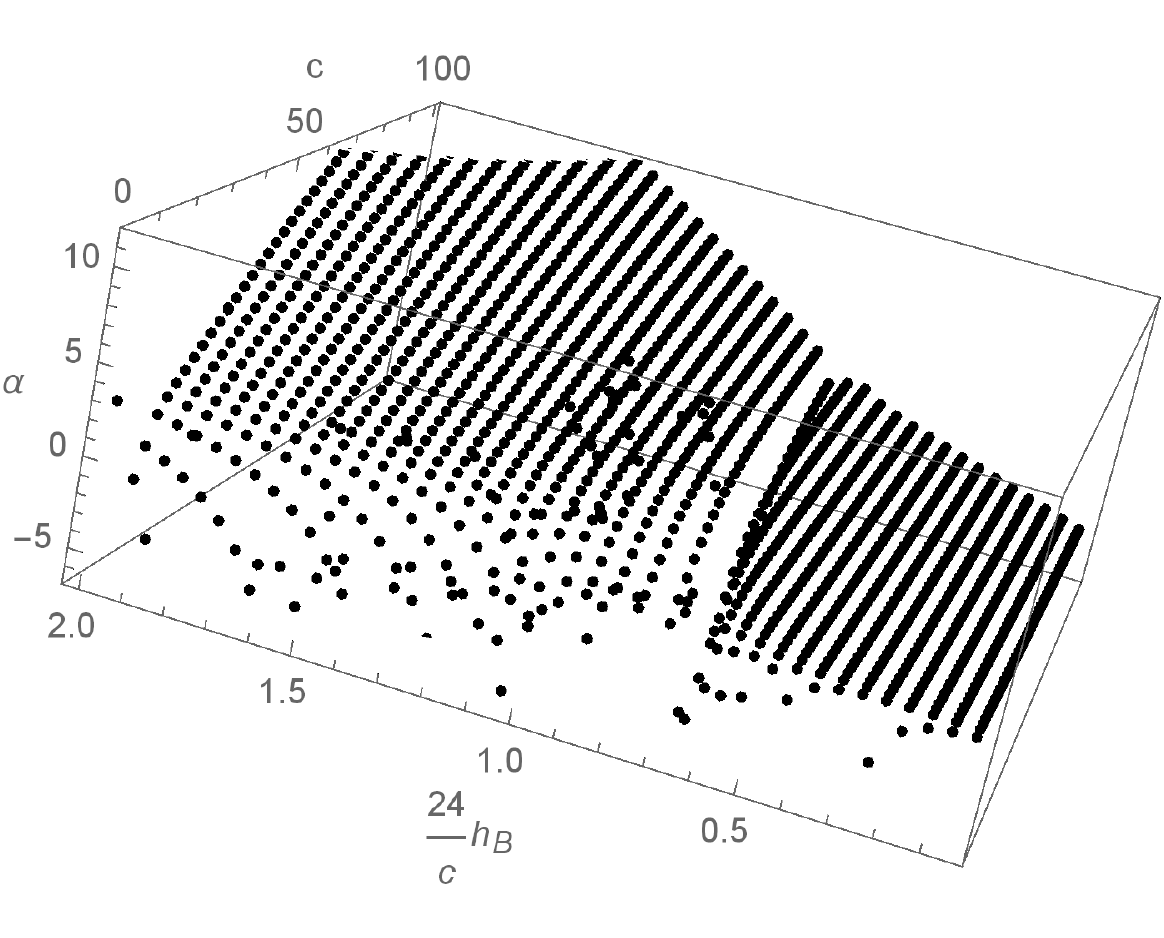}
  \end{center}
\caption{The plots of $A$ (left) and $\a$ (right) of AABB blocks for various values of ($c$, $h_B$) with $h_A=\fr{c}{24}$ and $h_p=0$.}
\label{fig:A-hphBdep.pdf}
\end{figure}

%%%%%%%%%%%%%%%%%%%%%%%%%%%%%%%%%%%%%%%%%%%
%%%%%%%%%%%%%%%%%%%%%%%%%%%%%%%%%%%%%%%%%%%
One might ask whether the transition at $\fr{c}{32}$ occurs for $c>25$ and we could answer ``yes''  at least based on our numerical computations. Take a look at Figure \ref{fig:A-hphBdep.pdf}, which shows the values of A of AABB blocks for various values of ($c,h_B$) with $h_A=\fr{c}{24}$ and $h_p=0$.  It suggests that the transition point is always at $\fr{c}{32}$.

Note that in our numerical computation, we approximate $\s{c}$ at 500 digits of accuracy, which is in particular a rational number. Therefore we always encounter the problem of the divergence of the denominator of (\ref{eq:ckrec}). In our setup, this problem occurs at very high $n$ and therefore we can neglect it. However, if one might try to set $c$ small, then the denominator of (\ref{eq:ckrec}) approaches zero many times, which leads to singular behaviors of $c_n$ in Figure \ref{fig:A-hphBdep.pdf}. Therefore, in our computation, we can not see universal asymptotic form of $c_n$ for small $c$.

\begin{figure}[H]
  \begin{center}
   \includegraphics[width=70mm]{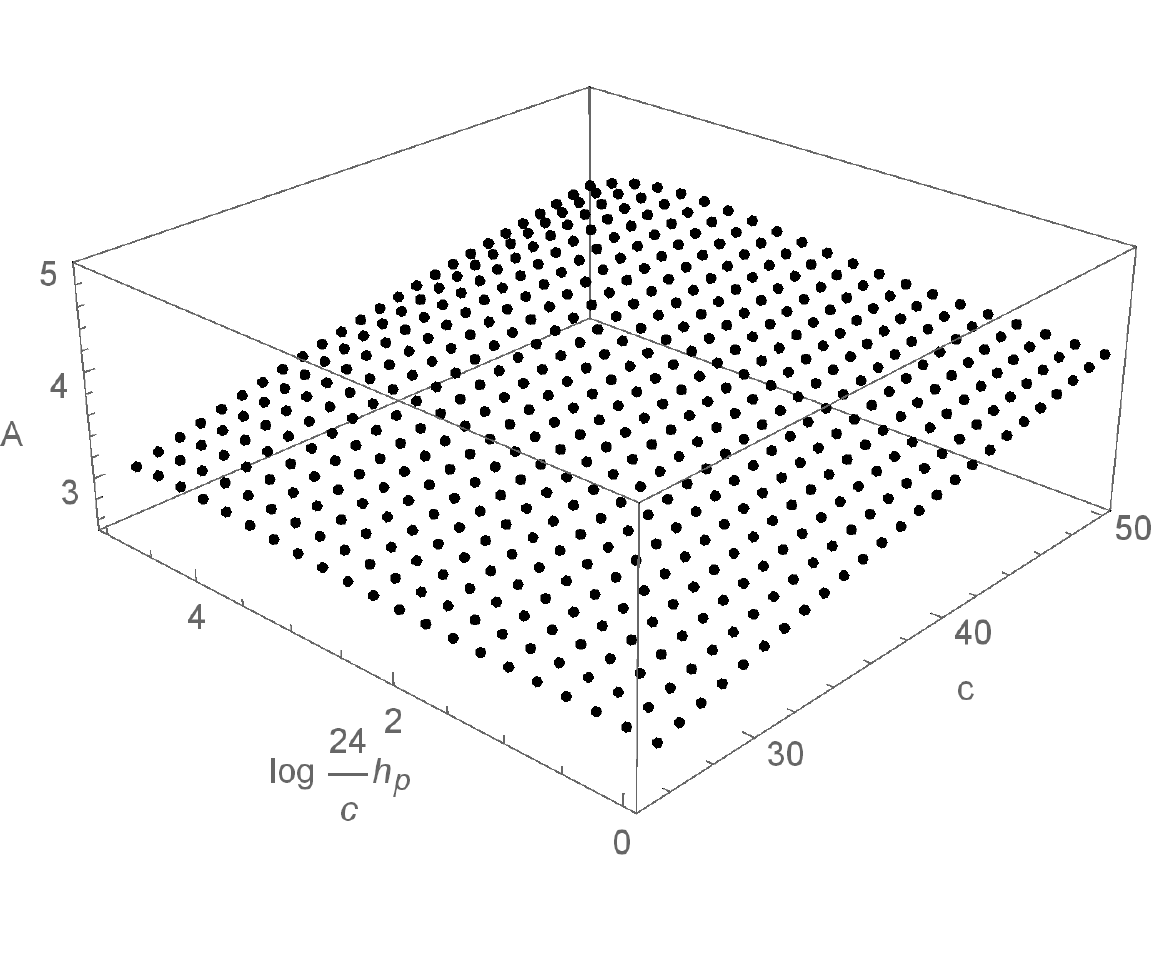}
   \includegraphics[width=70mm]{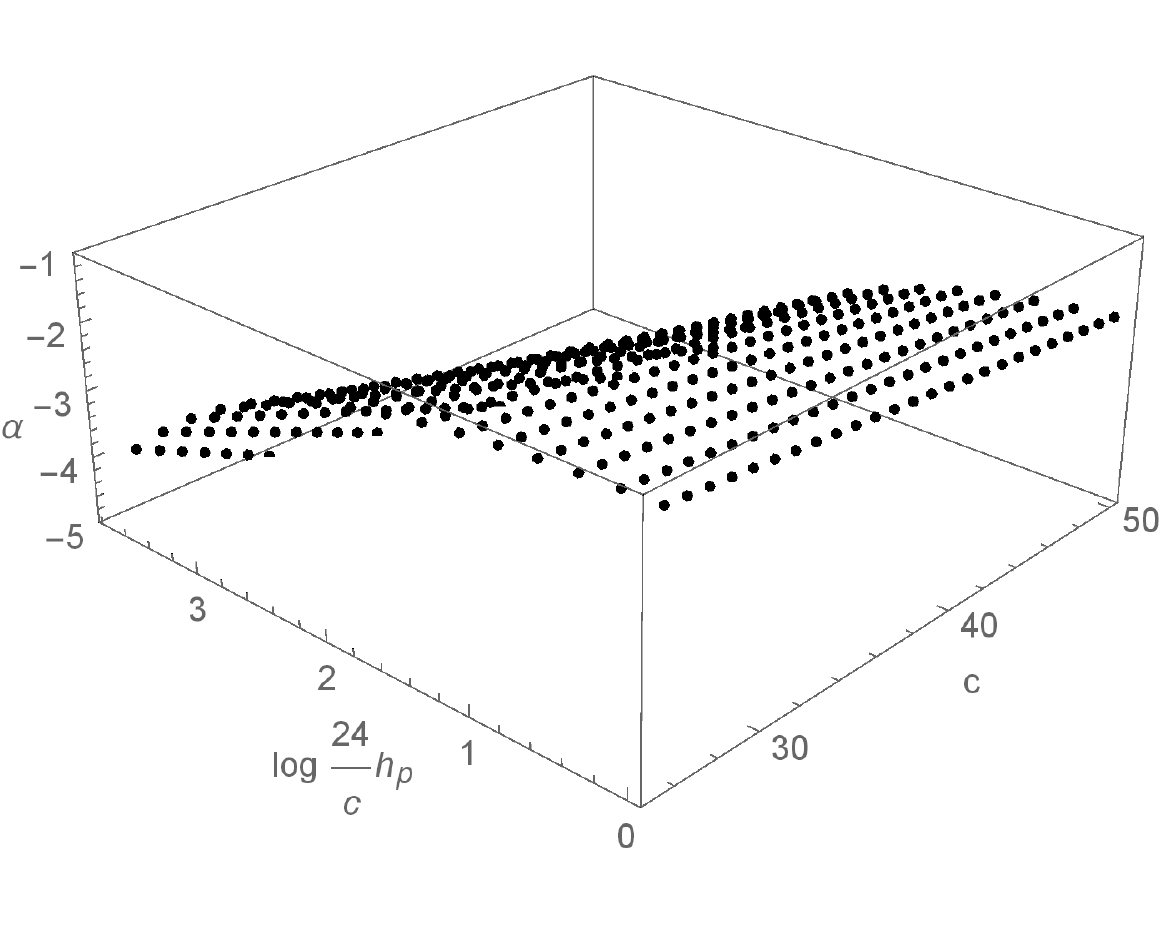}
   \end{center}
 \caption{The plot of the values of $A$ (left) and $\a$ (right) for various values of
 ($c,h_p$).}
\label{fig:chpdep}
\end{figure}

Figure \ref{fig:chpdep} shows the values of $A$ of AABB blocks for various values of ($c,h_p$)  with $h_A=\fr{c}{24}$ and $h_B=\fr{c}{240}$. One can see that the slope of the values of $A$ vs. $c$ is independent of $h_p$, which suggests that the coefficients $c_n$ have no product term $h_p \times c$.

Note that the fits of $A$ and $\a$ by using the values $c_n$, ($n=1,2,...,N$)  are not valid for $h_p\sim N$ because it is expected that the universal behavior arises from $n \gg h_p$ as discussed in Section \ref{subsec:TOC}. The steep slope in Figure \ref{fig:chpdep} for large $h_p$ is caused by this problem, and consequently, it is meaningless. We have to see only the region $h_p \ll N$. In Figure \ref{fig:chpdep}, we set $N=300$.

\begin{figure}[h]
 \begin{minipage}{0.5\hsize}
  \begin{center}
   \includegraphics[width=80mm]{h2-1-AABBcnhp1.pdf}
  \end{center}
 \end{minipage}
 \begin{minipage}{0.5\hsize}
  \begin{center}
   \includegraphics[width=80mm]{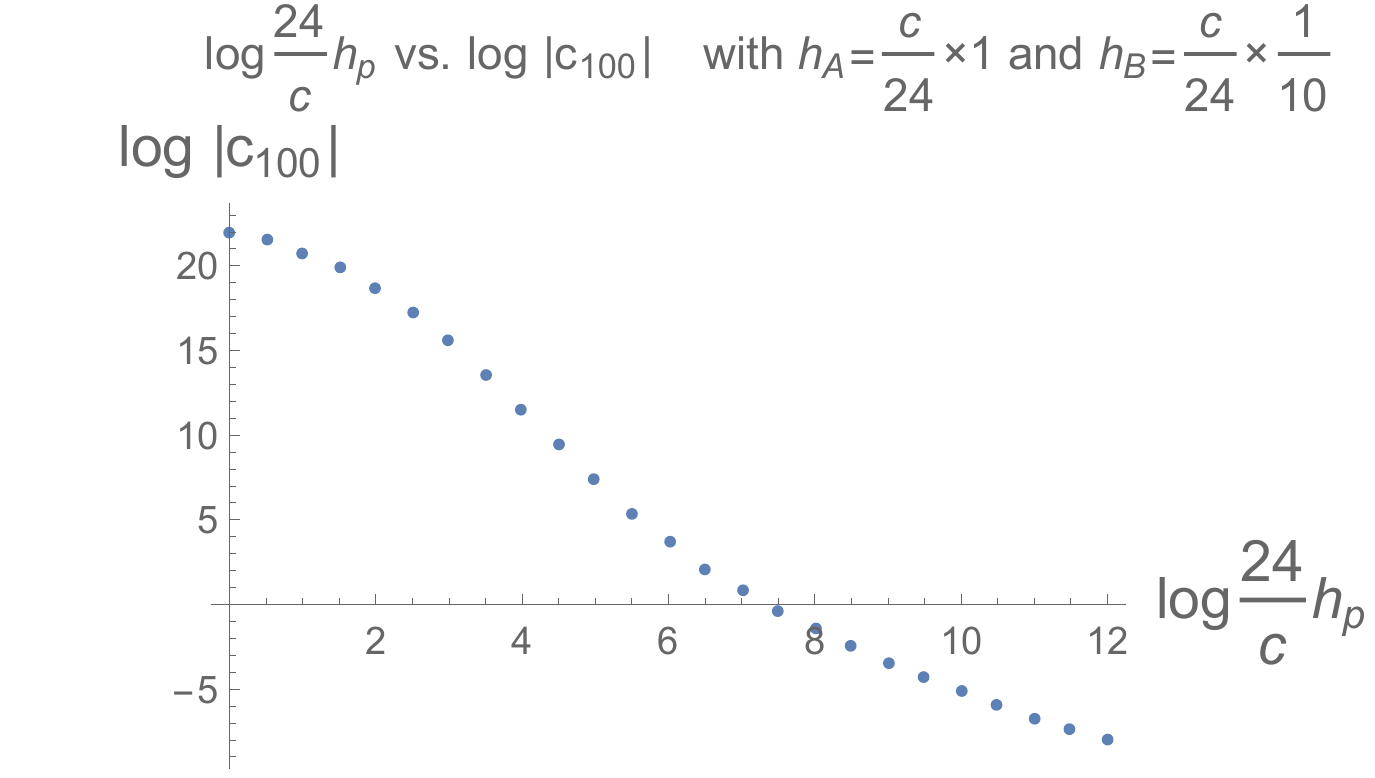}
  \end{center}
 \end{minipage}
 \begin{minipage}{0.5\hsize}
  \begin{center}
   \includegraphics[width=80mm]{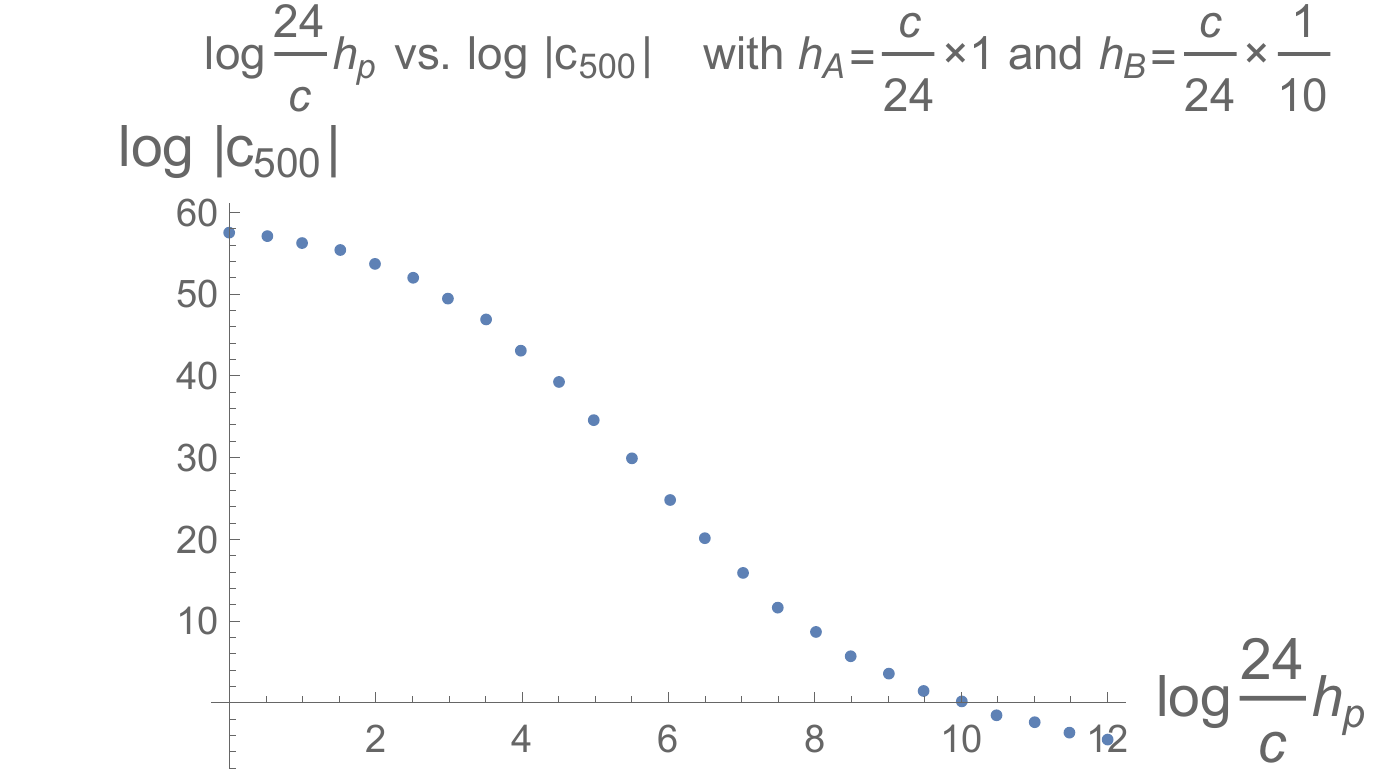}
  \end{center}
 \end{minipage}
 \begin{minipage}{0.5\hsize}
  \begin{center}
   \includegraphics[width=80mm]{h2-1-AABBcnhp4.pdf}
  \end{center}
 \end{minipage}
\caption{The $h_p$ dependence of $c_n(h_p)$ with fixed $n=10,100,500,1000$ for AABB blocks with $(h_A, h_B) = (\fr{c}{24}, \fr{c}{240})$, which is in the {\it heavy-light} region. }
\label{fig:AABBcnhp2}
\end{figure}

\begin{figure}[H]
 \begin{minipage}{0.5\hsize}
  \begin{center}
   \includegraphics[width=80mm]{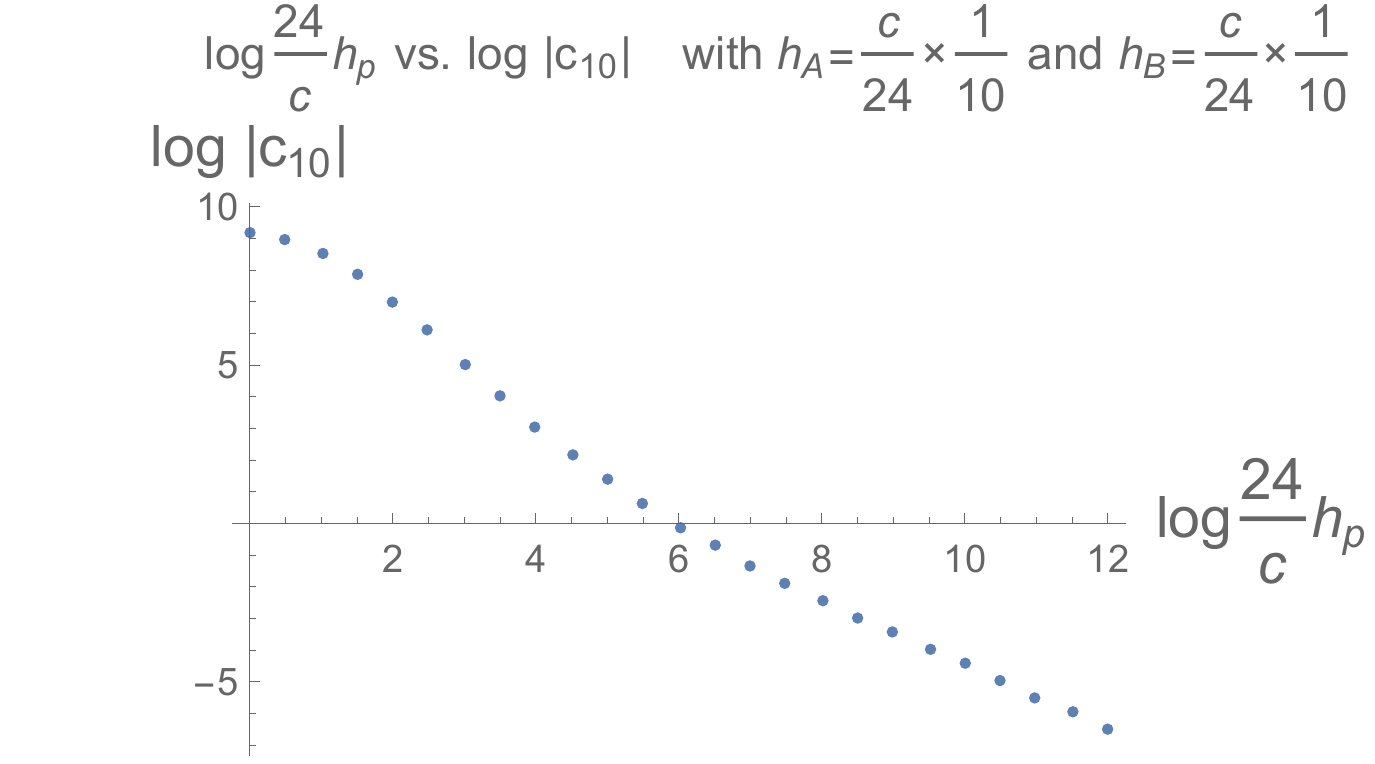}
  \end{center}
 \end{minipage}
 \begin{minipage}{0.5\hsize}
  \begin{center}
   \includegraphics[width=80mm]{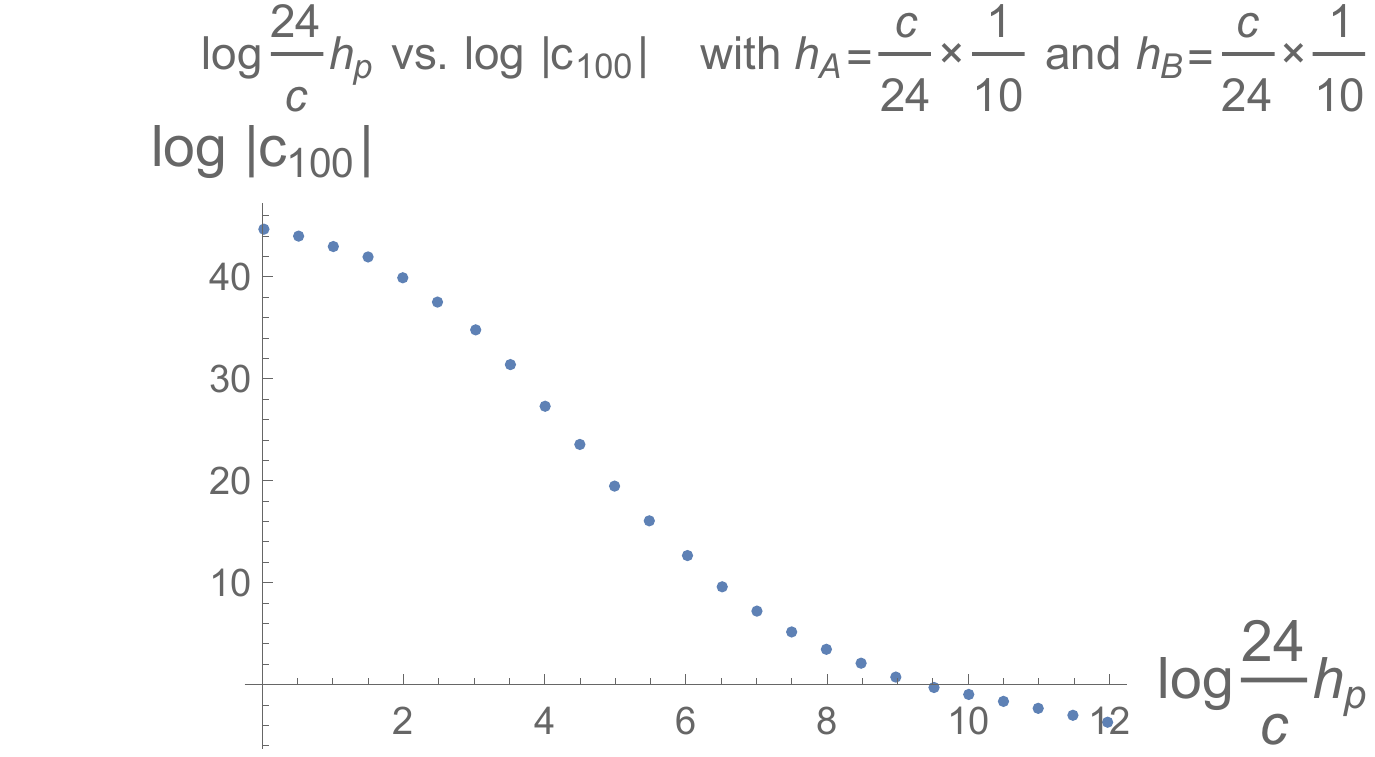}
  \end{center}
 \end{minipage}
 \begin{minipage}{0.5\hsize}
  \begin{center}
   \includegraphics[width=80mm]{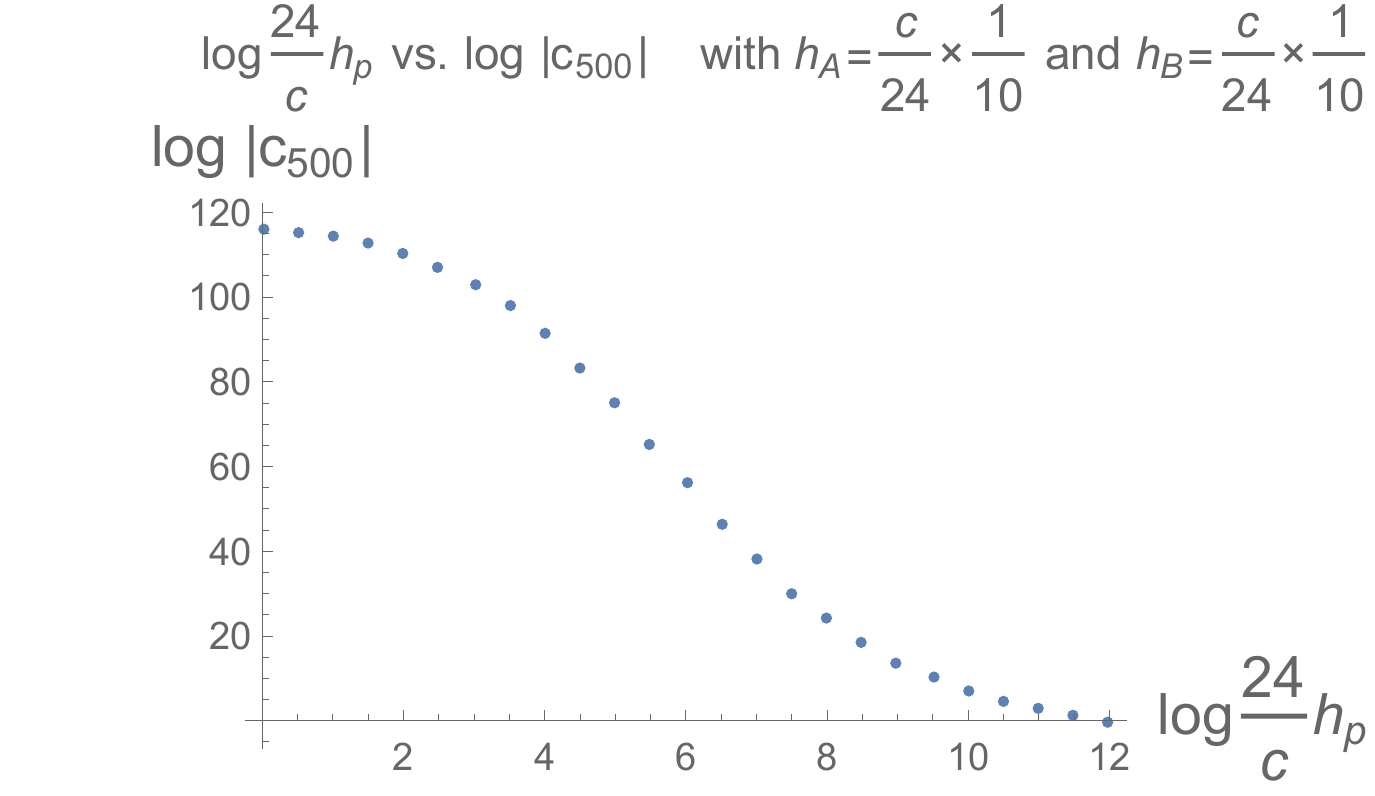}
  \end{center}
 \end{minipage}
 \begin{minipage}{0.5\hsize}
  \begin{center}
   \includegraphics[width=80mm]{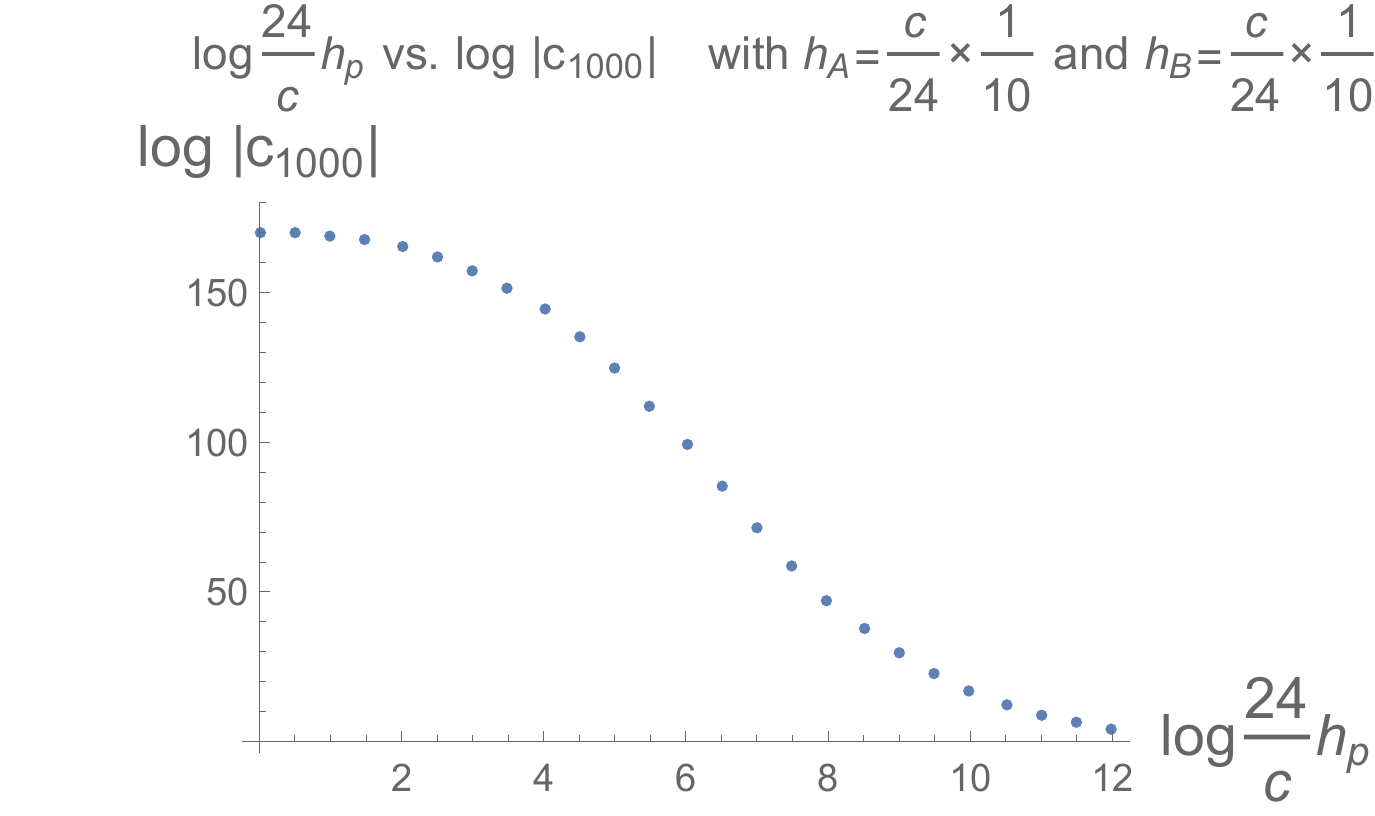}
  \end{center}
 \end{minipage}
\caption{The $h_p$ dependence of $c_n(h_p)$ with fixed $n=10,100,500,1000$ for ABBA blocks with $(h_A, h_B) = (\fr{c}{24}, \fr{c}{240})$, which is in the {\it heavy-light} region. }
\label{fig:ABBAcnhpL}
\end{figure}

\begin{figure}[H]
 \begin{minipage}{0.5\hsize}
  \begin{center}
   \includegraphics[width=80mm]{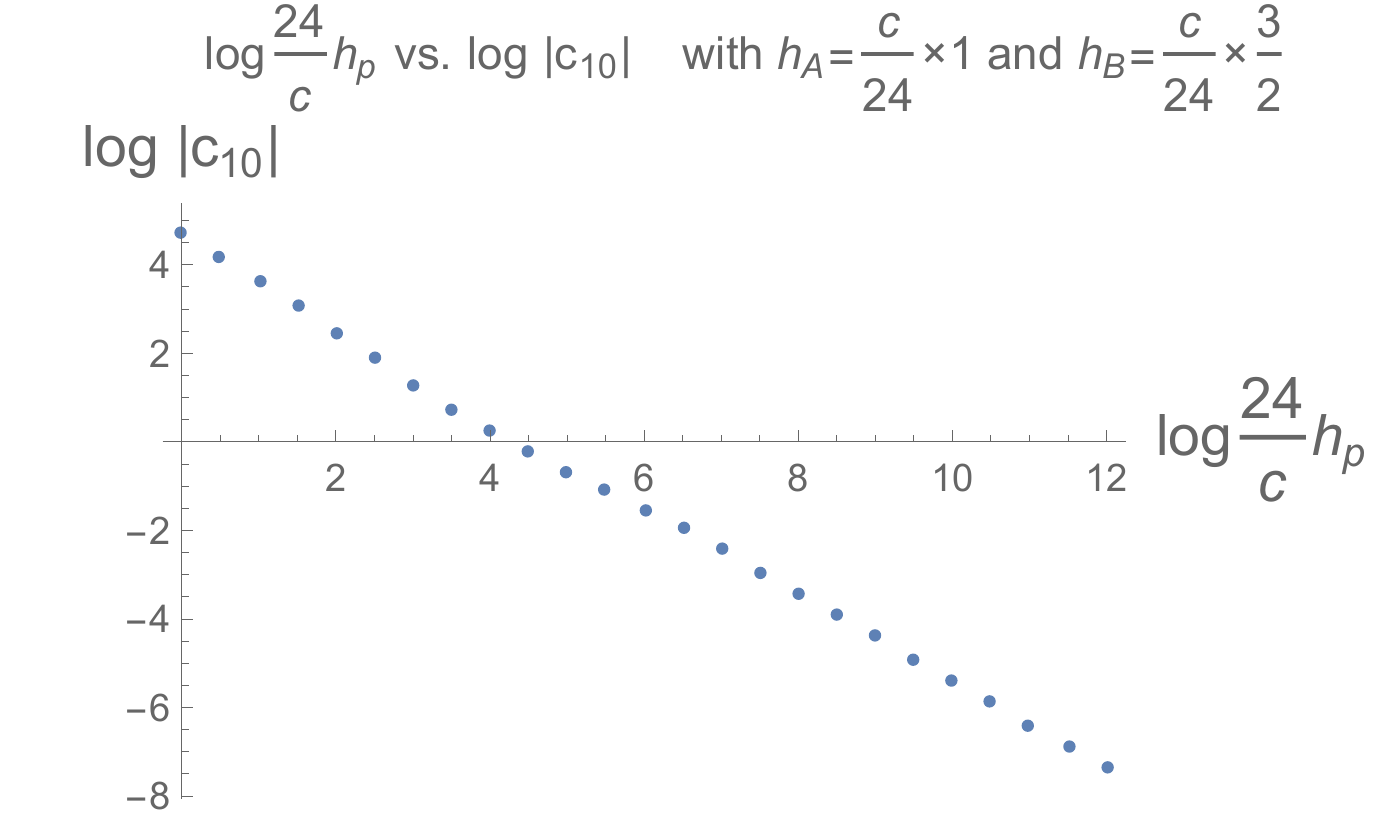}
  \end{center}
 \end{minipage}
 \begin{minipage}{0.5\hsize}
  \begin{center}
   \includegraphics[width=80mm]{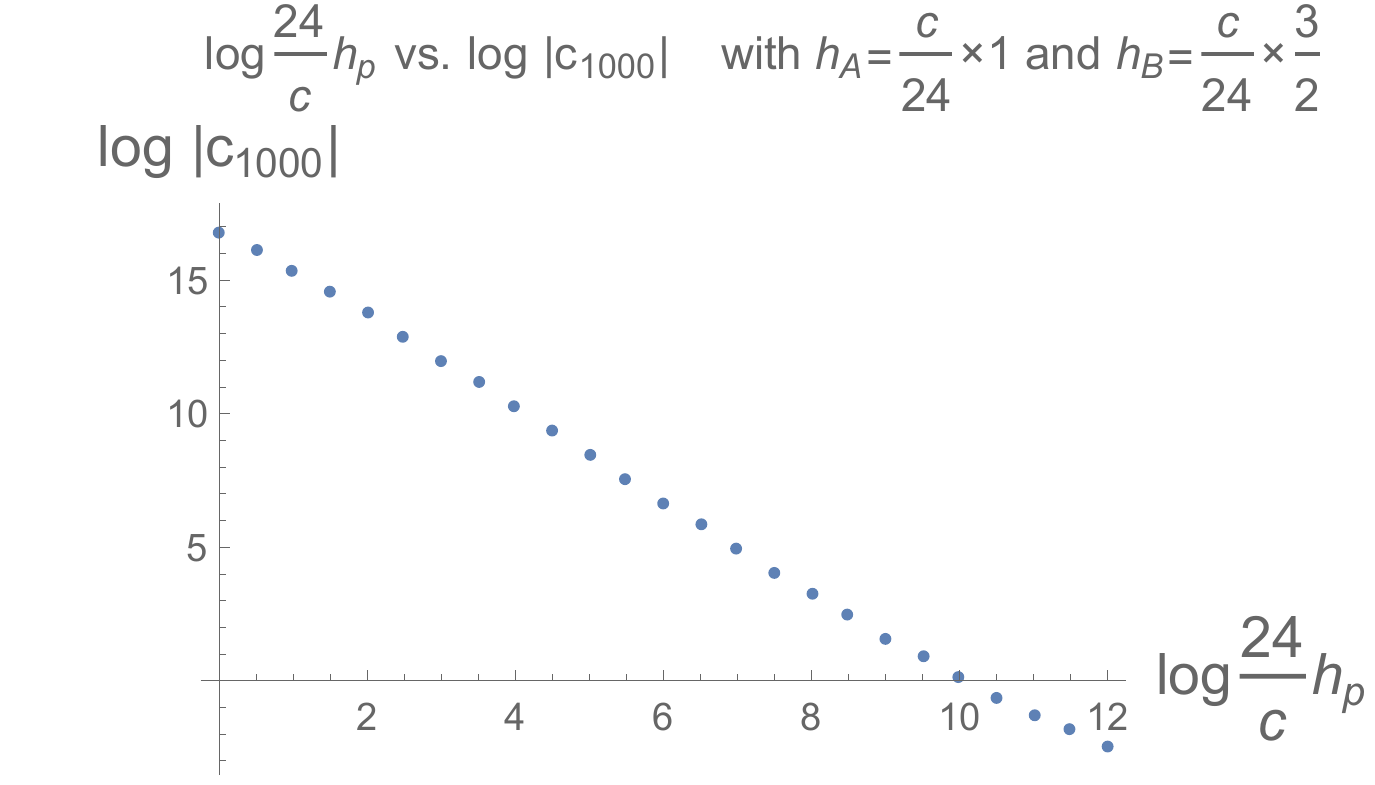}
  \end{center}
 \end{minipage}
\caption{The $h_p$ dependence of $c_n(h_p)$ with fixed $n=10$ (left) and $1000$ (right) for ABBA blocks with $(h_A, h_B) = (\fr{c}{24}, \fr{c}{16})$, which is in the {\it heavy-light} region. }
\label{fig:ABBAcnhpH}
\end{figure}

\begin{figure}[H]
  \begin{center}
   \includegraphics[width=80mm]{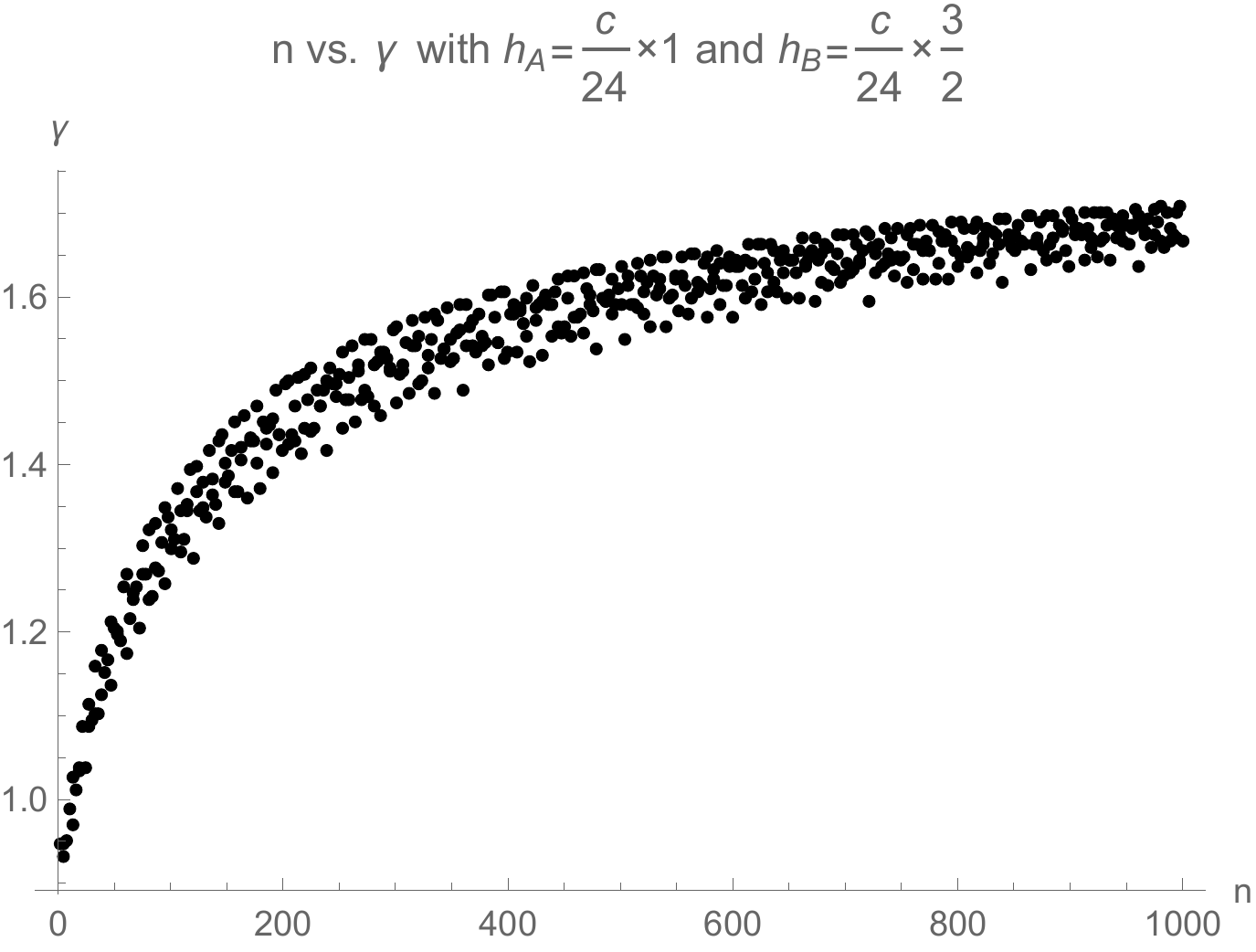}
  \end{center}
\caption{The $n$ dependence of $\g(n)$ for ABBA blocks, which is the power of (\ref{eq:veryH2}). We can see that the growth of $\g(n)$ with $n$ is slower and slower as $n$ approaches infinity.}
\label{fig:gammaABBA}
\end{figure}

Figure \ref{fig:AABBcnhp2} shows the $h_p$ dependence of $c_n(h_p)$ with fixed $n=10,100,500,1000$ for AABB blocks with $(h_A, h_B) = (\fr{c}{24}, \fr{c}{240})$, which is in the {\it heavy-light} region. One can see that the point of the transition form the gentle slope to steep slope shifts to the right as we take $n$ larger. This means that the value of $h_p$ where our Cardy-like formula breaks down  is of order $O(n)$, in that, 
\begin{equation}
\begin{aligned}
c_n(h_p)& \sim \left\{
    \begin{array}{ll}
     const. \ \   ,& \text{if } h_p \lesssim n  ,\\
     \pa{\fr{1}{h_p}}^{const.}  ,& \text{if } h_p \gtrsim n  .\\
    \end{array}
  \right.\\
\end{aligned}
\end{equation}
Moreover, this property can be also seen for ABBA blocks. It is shown in Figure \ref{fig:ABBAcnhpL}, \ref{fig:ABBAcnhpH} and \ref{fig:gammaABBA}.
We can immediately see from these figures that the relations (\ref{eq:veryH}), (\ref{eq:veryH2}) and (\ref{eq:constgamma}) hold not only for AABB blocks but also ABBA blocks.

%%%%%%%%%%%%%%%%%%%%%%%%%%%%%%%%%%%%%%%%%%%%%%%%%%%%%%%%%%%%%%%%%%%%%%%%%%%%%%%%%%%%%%%%%%%%%%
%%%%%%%%%%%%%%%%%%%%%%%%%%%%%%%%%%%%%%%%%%%%%%%%%%%%%%%%%%%%%%%%%%%%%%%%%%%%%%%%%%%%%%%%%%%%%%
\section{Comparing with the Semiclassical Limit} \label{app:comparing}
%%%%%%%%%%%%%%%%%%%%%%%%%%%%%%%%%%%%%%%%%%%%%%%%%%%%%%%%%%%%%%%%%%%%%%%%%%%%%%%%%%%%%%%%%%%%%%
%%%%%%%%%%%%%%%%%%%%%%%%%%%%%%%%%%%%%%%%%%%%%%%%%%%%%%%%%%%%%%%%%%%%%%%%%%%%%%%%%%%%%%%%%%%%%%

%%%%%%%%%%%%%%%%%%%%%%%%%%%%%%%%%%%%%%%%%%%
%%%%%%%%%%%%%%%%%%%%%%%%%%%%%%%%%%%%%%%%%%%
\subsection{Heavy-Light Limit}\label{subsec:HHLL}
%%%%%%%%%%%%%%%%%%%%%%%%%%%%%%%%%%%%%%%%%%%
%%%%%%%%%%%%%%%%%%%%%%%%%%%%%%%%%%%%%%%%%%%

In the region $h_A>\fr{c}{32} h_B<\fr{c}{32}$, AABB blocks have the sign pattern
\begin{equation}
\text{Sign} (c_{2n}) = (-1)^n,
\end{equation}
therefore, the function $H(h_p|q)$ is described by
\begin{equation}\label{eq:HHLLsum}
\sum_{n=0}^{\infty} (-1)^{n} (2n)^\a \ex{A\s{2n}}q^{2n}.
\end{equation}
Let us consider the limit $q \to i$, which is corresponding to the limit $z \to 0$ after the transformation $(1-z) \to \ex{-2\pi i} (1-z)$.
We know the expression for the large $c$ block in the heavy light limit,
\begin{equation}{eq:HHLLexpression}
\ca{F}^{HH}_{LL}(h_p|z) = (1-z)^{h_L(\a-1)}\pa{\fr{1-(1-z)^\a}{\a}}^{h_p-2h_L} ~_2F_1(h_p,h_p,2h_p|1-(1-z)^a),
\end{equation}
where $\a=\s{1-\fr{24}{c}h_H}$. In the limit $q \to i$, the asymptotic behavior is
\begin{equation}
\ca{F}^{HH}_{LL,mono}(h_p|z) \ar{z \to 0} O(z^0),
\end{equation}
which leads to 
\begin{equation}\label{eq:appHHLL}
H^{HH}_{LL,mono}(h_p|q) \ar{z \to 0} z^{-\fr{c-1}{24}+2h_L} \pa{\log z}^{-\fr{c-1}{4}+4(h_H+h_L)}.
\end{equation}
And also in the limit $q \to i$, we can approximate the sum (\ref{eq:HHLLsum}) as
\begin{equation}\label{eq:appsum2}
z^{-\fr{A^2}{\pi^2}} (\log z)^{\fr{3}{2}+2 \a},
\end{equation}
where we used the asymptotic behavior of $\tau$,
\begin{equation}
\tau_{mono}(z)\ \ar{z \to 0} \fr{1}{2}\pa{1-\fr{\pi i}{2} \fr{1}{\log \fr{z}{16}}}.
\end{equation}
Substituting our numerical result  (\ref{eq:AdepHL}), that is
\begin{equation}\label{eq:appsum3}
\begin{aligned}
A&=\pi\s{\fr{c-1}{24}-2h_L},\\
\a&=2(h_H+h_L)-\fr{c+5}{8},
\end{aligned}
\end{equation}
into the sum approximation (\ref{eq:appsum2}), then we can exactly reproduce the asymptotic behavior of the HHLL block (\ref{eq:appHHLL}).

We can also consider the limit $q \to 1$, which is corresponding to $z \to 1$. In this case, one can naively expect that the summation (\ref{eq:HHLLsum}) could be approximated by
\begin{equation}\label{eq:alter}
\begin{aligned}
\sum_{n=0}^{\infty} (-1)^{n} (2n)^\a \ex{A\s{2n}}q^{2n}
&=\sum_{k=0}^{\infty}(4k)^\a \ex{A\s{4k}}q^{4k}-\sum_{k=0}^{\infty}(4k+2)^\a \ex{A\s{4k+2}}q^{4k+2}\\
&\simeq \sum_{k=0}^{\infty}(4k)^\a \ex{A\s{4k}}q^{4k}\pa{1-\pa{1+\fr{A}{\s{4k}}}q^2}\\
&\ar{z \to 1}  (1-z)^{-\fr{A^2}{4\pi^2}}.
\end{aligned}
\end{equation}
If this is correct, then  substituting the value of A in (\ref{eq:appsum3}) leads to
\begin{equation}
\ar{z \to 1} (1-z)^{-\fr{1}{4}\s{\fr{c-1}{24}-2h_L}}.
\end{equation}
However, this is different from the behavior from the HHLL block (\ref{eq:HHLLexpression}) in the limit $z \to 1$ as
\begin{equation}
H^{HH}_{LL}(h_p|q) \ar{z \to 1} (1-z)^{-\fr{c-1}{24}+h_H+h_L\s{1-\fr{24}{c}h_H}}.
\end{equation}
Actually, it is not allowed to approximate the alternating series by the saddle point method as (\ref{eq:alter}). One can see this breakdown easily , for example, if one tries to approximate the following alternating series in the limit $x\to\infty$,
\begin{equation}
\sum_n \fr{(-x)^n}{n!}=\sum_k \fr{x^{2k}}{2k!}\pa{1-\fr{x}{2k+1}}.
\end{equation}
We know that the exact behavior of this series as $\simeq \ex{-x}$, however  if one approximates the alternating series by the saddle point approximation, then one gets a wrong behavior $\simeq \ex{x}$.
\footnote{
If the signs are not alternating, we can get the correct answer by using the saddle point approximation, in that,
\begin{equation}
\sum_n \fr{x^n}{n!} \simeq \int \dd n \ \ \ex{n \log x- n \log n + n} \simeq \ex{x}.
\end{equation}
}
In fact, this is obvious because when the alternating series is split into two parts as (\ref{eq:alter}) and approximated by the saddle point approximation, only the dominant contributions for each of two parts are extracted and other contributions are removed, but the dominant contributions cancel each other and the correct asymptotic behavior consists of the subleading contributions, rather than the dominant contributions. Therefore, the saddle point approximation for each of the two terms leads to the wrong asymptotic behavior.
We hope to know how to obtain the correct asymptotic behavior of (\ref{eq:alter}), in other words, how to evaluate an alternating series by approximation.

%%%%%%%%%%%%%%%%%%%%%%%%%%%%%%%%%%%%%%%%%%%
%%%%%%%%%%%%%%%%%%%%%%%%%%%%%%%%%%%%%%%%%%%
\subsection{Light-Light Region}\label{subsec:LLLL}
%%%%%%%%%%%%%%%%%%%%%%%%%%%%%%%%%%%%%%%%%%%
%%%%%%%%%%%%%%%%%%%%%%%%%%%%%%%%%%%%%%%%%%%

In a pat of the {\it light-light} region (displayed by the blue region in Figure \ref{fig:HHLL-hAhBdep}) , the semiclassical blocks in the limit $\e\ll1$ can be obtained by using the monodromy method near $z=1$ \cite{Fitzpatrick2017}, which is given by
\begin{equation}\label{eq:exactHHLL}
\ca{F}^{AA}_{BB}(h_p|1-\e)\sim \e^{-\fr{c}{12}\pa{1-\s{1-\fr{24h_A}{c}}}\pa{1-\s{1-\fr{24h_B}{c}}}}
\end{equation}
and therefore,
\begin{equation}
H^{AA}_{BB}(h_p|1-\e)\sim \e^{-\fr{c-1}{24}+h_A+h_B-\fr{c}{12}\pa{1-\s{1-\fr{24h_A}{c}}}\pa{1-\s{1-\fr{24h_B}{c}}}}.
\end{equation}
Comparing this with (\ref{eq:A=!0}), we obtain the theoretical value of $A$ as
\footnote{
We are very much grateful to Henry Maxfield for pointing out this to us
}
\begin{equation}
A=2\pi \s{\fr{c-1}{24}-h_A-h_B+\fr{c}{12}\pa{1-\s{1-\fr{24h_A}{c}}}\pa{1-\s{1-\fr{24h_B}{c}}}}.
\end{equation}
The block (\ref{eq:exactHHLL}) is derived under the limit $c \to \infty$. It's natural that the exact expression can be obtained by a shift of $c \to c-1$. Therefore, we expect that the explicit $A$ is given by
\begin{equation}
A=2\pi \s{\fr{c-1}{24}-h_A-h_B+\fr{c-1}{12}\pa{1-\s{1-\fr{24h_A}{c-1}}}\pa{1-\s{1-\fr{24h_B}{c-1}}}}.
\end{equation}
This value is perfectly match our numerical computations. Note that if expanding $A$ at small $\fr{h_B}{c}$, we obtain the heavy-light limit of $A$ by
\begin{equation}
A=2\pi \s{\fr{c-1}{24}-h_A-h_B\s{1-\fr{24h_A}{c-1}}},
\end{equation}
which has been derived by the heavy-light blocks in our previous paper \cite{Kusuki2018}.

%%%%%%%%%%%%%%%%%%%%%%%%%%%%%%%%%%%%%%%%%%%%%%%%%%%%%%%%%%%%%%%%%%%%%%%%%%%%%%%%%%%%%%%%%%%%%%
%%%%%%%%%%%%%%%%%%%%%%%%%%%%%%%%%%%%%%%%%%%%%%%%%%%%%%%%%%%%%%%%%%%%%%%%%%%%%%%%%%%%%%%%%%%%%%
\section{The Asymptotics of Heavy-Light-Light Coefficients} \label{sec:descendants}
%%%%%%%%%%%%%%%%%%%%%%%%%%%%%%%%%%%%%%%%%%%%%%%%%%%%%%%%%%%%%%%%%%%%%%%%%%%%%%%%%%%%%%%%%%%%%%
%%%%%%%%%%%%%%%%%%%%%%%%%%%%%%%%%%%%%%%%%%%%%%%%%%%%%%%%%%%%%%%%%%%%%%%%%%%%%%%%%%%%%%%%%%%%%%
In this section, we will show the heavy-light-light three point function for arbitrary states.
A four point functions can be expanded by
\footnote{
Here, the convention is different from the usual $2d$ CFT convention. We consider normalized eigenstates of $L_0$ as in the former of \cite{Kraus2016}.
}
\begin{equation}
\braket{O(\infty)O(1)O(x)O(0)}=  x^{-2\D_O} \sum_p C_{OOp}^2 x^{\D_p},
\end{equation}
where we set $z=\bar{z}=x$ and the sum is over all operators in the CFT. This expansion leads to the bootstrap equation in the limit $x \to 1$,
\begin{equation}
 \sum_p C_{OOp}^2 x^{\D_p} \sim \pa{1-x}^{-2\D_O}.
\end{equation}
From this equation, we can immediately obtain
\begin{equation}
 \sum_{\D_p \text{ fixed}} C_{OOp}^2\ar{\D_p \to \infty}\pa{\D_p}^{2\D_O-1}
\end{equation}
by using the inverse Laplace transformation. This means that the heavy-light-light three point function for arbitrary states is given by
\begin{equation}
\s{\overline{C_{OOp}^2}} \ar{\D_p \to \infty} \ex{-\fr{S(\D_p)}{2}}.
\end{equation}
The result (\ref{eq:HLLcoef}) for only primaries does not  reduce to the above three point function even if $c \to \infty$. However, it's natural because for general four point conformal blocks, we can not neglect descendants even if $c \to \infty$.

\clearpage
\bibliographystyle{JHEP}
\bibliography{ABAB}

\end{document}